\documentstyle[12pt,twoside,fancyhdr]{report}

\renewcommand{\chaptername}{Cap{\'\i}tulo}

\renewcommand{\baselinestretch}{1.50}
\renewcommand{\theequation}{\arabic{chapter}.\arabic{equation}}

\newlength{\captsize}       \let\captsize=\normalsize
\newlength{\captwidth}      
\newlength{\beforetableskip}    
\newcommand{\capt}[1]{
    \renewcommand{\baselinestretch}{0.9}
        \begin{minipage}{\captwidth}
    \let\normalsize=\captsize
    \caption[#1]{\sf #1}
    \end{minipage}\\ \vspace{\beforetableskip}}

\makeatletter
    \long\def\@makecaption#1#2{\vskip 10\p@
    \setbox\@tempboxa\hbox{{\bf #1:} #2} 
    \ifdim \wd\@tempboxa >\hsize
           {\bf #1:} #2\par
        \else
           \hbox to\hsize{\box\@tempboxa\hfill}
        \fi}
\makeatother

\begin{document}



\newcommand\ds{\displaystyle}
\newcommand\pl{{\partial}^L}
\newcommand\pab{\partial\eta^{a_k}\partial\eta^{b_j}}
\newcommand\pba{\partial\eta^{b_j}\partial\eta^{a_k}}
\newcommand\oc{\hat\Omega^{c_i}}
\newcommand\ob{\hat\Omega^{b_j}}
\newcommand\oa{\hat\Omega^{a_k}}
\newcommand\pa{\partial\eta^{a_k}}
\newcommand\pb{\partial\eta^{b_j}}
\newcommand\pc{\partial\eta^{c_i}}
\newcommand\Pa{\hat{\mathcal  P}_{a_k}}
\newcommand\Pb{\hat{\mathcal  P}_{b_j}}
\newcommand\Pc{\hat{\mathcal  P}_{c_i}}
\newcommand\oab{\hat\Omega^{a_kb_j}}
\newcommand\oac{\hat\Omega^{a_kc_i}}
\newcommand\ocb{\hat\Omega^{c_ib_j}}
\newcommand\obc{\hat\Omega^{b_jc_i}}
\newcommand\fm{f^{1\over 2}}
\newcommand\fmm{f^{-{1\over 2}}}



\thispagestyle{empty}
\begin{center}
\vspace{-4.5cm} {\huge Universidad de Buenos Aires} \\
 \vspace{0.3cm} {\Large Facultad de Ciencias Exactas y Naturales} \\
 \vspace{0.3cm} {\Large Departamento de F{\'\i}sica} \\
 \vspace{3.0cm} {\Huge\bf Cuantificaci\'on de Sistemas} \\
 \vspace{0.5cm} {\Huge\bf con Covariancia General} \\
 \vspace{3.0cm} {\Large Autor: {\bf Daniel Marcelo Sforza}} \\
 \vspace{0.7cm} {\Large Director: {\bf Dr. Rafael Ferraro}}\\
 \vspace{3.0cm} \normalsize{Lugar de trabajo} \\
 \normalsize{Grupo de Teor\'{\i}as Cu\'{a}nticas Relativistas y
 Gravitaci\'{o}n}\\
 \normalsize{Instituto de Astronom\'{\i}a y F\'{\i}sica del Espacio} \\
 \normalsize{Casilla de Correo 67, Sucursal 28, 1428 Buenos Aires, Argentina.}

 \vspace{0.5cm} {\normalsize Trabajo de Tesis para optar por el t\'{\i}tulo de
 Doctor en Ciencias F\'{\i}sicas} \\
 \vspace{0.3cm} {\normalsize Octubre 2000}

\end{center}

\newpage

\thispagestyle{empty} ~
\newpage


\thispagestyle{empty}

\rightline{\it A Eliana.}

\newpage

\thispagestyle{empty}

~\newpage


\thispagestyle{empty}

\vspace{-5.5cm}

\noindent {\Huge\bf Resumen}

\vspace{0.1cm}

En esta tesis se estudia la cuantificaci\'{o}n de sistemas con covariancia
general. En el marco de los formalismos BRST y de cuantificaci\'{o}n
can\'{o}nica de Dirac, se estudian modelos de dimensi\'{o}n finita que emulan
la estructura de v\'{\i}nculos de la Relatividad General.

Se comienza con el estudio de un sistema sujeto a un v\'{\i}nculo
cuadr\'{a}tico en los momentos y un conjunto de v\'{\i}nculos lineales en los
momentos (correspondientes respectivamente a los v\'{\i}nculos
``super-Hamiltoniano" y de ``supermomentos"  de Relatividad General).  El
punto de partida es discernir que las contribuciones no herm\'{\i}ticas de los
fantasmas a los v\'{\i}nculos de supermomento pueden leerse en t\'{e}rminos
del volumen natural inducido por los v\'{\i}nculos sobre las \'{o}rbitas. Este
volumen juega luego un papel fundamental en la cons-\break trucci\'{o}n del
sector cuadr\'{a}tico de la carga BRST nilpotente. A nivel cu\'{a}ntico, la
teor\'{\i}a permanece invariante ante transformaciones de escala del
v\'{\i}nculo super-Hamiltoniano. En el caso en que el sistema posee un tiempo
intr\'{\i}nseco, esta propiedad se traduce en una contribuci\'{o}n del
potencial al t\'{e}rmino cin\'{e}tico. En este aspecto el resultado difiere
sustancialmente del tratamiento usual, donde la invariancia ante
transformaciones de escala se fuerza con la introducci\'{o}n de un
acoplamiento con la curvatura. Dicha contribuci\'{o}n lejos de ser
antinatural, se justifica elegantemente a la luz del principio variacional de
Jacobi.

Luego, el tratamiento se extiende al caso de sistemas con tiempo
extr\'{\i}nseco. En este caso, dado que la m\'{e}trica posea un vector de
Killing temporal conforme y el potencial se comporte de manera adecuada
respecto al mismo, el rol jugado por el potencial en el caso con tiempo
intr\'{\i}nseco aqu\'{\i} es tomado por el m\'{o}dulo del vector de Killing de
la teor\'{\i}a.

Finalmente, los resultados obtenidos se extienden para un sistema con dos
v\'{\i}nculos super-Hamiltonianos. Este paso es sumamente importante ya que la
Relatividad General posee una infinidad de tales v\'{\i}nculos, con un
\'{a}lgebra entre ellos no trivial.

\noindent {\bf Palabras claves:} {\it Sistemas hamiltonianos con v\'{\i}nculos
- Covariancia general - Relatividad general - Cosmolog\'{\i}a cu\'{a}ntica -
Formalismo BRST.}


\newpage

\thispagestyle{empty} ~
\newpage


\thispagestyle{empty}

\vspace{-5.5cm}

\noindent {\Huge\bf Abstract}

\vspace{0.1cm}

In this Thesis it is studied the quantization of generally covariant systems.
Finite dimensional models which mimic the constraint structure of Einstein's
General Relativity theory are studied in the framework of BRST formalism and
Dirac's canonical quantization.

First, it is studied a system featuring a quadratic constraint in the momenta
and a set of linear constraints in the momenta (the ``super-Hamiltonian" and
``supermomenta" constraints of General Relativity respectively). The starting
point is to realize  that the ghost contributions to the supermomenta
constraint operators can be read in terms of the natural volume induced by the
constraints in the orbits. This volume plays a fundamental role in the
construction of the quadratic sector of the nilpotent BRST charge. It is shown
that the quantum theory is invariant under scaling of the super-Hamiltonian
constraint.  As long as the system has an intrinsic time, this property
translates in a contribution of the potential to the kinetic term. In this
aspect, the results substantially differ from other works where the scaling
invariance is forced by introducing a coupling to the curvature. The
contribution of the potential, far from being unnatural, is beautifully
justified in light of the Jacobi's principle.

Then, it is shown that the obtained results can be extended to systems with
extrinsic time. In this case, if the metric has a conformal temporal Killing
vector and the potential exhibits a suitable behavior with respect to it, the
role played by the potential in the case of intrinsic time is now played by
the norm of the Killing vector.

Finally, the results for the previous cases are extended to a system featuring
two super-Hamiltonian constraints. This step is extremely important due to the
fact that General Relativity features an infinite number of such constraints
satisfying a non trivial algebra among themselves.


\noindent {\bf Keywords:} {\it Hamiltonian constrained systems - General
covariance - General relativity - Quantum cosmology - BRST formalism.}

\newpage

\thispagestyle{empty} ~
\newpage


\pagenumbering{roman} \fancyhead[LE]{\thepage} \fancyhead[RE]{\sl \'{I}ndice}
\fancyhead[RO]{\bf \thepage} \fancyhead[LO]{} \fancyfoot[CE,CO]{}

\tableofcontents

\clearpage

\thispagestyle{empty}
 ~\newpage           


\pagenumbering{arabic}

\fancyhead{} \fancyhead[LE]{\bf \thepage} \fancyhead[RE]{\sl Introducci\'{o}n}
\fancyhead[RO]{\bf \thepage}\fancyhead[LO]{}

\addcontentsline{toc}{chapter}{Introducci\'{o}n}

\noindent \Huge{\bf Introducci\'{o}n}

\thispagestyle{empty}

\hyphenation{in-va-rian-tes coor-de-na-das va-ria-bles ve-re-mos re-fe-ren-cia
ha-mil-to-nia-nos ne-ce-sa-ria-men-te de-pen-dien-te de-no-mi-na par-ti-cu-lar
o-pues-ta pro-pues-to Ha-lla-re-mos mo-de-los e-sen-cial Ha-mil-to-nia-no
va-ria-ble pro-pie-da-des pro-pues-tos ge-ne-ra-li-zar o-pe-ra-do-res}

\bigskip

\noindent \Large{\bf Antecedentes existentes sobre el tema}

\medskip

\normalsize

En F\'\i sica relativista se trabaja a menudo con sistemas descriptos por un
n\'umero de variables que excede el n\'umero de grados de libertad del
sistema. Los casos m\'as familiares son la part\'\i cula relativista (tres
grados de libertad), descripta por cuatro coordenadas $x^{\mu}$, y el campo
electromagn\'etico (dos grados de libertad en cada punto del espacio),
descripto por un campo cuadrivectorial $A_{\mu}(x)$. De esta manera es posible
mantener expl\'\i cita la covariancia de la teor\'\i a. La presencia de
variables espurias, que no corresponden a aut\'enticos grados de libertad,
implica que un mismo estado f\'\i sico puede ser descripto por distintos
conjuntos de variables, lo cual se refleja en la invariancia de la funcional
acci\'on ante un grupo de transformaciones usualmente llamadas ``de gauge''.
Debido al exceso de variables, las relaciones $p=p(q,\dot{q})$ no pueden ser
todas invertidas para expresar las velocidades en funci\'on de las coordenadas
y los momentos, dando lugar a relaciones de v\'\i nculo entre las variables
can\'onicas. Estos v\'\i nculos son precisamente los generadores de las
transformaciones de simetr\'\i as {\it locales} ante las cuales es invariante
la acci\'on (invariancia de gauge).

Al cuantificar este tipo de sistemas, resultar\'{\i}a  natural proceder a
identificar  los grados genuinos  de  libertad  y  cuantificar la teor\'\i a
en ese espacio de fases reducido (m\'etodo del espacio de fases reducido). Sin
embargo, este procedimiento puede conducir a la p\'erdida de la covariancia y
la localidad expl\'\i citas de la teor\'\i a [\cite{dirac2,lusanna3}]. En
muchos casos ni siquiera es posible  tal identificaci\'on  o \'{e}sta se torna
inmanejable. Debido a estos inconvenientes, suele tomarse alguno de los
siguientes caminos alternativos:

I)  El  m\'etodo  de  \cite{dirac}  que  considera  a  todas  las variables
din\'amicas (las invariantes de gauge y las que no lo son) y las convierte en
operadores lineales que act\'uan en cierto espacio de estados, y selecciona
los estados  f\'\i sicos  por  medio de  una  condici\'on  auxiliar (deben ser
aniquilados por los operadores de v\'\i nculo).

II) El formalismo BRST [\cite{brs,t,h85,ht92}] que extiende el espacio de
fases mediante el agregado de variables fermi\'onicas (fantasmas) a  nivel  de
la teor\'\i a cl\'asica, incrementando as\'\i ~a\'un m\'as la redundancia en
la descripci\'on del sistema. El nuevo sistema no tiene v\'\i nculos  y la
acci\'on extendida se construye de tal manera que ahora posee una simetr\'\i a
{\it  global}  (en general, asociada a la conservaci\'on de cierta cantidad,
la carga BRST en este caso) en lugar de la simetr\'{\i}a local del sistema
original. A nivel cu\'antico,  los estados f\'\i sicos son seleccionados,
nuevamente, por medio de una condici\'on auxiliar (deben ser aniquilados por
el operador BRST, realizaci\'on herm\'\i tica de la carga BRST).

Es f\'acil verificar  que los tres caminos indicados son equivalentes en el
caso de v\'\i nculos sencillos.  La cuesti\'on de su equivalencia para
sistemas arbitrarios  es  m\'as sutil, y recientemente ha generado un
creciente inter\'es [\cite{k86,kt,mp89,hk90,b90,kuns92,bk93,b93,fhp93}]. En
particular, en publicaciones recientes [\cite{kuns92,fhp93}]  se ha estudiado
la aplicaci\'{o}n de estos m\'{e}todos a la cuantificaci\'on de sistemas con
v\'\i nculos que son lineales y homog\'eneos en los momentos. Si  bien  este
caso  incluye numerosos ejemplos de inter\'es f\'\i  sico (teor\'\i  as de
Yang-Mills, campos de p-formas, etc.), no incluye el  caso  de  la
gravitaci\'on pues \'esta posee adem\'as un v\'\i nculo cuadr\'atico en los
momentos.

El campo gravitatorio es un sistema covariante ante transformaciones generales
de coordenadas ({\it covariancia general}) y su Hamiltoniano se  anula sobre
la superficie de v\'\i  nculo.  Esta es una caracter\'\i stica t\'\i pica de
los sistemas que poseen invariancia  ante  reparametrizaciones  (sistemas
``parametrizados''),  esto significa que el par\'ametro de las trayectorias en
el espacio de fases no es el tiempo f\'\i sico  sino que  es una magnitud
f\'\i sicamente irrelevante. El Hamiltoniano resulta ser una combinaci\'on
lineal de cuatro v\'\i nculos (en cada punto del espacio); tres de ellos son
lineales y homog\'eneos en los momentos, y el restante es cuadr\'atico en los
momentos.

La cuantificaci\'on del sistema requiere la b\'usqueda de un ordenamiento en
los operadores de v\'\i nculo que preserve el \'algebra de los v\'\i nculos,
reteniendo as\'{\i} la calidad de primera clase que caracteriza a los
generadores de las transformaciones de gauge (ausencia de ``anomal\'{\i}as''
en el \'{a}lgebra de v\'{\i}nculos). Este es a\'un un problema abierto
[\cite{a59,s63,dw67,k73,k92}] y presenta dos aspectos a resolver: uno asociado
a la regularizaci\'on de los operadores, y el otro es el dificultoso
tratamiento que provoca la presencia de funciones de estructura  en las
relaciones de clausura. Es posible analizar este \'ultimo aspecto eludiendo la
complicaci\'{o}n del primero si se estudian modelos de dimensi\'on finita.

\bigskip
\noindent {\Large{\bf Naturaleza del aporte original realizado}}

\medskip

El inter\'es del trabajo fue extender el estudio de la cuantificaci\'on de
sistemas con covariancia general (en particular, modelos en dimensi\'on finita
que emulan la estructura de v\'{\i}nculos de la teor\'{\i}a de la Relatividad
General) en el marco del formalismo BRST y del formalismo de Dirac. Este tipo
de sistemas contienen al menos un v\'\i nculo que es cuadr\'atico en los
momentos, y hasta la fecha no exist\'{\i}an trabajos que hubiesen aplicado los
m\'etodos del formalismo BRST en tal caso y lo hayan relacionado con la
cuantificaci\'on de Dirac.

El estudio de la cuantificaci\'on de este tipo de sistemas resulta de sumo
inter\'es debido a que su estructura de v\'\i nculos y \'algebra imitan a la
de la gravedad, y por lo tanto pueden dar pautas para su posterior
aplicaci\'on en Gravedad Cu\'antica.

El modelo de  dimensi\'on  finita  m\'as simple que captura la estructura del
campo gravitatorio, es el de una part\'\i cula relativista movi\'endose en un
espacio-tiempo curvo que posee adem\'as grados de libertad espurios
[\cite{hk90}]. Dicho modelo contiene un  v\'\i  nculo cuadr\'atico en los
momentos  (el ``super-Hamiltoniano") y un conjunto finito de v\'\i nculos
lineales y homog\'eneos en los momentos (los ``supermomentos"). Se parti\'{o}
del estudio de la cuantificaci\'on de este sistema en los marcos del
formalismo BRST y de Dirac (cap\'{\i}tulo 4), y luego se extendieron los
resultados a casos a\'un m\'as generales (cap\'{\i}tulos 5 y 6).

Como la misma superficie de v\'\i nculo puede ser descripta por distintos
conjuntos de funciones de v\'\i nculos, el m\'etodo de Dirac tiene sentido
s\'olo si las funciones de onda f\'\i sicas se transforman frente a un cambio
de funciones de v\'\i nculo de tal manera que el producto interno f\'\i sico
entre las funciones de onda permanezca invariante. Estas propiedades de
transformaci\'on, conjuntamente con las transformaciones ante cambio general
de coordenadas, define el tipo de objeto geom\'etrico que debe ser la
funci\'on de onda. Este conocimiento deber\'\i  a conducir a operadores
diferenciales naturales para realizar el \'algebra de los v\'\i nculos, que
son los mismos operadores que definir\'an el conjunto de funciones de onda
f\'\i sicas.

En cambio el formalismo BRST, requiere la ampliaci\'on del espacio de fases
mediante el agregado de variables fermi\'onicas (``fantasmas'').  El objeto
central de la  teor\'\i a se construye partiendo del conjunto de funciones de
v\'\i nculo, la ``carga  BRST'', una funci\'on fermi\'onica que captura toda
la informaci\'on acerca de  la  invariancia  de gauge  del sistema original, y
que genera una simetr\'\i a  r\'\i gida  del sistema extendido asociada con la
conservaci\'on de la propia carga. Para cuantificar el sistema, la carga BRST
es promovida a un operador herm\'\i tico y  nilpotente  que  define  estados
cu\'anticos f\'\i sicos a trav\'es de su cohomolog\'\i a.

La dificultad para realizar la cuantificaci\'on de Dirac de sistemas del tipo
estudiado, consiste en obtener el ordenamiento adecuado de los operadores de
v\'\i nculo y de las funciones de estructura de manera tal que se preserve a
este nivel el \'algebra que satisfacen a nivel cl\'asico (ausencia de
anomal\'\i as). En el marco del formalismo BRST esta dificultad se traslada a
la realizaci\'on cu\'antica adecuada de la carga BRST, proceso cuyo \'exito no
est\'a garantizado (la teor\'\i a podr\'\i a presentar anomal\'\i as  en  el
\'algebra de  los v\'\i nculos) y no existe un m\'etodo general a seguir.

Sin embargo, lo que convierte al formalismo BRST en una herramienta poderosa
al cuantificar sistemas con libertad de gauge, es que todas las relaciones que
provienen del \'algebra de los v\'\i nculos quedan capturadas en la carga
BRST, por lo tanto si se logra hallar su realizaci\'on cu\'antica herm\'\i
tica y nilpotente, queda autom\'aticamente garantizado el ordenamiento
adecuado de los operadores de v\'\i nculo y sus correspondientes funciones de
estructura. Otra ventaja del formalismo consiste en la invariancia
autom\'atica de la cuantificaci\'on bajo combinaciones lineales de los v\'\i
nculos, ya que dichas combinaciones son equivalentes a cambios de coordenadas
en el sector fermi\'onico.

Aprovechando estas cualidades del formalismo, teniendo en cuenta las
transformaciones de invariancia de la teor\'\i a y considerando seriamente en
pie de igualdad a los pares de variables can\'onicas originales y
fermi\'onicas, y a partir de una reinterpretaci\'{o}n de los resultados
conocidos para un sistema con v\'\i nculos lineales solamente [\cite{fhp93}]
result\'{o} factible extenderlos al tipo de sistemas estudiados en esta tesis
(es decir, incluyendo adem\'as uno o m\'{a}s v\'\i nculos cuadr\'aticos).

Hasta la fecha, todos los modelos de dimensi\'on finita en espacio-tiempo
curvos que se hallan en la literatura consideran sistemas con un \'unico v\'\i
nculo hamiltoniano. Sin embargo, la Relatividad General tiene una infinitud de
v\'\i nculos hamiltonianos, con un \'algebra  no trivial entre ellos y los
v\'\i nculos lineales en los momentos. Result\'{o} sumamente importante pues,
poder tratar exitosamente el caso de un modelo en dimensi\'on finita pero con
m\'as de un v\'\i nculo superhamiltoniano. La generalizaci\'{o}n de los
resultados que se hab\'{\i}an obtenido para los casos con un \'{u}nico
v\'{\i}nculo Hamiltoniano result\'{o} natural (cap\'{\i}tulo 6).

Al final del trabajo, en un ap\'{e}ndice se demuestra la nilpotencia de la
carga BRST cu\'{a}ntica para los casos que exhiben s\'{o}lo un
super-Hamiltoniano. Dicha demostraci\'{o}n es central, pero debido a su
extensi\'{o}n fue inclu\'{\i}da all\'{\i} para no perder fluidez en la
lectura.

\newpage

\thispagestyle{empty}

 ~\newpage


\renewcommand{\chaptermark}[1]{\markboth{\sl #1}{}}
\renewcommand{\sectionmark}[1]{\markright{\sl \thesection . \hspace{0.05cm}
#1}}

\fancyhead[LE,RO]{\bf \thepage} \fancyhead[RE]{\leftmark}
\fancyhead[LO]{\rightmark}

\chapter{Sistemas Hamiltonianos con V\'{\i}nculos}

\bigskip

{\it ~~~~Los sistemas con v\'{\i}nculos son sumamente frecuentes en
f\'{\i}sica. Teor\'{\i}as como el electromagnetismo de Maxwell, la
gravitaci\'{o}n de Einstein y numerosos sistemas mec\'{a}nicos que son
manifiestamente invariantes ante transformaciones de Lorentz exhiben
v\'{\i}nculos que invalidan la aplicaci\'{o}n directa del formalismo
can\'{o}nico cl\'{a}sico. Es claro tambi\'{e}n la importancia de poseer una
formulaci\'{o}n Hamiltoniana apropiada para sistemas con v\'{\i}nculos si se
desea desarrollar un procedimiento v\'{a}lido de cuantificaci\'{o}n
can\'{o}nica. El prop\'{o}sito de este cap\'{\i}tulo es introducir el
tratamiento cl\'{a}sico y cu\'{a}ntico de sistemas con v\'{\i}nculos
desarrollado inicialmente por \cite{dirac}.}

\section{Formalismo Hamiltoniano para sistemas con v\'\i nculos}
\medskip

~~~~El movimiento real de un sistema cl\'{a}sico entre dos puntos dados es
aquel que hace que la acci\'{o}n
\begin{equation}
S[q^i(t)]=\int L(q^i,\dot q^i) dt.
\end{equation}
sea estacionaria.

La soluci\'{o}n que hace estacionaria la acci\'{o}n satisface las ecuaciones
de Euler-Lagrange
\begin{equation}
{d\over dt}\left({\partial L\over\partial\dot q^i}\right) -{\partial
L\over\partial q^i}=0,~~~~~~~~~~~~i=1,...,N.
\end{equation}
Estas ecuaciones pueden ser reescritas como
\begin{equation}
\ddot  q^k  {\partial^2 L\over\partial\dot q^k\dot  q^i} ={\partial
L\over\partial q^i}-\dot q^k{\partial^2 L\over\partial  q^k\dot q^i}.
\end{equation}
Luego, puede verse de esta \'{u}ltima ecuaci\'{o}n que las aceleraciones a un
dado tiempo est\'{a}n determinadas un\'{\i}vocamente por las posiciones y
velocidades a ese tiempo si y solo si la matriz ${\partial^2 L\over
\partial\dot q^k\dot q^i}$ es inversible, es decir, si el determinante
\begin{equation}\label{det}
det\left( {\partial^2 L\over\partial\dot q^k\dot q^i} \right)
\end{equation}
no se anula.

Por otra parte, si dicho determinante se anula, las aceleraciones no
estar\'{a}n determinadas un\'{\i}vocamente por las posiciones y velocidades y
la soluci\'{o}n de las ecuaciones de movimiento puede contener funciones
arbitrarias del tiempo. A nivel Hamiltoniano, esto implicar\'{a} que las
variables can\'{o}nicas no ser\'{a}n todas independientes.

El punto de partida para el formalismo Hamiltoniano es definir, como es usual,
los momentos can\'{o}nicos
\begin{equation}
p_i={\partial L\over\partial\dot q^i}
\end{equation}
con esta definici\'on la Ec. (\ref{det}) equivale a
\begin{equation}
det\left( {\partial p_i\over\partial \dot q^k}\right),
\end{equation}
que nos dice que en el caso en que se anule, las velocidades no pueden
obtenerse de forma \'unica a partir de las coordenadas y los momentos. Es
decir, la anulaci\'on del determinante refleja la existencia de v\'\i nculos
entre los momentos [\cite{hrt,sm,ht92}]. Si el n\'{u}mero de grados de
libertad del sistema es $N$, el rango de la matriz  es $R<N$, y existe
entonces un menor principal de orden $R$. Reetiquetando apropiadamente las
coordenadas tendremos
\begin{equation}\label{2.6}
det\left({\partial^2 L\over\partial\dot q^r\dot q^s} \right) \not= 0
\end{equation}
para $r,s=1....R$.   Esto significa que las velocidades $\dot q^r$ pueden ser
despejadas en funci\'on de $q^i$, $p_s$ y $\dot q^m$ $(m=R+1....N):$
\begin{equation}\label{qr}
\dot q^r=\dot q^r (q^i,p_s,\dot q^m)
\end{equation}
Si reemplazamos las $\dot q^r$ en la definici\'on para los restantes momentos
$p_m$ tenemos
\begin{equation}\label{2.7}
p_m={\partial L\over\partial \dot q^m}= \tilde p_m (q^i,\dot q^r(q^i,p_s,\dot
q^n),\dot q^n)=p_m (q^i, p_s, \dot q^n).
\end{equation}
Ahora bien, $p_m$ no puede depender de las velocidades $\dot q^n$, pues si
as\'\i \ fuera se podr\'\i a despejar al menos una $\dot q^n$ en funci\'{o}n
de las restantes. Pero esto no puede ser, pues que el rango de la matriz sea
$R$ implica que no pueden ser despejadas m\'as velocidades. Luego, $p_m
=p_m(q^i,p_s)$, o, de manera equivalente, existen funciones $\phi_m$ tales que
\begin{equation}
\phi_m(q^i,p_i)=0.
\end{equation}
Estas ecuaciones definen $M=N-R$  {\it v\'\i nculos primarios} (supondremos
que son independientes): nos dicen  que s\'olo las cantidades $q^i$ y $p_s$
pueden elegirse arbitrariamente. Por lo tanto, los v\'{\i}nculos definen una
superficie de dimensi\'on $2N-M=N+R$ en el espacio de las  fases, que podemos
describir con las $N$ coordenadas $q^i$ y los  $R$ momentos  $p_r$. Dado un
punto  sobre esta superficie  de  v\'\i  nculo, podemos ver  que  el punto
correspondiente en el espacio  de  velocidades  no est\'a completamente
determinado:    en las $R$ ecuaciones  (\ref{qr})  las $\dot  q^m$  no est\'an
determinadas porque, para valores  dados de $q^i$ y $p_r$, no queda definido
un punto $(q^i,\dot q^i)$, sino  una  variedad  en la cual las $M$ variables
arbitrarias $\dot q^m$ juegan el papel de par\'ametros. Luego, para que la
transformaci\'{o}n entre coordenadas y velocidades, y coordenadas y momentos
sea {\it biun\'\i voca} ser\'{\i}a necesario introducir al menos $M$
par\'{a}metros para poder as\'{\i} precisar la localizaci\'{o}n de $\dot q$ en
la variedad. Estos par\'{a}metros aparecer\'{a}n como multiplicadores de
Lagrange cuando definamos el Hamiltoniano y estudiemos sus propiedades.

El pr\'{o}ximo paso en el an\'{a}lisis es entonces, introducir el
Hamiltoniano. Para un sistema sin v\'\i nculos, el principio variacional
$\delta S=0$ escrito en la forma Hamiltoniana
\begin{equation}
\delta \int (p_i \dot q^i-H)dt =0,
\end{equation}
conduce  a las ecuaciones can\'onicas
\begin{equation}
\dot q^i={\partial H\over  \partial p_i},
\end{equation}
\begin{equation}
\dot p_i=-{\partial H\over\partial q^i},
\end{equation}
la primera de  las cuales permite obtener las velocidades  en t\'erminos de
las coordenadas y los momentos, dado que $H$ es una funci\'on de $q^i$ y $p_i$
que se  obtiene mediante  una transformaci\'on de  Legendre de  la
lagrangiana. En un sistema con v\'\i nculos, en cambio, los momentos no son
independientes, sino que su variaci\'on debe restringirse a  la superficie
definida por $\phi_m =0$.  Por lo tanto, ahora se debe encontrar un extremo de
la funcional $S=\int (p_i \dot q^i-H (q^i,p_i))dt$ sujeta a  las restricciones
$\phi_m(q^i,p_i)=0$, lo cual puede hacerse variando
\begin{equation}\label{pvcv}
\delta \int_{t_1}^{t_2}\left(p_i\dot q^i-H(q^i,p_i)-u^m \phi_m(q^i,p_i)\right)
dt=0,
\end{equation}
para variaciones arbitrarias de $\delta q^i$, $\delta p_i$, $\delta u^m$,
sujetas solamente a la restricci\'{o}n  $\delta q^i(t_1)=\delta q^i(t_2)=0$.
Las nuevas variables $u^m$ son multiplicadores de Lagrange (en principio,
arbitrarios) que conducen a los v\'{\i}nculos primarios. As\'{\i}, las
ecuaciones resultantes de este principio variacional son
\begin{equation}\label{2.11}
\dot q^i={\partial H\over\partial p_i}+u^m {\partial \phi_m\over\partial p_i}
\end{equation}
\begin{equation}\label{2.11'}
 \dot   p_i=-{\partial H\over\partial    q^i} -u^m
{\partial \phi_m\over\partial q^i}.
\end{equation}
\begin{equation}\label{2.11"}
\phi_m(q,p)=0
\end{equation}

Dado un punto sobre la superficie $\phi_m =0$, podemos  utilizar  la primera
ecuaci\'on para obtener la velocidad correspondiente. Las funciones
arbitrarias $u_m$ act\'uan como coordenadas sobre la variedad de velocidades
$\dot q^m$, haciendo que la transformaci\'on sea ahora biun\'\i voca.

Las ecuaciones de movimiento pueden ser expresadas en t\'{e}rminos del
formalismo de corchetes de Poisson\footnote{Para dos funciones arbitrarias de
las variables can\'{o}nicas, $F(q,p)$ y $G(q,p)$,  el corchete de Poisson se
define como es usual: $\{F,G\}={\partial F\over\partial q^i}{\partial
G\over\partial p_i}-{\partial F\over\partial p_i}{\partial G\over\partial q^i}
$.}; por ejemplo, para una magnitud f\'\i sica $F$ podemos escribir, en
general,
\begin{equation}\label{2.12}
\dot{F}={\partial F\over\partial q^i} \dot{q}^i+ {\partial F\over\partial p_i}
\dot{p}_i
\end{equation}
si $F$ no  depende expl\'\i citamente del tiempo.  A partir de las relaciones
(\ref{2.11})-(\ref{2.11'}) y la definici\'on de los corchetes de Poisson
obtenemos
\begin{equation}
 \dot{F}=\{F,H\}+u^m\{F,\phi_m\}.
\end{equation}
Pero, sobre la superficie de v\'\i nculo vale la igualdad
\begin{equation}
u^m\{F,\phi_m\}=\{F,u^m\phi_m\}
\end{equation}
ya que
\begin{equation}
\{F,u^m\phi_m\}=u^m\{F,\phi_m\}+\{F,u^m\}\phi_m
\end{equation}
y el  \'ultimo  t\'ermino se anula sobre la superficie $\phi_m=0$.

Por lo tanto,  podemos  escribir  la igualdad ``d\'ebil", es decir,
restringida a la superficie de v\'\i nculo,
\begin{equation}\label{din}
\dot{F}\approx \{F,H_T\}
\end{equation}
donde $H_T$ es el {\it Hamiltoniano Total}
\begin{equation}\label{ht}
H_T=H+u^m\phi_m.
\end{equation}

Examinemos ahora, las consecuencias de las ecuaciones de movimiento. En primer
lugar, existen ciertas condiciones de consistencia. Por ejemplo, la
evoluci\'{o}n del sistema debe preservar los v\'\i nculos en el tiempo, es
decir que las funciones $\phi_m$ deben ser siempre nulas. Luego, podemos
aplicar la ecuaci\'on (\ref{din}) a los v\'\i nculos, tomando $F$ como una de
las funciones $\phi_m$. Esto nos lleva a
\begin{equation}\label{2.15}
\dot{\phi}_m=\{\phi_m,H\}+u^{m'}\{\phi_m,\phi_{m'}\}\approx 0.
\end{equation}
Obtenemos as\'\i\ $m$ condiciones de consistencia, una para cada v\'\i nculo.
Si descartamos  la posibilidad   de  que  las  ecuaciones  de  movimiento sean
inconsistentes, las $m$ condiciones (\ref{2.15}) pueden ser separadas en tres
tipos:
\begin{description}
\item[{\it 1.}] Un tipo que se satisface id\'{e}nticamente.
\item[{\it 2.}] Otro tipo que puede dar origen a nuevas ecuaciones de la
forma (supondremos que existen K de ellas)
\begin{equation}\label{sec}
\phi_k(q^i,p_i)=0,~~~~~~~~~~~~~~~~~k=M+1,...,M+K;
\end{equation}
estas restricciones, que resultan  de aplicar las ecuaciones de movimiento, se
denominan {\it v\'\i nculos secundarios}.
\item[{\it 3.}] Finalmente, una ecuaci\'{o}n de consistencia puede que no
se reduzca a ninguno de los dos tipos anteriores, e imponga condiciones sobre
las funciones $u^m$.
\end{description}

Consideremos detalladamente esta \'{u}ltima posibilidad; tenemos las
ecuaciones
\begin{equation}
\{\phi_j,H\}+u^m\{\phi_j,\phi_m\}\approx 0
\end{equation}
donde $\phi_j$ son todos los v\'\i nculos, primarios $\phi_m$ y secundarios
$\phi_k$ (es decir $j=1,...,M+K=J$), dado que todos los v\'\i nculos deben ser
preservados por la evoluci\'{o}n del sistema; su soluci\'on general es
\begin{equation}
u^m=U^m+\lambda^a{V_a}^m
\end{equation}
con $a=1,...,A$, donde los coeficientes $\lambda^a$ son arbitrarios, las
funciones ${V_a}^m$ son todas las soluciones independientes de las ecuaciones
homog\'eneas
\begin{equation}\label{2.19}
{V_a}^m\{\phi_j,\phi_m\}=0,
\end{equation}
y $U^m$ son soluciones particulares. Sustituyendo estas expresiones para las
$u^m$ en el Hamiltoniano Total de la teor\'{\i}a, Ec. (\ref{ht}),
\begin{equation}
H_T=H+U^m\phi_m+\lambda^a{V_a}^m\phi_m.
\end{equation}

Finalmente, podemos reescribir el Hamiltoniano Total seg\'{u}n
\begin{equation}\label{htf}
H_T=H'+\lambda^a \phi_a,
\end{equation}
donde
\begin{equation}\label{2.21}
H'=H+U^m\phi_m
\end{equation}
y
\begin{equation}\label{vpc}
\phi_a={V_a}^m\phi_m.
\end{equation}
Ahora las $U^m$ y ${V_a}^m$ son funciones determinadas de las variables $q^i$
y $p_i$ (debido a las ecuaciones de consistencia), pero los $\lambda^a$ son
coeficientes totalmente arbitrarios; su n\'umero, sin embargo, es menor o
igual que el de los coeficientes originales $u^m$.

Es necesario definir cierta terminolog\'{\i}a para apreciar las relaciones
entre las cantidades que aparecen en el formalismo. Cualquier magnitud que
tenga corchete d\'ebilmente nulo con todos los v\'\i nculos $\phi_j$  se
denomina de {\it primera clase}. Las magnitudes que no tienen esta propiedad
se denominan de {\it segunda clase}. En general, como los v\'\i nculos
$\phi_j$ son las \'unicas funciones independientes d\'ebilmente nulas, para
cualquier magnitud $R$  que sea  de primera clase su corchete de Poisson
$\{R,\phi_j\}$ debe ser estrictamente igual a una combinaci\'on lineal de los
$\phi_j$:
\begin{equation}
\{R,\phi_j\}=r_{jj'}\phi_{j'}.
\end{equation}

Tenemos por la tanto, cuatro tipo diferentes de v\'{\i}nculos. Pueden ser
divididos entre los de primera clase y los de segunda clase, que es una
propiedad independiente de la divisi\'{o}n entre primarios y secundarios. En
realidad, puede verse que la distinci\'{o}n verdaderamente significativa es la
condici\'{o}n de primera o segunda clase (m\'{a}s a\'{u}n a nivel cu\'{a}ntico
como veremos luego). La distinci\'{o}n entre primarios y secundarios es
fundamentalmente dependiente de la formulaci\'{o}n Lagrangiana original de
cual se parti\'{o}, y pierde significado una vez que se ha avanzado a la
formulaci\'{o}n Hamiltoniana.

Es importante notar que los $\phi_a$ definidos por la Ec. (\ref{vpc}) son de
primera clase. En efecto,
\begin{equation}
\{\phi_a,\phi_j\}=\{{V_a}^m\phi_m,\phi_j\}
={V_a}^m\{\phi_m,\phi_j\}+\{{V_a}^m,\phi_j\}\phi_m \approx
{V_a}^m\{\phi_m,\phi_j\},
\end{equation}
y como  las  funciones  ${V_a}^m$ son soluciones de las ecuaciones
homog\'eneas (\ref{2.19}), se tiene
\begin{equation}
\{\phi_a,\phi_j\}\approx 0.
\end{equation}

En forma similar, puede demostrarse que el Hamiltoniano $H'$ definido en la
Ec. (\ref{2.21}) tambi\'{e}n es de primera clase.

Las  funciones  $\phi_a$, adem\'{a}s de ser de primera clase, son  v\'\i
nculos primarios, ya que son combinaciones lineales de  los  v\'\i  nculos
primarios $\phi_m$. Luego, vemos que la situaci\'{o}n es tal que el
Hamiltoniano Total est\'{a} expresado como la suma de un Hamiltoniano de
primera clase m\'{a}s una combinaci\'{o}n lineal de v\'{\i}nculos primarios de
primera clase.

Calculemos la evoluci\'{o}n de una variable din\'{a}mica para analizar el
papel de los v\'{\i}nculos primarios $\phi_a$. Si la variable din\'{a}mica $F$
no depende expl\'{\i}citamente del tiempo , su evoluci\'{o}n estar\'{a}
determinada por
\begin{equation}\label{ed}
\dot F=\{F,H'\}+\lambda^a\{F,\phi_a\}
\end{equation}
La arbitrariedad en la elecci\'on de las funciones $\lambda^a$ hace que la
evoluci\'on de la variable $F$ no quede totalmente determinada: para una misma
condici\'on inicial $F(t_0)= F_0,$ en un instante $t$ posterior a $t_0$, $F$
puede tomar diversos valores seg\'un la elecci\'on que se haga de los
coeficientes arbitrarios $\lambda^a$. Para ver esto, hallemos el valor de $F$
en un intervalo de tiempo peque\~{n}o $\Delta t$ posterior a $t_0$
\begin{equation}\label{eva}
F(\Delta t)=F_0+\dot F\Delta t = F_0+\{F,H'\}\Delta t+\lambda^a\Delta
t\{F,\phi_a\},
\end{equation}
tomemos otro conjunto de valores diferentes $\lambda^{'a}$, que puede dar un
valor distinto para $F(\Delta t)$, siendo la diferencia
\begin{equation}\label{difa}
\delta F(\Delta t)=(\lambda^a-\lambda^{'a})\Delta t\{F,\phi_a\}.
\end{equation}
Finalmente, podemos reescribir el cambio de $F(\Delta t)$ como
\begin{equation}\label{3.4}
\delta_{\epsilon}F=\epsilon^a \{F,\phi_a\},
\end{equation}
donde $\epsilon^a =(\lambda^a-\lambda^{'a})\Delta t$ es un n\'{u}mero
arbitrariamente peque\~{n}o. Ahora bien, el estado f\'\i sico de un sistema no
puede depender de la elecci\'on, arbitraria, de  los coeficientes; por lo
tanto, ambos valores de la variable $F(\Delta t)$ deben corresponden al mismo
estado f\'\i sico. Vemos entonces, que la transformaci\'on infinitesimal
generada por la funci\'on $\epsilon^a \phi_a$ relaciona descripciones
equivalentes de un mismo estado f\'{\i}sico. Tal tipo de transformaci\'{o}n se
denomina de {\it medida} o {\it gauge}, y por esto se dice que los v\'\i
nculos primarios de primera  clase $\phi_a$ son {\it generadores} de la
transformaci\'on de gauge.

Los v\'\i nculos $\phi_a$ no son los \'unicos generadores de transformaciones
de gauge; puede probarse que tambi\'en los v\'\i nculos secundarios de primera
clase pueden generar transformaciones que no modifiquen el estado f\'\i sico
del sistema [\cite{hrt,sm,ht92}]. Si llamamos $G_a$ a todos los v\'\i nculos
de primera clase, primarios y secundarios, la transformaci\'on infinitesimal
de gauge m\'as general puede escribirse
\begin{equation}\label{tgg}
\delta_{\epsilon}F=\epsilon^a \{F,G_a\}.
\end{equation}
Como los v\'\i nculos $G_a$ son de primera clase verifican que
\begin{equation}\label{fcg}
\{G_a,G_b\}=C_{ab}^c G_c\approx 0;
\end{equation}

Si existen, adem\'as, v\'\i nculos de segunda clase, puede redefinirse el
corchete entre dos magnitudes de manera tal que esta ecuaci\'{o}n  siga siendo
v\'alida ({\it corchete de Dirac}, \cite{dirac,ht92}). De acuerdo con la
(\ref{fcg}), se tiene que $\delta_{\epsilon}G_a\approx 0,$ lo cual indica que
las transformaciones de gauge no modifican a las funciones $G_a$, con lo cual
el sistema se mantiene sobre la superficie de v\'\i nculo.

Dirac conjetur\'{o} que todos los v\'{\i}nculos de primera clase eran
generadores de transformaciones de gauge, y propuso por lo tanto, escribir las
ecuaciones din\'{a}micas utilizando un Hamiltoniano extendido que contuviera a
todos los v\'{\i}nculos de primera clase
\begin{equation}\label{he}
H_E=H_T+\lambda^a G_a.
\end{equation}

Sin embargo, hoy se conocen ejemplos en donde no todos los v\'{\i}nculos de
primera clase generan invariancias de gauge [\cite{ht92}]. A pesar de esto, en
los sistemas f\'{\i}sicos usuales la conjetura de Dirac se satisface, y es
\'{u}til mantener la definici\'{o}n del Hamiltoniano extendido.

La descripci\'on de un sistema mediante variables cuya evoluci\'on depende de
par\'ametros arbitrarios    es,  claramente,  ambigua  y,  por  lo  tanto,
insatisfactoria. Es necesario, entonces, definir funciones ${\cal O}$ cuya
evoluci\'{o}n est\'{e} libre de ambig\"uedades, lo cual ocurrir\'{a} si su
corchete de Poisson con los v\'\i nculos es d\'ebilmente nulo:
\begin{equation}\label{obs}
\{{\cal O} ,G_a\}\approx 0.
\end{equation}
Estas funciones  se  denominan  {\it observables}: son invariantes  ante
transformaciones de gauge  y su evoluci\'on est\'a completamente determinada.
Si se describe un  sistema mediante obser\-vables \'unicamente, los valores
iniciales de estos permiten conocer su evoluci\'on sin ninguna ambig\"uedad.

Los conjuntos de puntos del espacio de las fases relacionados por
transformaciones de gauge se denominan \'orbitas;  las funciones que llamamos
observables tienen el mismo valor  a lo largo de los puntos de una \'orbita y
caracterizan  as\'\i \ el estado f\'\i sico  del  sistema,  mientras  que  las
funciones que no son invariantes de gauge contienen  informaci\'on acerca de
cu\'al punto de la \'orbita se considera, pero  esta informaci\'on  es  f\'\i
sicamente irrelevante, dado que una transformaci\'on de gauge aplicada  a  un
punto del espacio de las fases lo transforma en otro  punto del mismo espacio
que  corresponde al mismo estado f\'\i sico.  Si se considera  la evoluci\'on
de un  sistema  desde  la  configuraci\'on  $(q^0,p_0)$  a  dos
configuraciones $(q^1,p_1)$ y $(q^{1'} ,p_{1'})$  dadas  por  dos elecciones
distintas  de los coeficientes $\lambda^a$,    ambas  configuraciones finales
corresponden al mismo estado f\'\i  sico pues  se  encuentran  sobre la misma
\'orbita y est\'an, por lo tanto,  conectadas  por una  transformaci\'on de
gauge.

En principio, siempre es posible eliminar la  ambig\"uedad en la evoluci\'on
de las variables  eligiendo  {\it  una}  configuraci\'on  entre  todas  las
configuraciones f\'\i sicamente equivalentes.  Esto se logra seleccionando  un
\'unico punto de cada \'orbita mediante la imposici\'on de las  llamadas {\it
condiciones de gauge}, es decir, funciones de la forma
\begin{equation}\label{cg}
\chi_{_b}(q^i,p_i,t)=constante
\end{equation}
que corten una y s\'olo una vez a cada \'orbita. Si el n\'umero  de  v\'\i
nculos independientes $G_a$ es $M$, la dimensi\'on de cada \'orbita es
tambi\'en $M$,  pues cada punto  se alcanza eligiendo los $M$ par\'ametros de
gauge $\epsilon^a$. Las condiciones de gauge deben definir  una variedad de
dimensi\'on igual a $2N-M$ (N es el n\'{u}mero de coordenadas can\'{o}nicas)
para cortar una vez a cada \'orbita; por lo tanto, se requieren $M$
condiciones de gauge para seleccionar un representante de cada \'orbita  que
identifique el estado del sistema.

La  condici\'on  de  que  s\'olo  un  punto de  cada \'orbita pertenezca a la
variedad definida  por  las    condiciones de  gauge  significa  que  una
transformaci\'on de gauge mueve a  un  punto sobre una \'orbita de manera de
sacarlo de la superficie dada por $\chi_{_b}=0$, por lo tanto
\begin{equation}\label{condgauge}
\delta\chi_{_b}={\epsilon^a\{\chi_{_b},G_a\}}\approx 0~~~\Longleftrightarrow~~
\epsilon^a=0~.
\end{equation}
Esto significa que
\begin{equation}\label{condeq}
{det\left( \{\chi_{_b},G_a\}\right) }{\not\approx 0},
\end{equation}
(notar que la matriz $\left( \{\chi_{_b},G_a\}\right)$ es cuadrada) por lo
cual, el conjunto $(G_a,\chi_{_b})$ puede verse como un conjunto de v\'\i
nculos de segunda clase. Estrictamente, la condici\'on (\ref{condeq}) asegura
la posibilidad  de seleccionar un \'unico punto de cada \'orbita s\'olo
localmente (impide que las \'{o}rbitas sean tangentes a la superficie
$\chi_{_b}=0$). Sin embargo, no impide que la superficie $\chi_{_b}=0$ corte
m\'{a}s de una vez a alguna \'{o}rbita, dificultad que se denomina {\it
problema de Gribov}.

\bigskip

\section{Cuantificaci\'{o}n can\'{o}nica de sistemas con v\'{\i}nculos}
\medskip

~~~~ Uno de los m\'{e}todos tradicionales para cuantificar un sistema
f\'{\i}sico es el denominado como cuantificaci\'{o}n can\'{o}nica. Su
aplicaci\'{o}n involucra los siguientes pasos:
\begin{description}
\item[{\it 1.}] El estado del sistema ser\'{a} descripto por un elemento
$|\psi>$ de un espacio de Hilbert ${\cal H}$ con producto escalar
$<\psi|\psi'>$.

\item[{\it 2.}] Los observables corresponden a operadores lineales
herm\'{\i}ticos sobre ${\cal H}$. El resultado de una medici\'{o}n es
identificado con un autovalor del operador. Esto significa que si $|\psi>$ es
desarrollado en t\'{e}rminos de autovectores $|a_n>$ de una observable $\hat
A$, $$\hat A |\psi>=a_n|\psi>$$ $$|\psi>=\sum c_n|\psi>,$$ entonces $|c_n|^2$
es la probabilidad de que $\hat A$ tome el valor $a_n$. El valor medio de
$\hat A$ en el estado $|\psi>$ es $<\hat A>=<\psi|\hat A|\psi>$.

\item[{\it 3.}] La evoluci\'{o}n din\'{a}mica se obtiene a partir de la
ecuaci\'{o}n de Schr\"odinger $$i\hbar {d\over dt} |\psi>= \hat H |\psi>$$ que
implica que el valor medio de $\hat A$ $$i\hbar {d\over dt} <\hat A>= {i\over
\hbar} <(\hat H \hat A - \hat A \hat H)>$$ que resulta semejante a las
ecuaciones cl\'{a}sicas de Hamilton si se reemplaza el corchete de Poisson
cl\'{a}sico por el conmutador cu\'{a}ntico $[\hat A, \hat B]=\hat A\hat B-\hat
B\hat A$ de los operadores $\hat A$ y $\hat B$:
\begin{equation}\label{reecq}
\{A,B\} \longrightarrow {i\over \hbar} [\hat A, \hat B]
\end{equation}

\end{description}
Luego, es necesario que los operadores sean tales que se satisfaga esta
correspondencia en alg\'{u}n l\'{\i}mite, para reobtener los resultados
cl\'{a}sicos. De este modo, se elige $$[\hat p, \hat q]= -i \hbar.$$

Sin embargo, existe un problema y es que los observables cl\'{a}sicos no
tienen un \'{u}nico operador asociado. Existen ambig\"uedades de
``ordenamiento". Por ejemplo, el observable $p^2q$ pude asociarse con los
operadores $\hat p^2 \hat q,~\hat p\hat q\hat p,~\hat q\hat p^2$, que son
diferentes pues $\hat p$ y $\hat q$ no conmutan (difieren entre s\'{\i} por
t\'{e}rminos de orden $\hbar$).

M\'{a}s a\'{u}n, cualquier observable cl\'{a}sico puede escribirse como
$$A(q,p)=A(q,p)+B(q,p)(pq-qp),$$ pero la realizaci\'{o}n del segundo miembro
como operador da $$\hat A (q,p)-i\hbar \hat B(q,p) \neq \hat A(q,p)$$ De modo
que los operadores son conocidos a menos de t\'{e}rminos de orden $\hbar$. Si
bien, ciertas condiciones sobre los operadores pueden limitar el ordenamiento
(como por ejemplo, el requisito de hermiticidad), en general, las
ambig\"uedades persisten.

Al tratar la cuantificaci\'{o}n de sistemas con v\'{\i}nculos, lo primero que
uno intentar\'{\i}a ser\'{\i}a aislar los verdaderos grados de libertad y
aplicar a este sistema ``reducido" las reglas de cuantificaci\'{o}n
can\'{o}nica {\it (1.)-(3.)}. Sin embargo, el conocimiento del espacio de
fases reducido, que es el subespacio que se obtiene de retener solamente los
momentos no afectados a los v\'{\i}nculos y sus variables conjugadas, muchas
veces es s\'{o}lo impl\'{\i}cito (a trav\'{e}s de las ecuaciones de
v\'{\i}nculo), y a\'{u}n en caso de conocerse expl\'{\i}citamente, el proceso
de reducci\'{o}n puede tornarse en la pr\'{a}ctica, inmanejable. Mas a\'{u}n,
el proceso de reducci\'{o}n puede llevar a la p\'{e}rdida  de la covariancia y
localidad expl\'{\i}citas de la teor\'{\i}a [\cite{dirac2,lusanna3}]. Un
camino alternativo fue propuesto por \cite{dirac}: Supongamos primero que
todos los v\'{\i}nculos son de primera clase, entonces el m\'{e}todo consiste
en resolver la ecuaci\'{o}n de Schr\"{o}dinger
\begin{equation}\label{ecsch}
i\hbar {d\over dt} |\psi>= \hat H' |\psi>
\end{equation}
donde $\hat H'$ es el Hamiltoniano de primera clase de la teor\'{\i}a. Luego,
se imponen los v\'{\i}nculos como condiciones suplementarias sobre los
estados:
\begin{equation}\label{csup1}
\hat G_a |\psi>=0
\end{equation}
Tenemos por lo tanto, tantas condiciones suplementarias como v\'{\i}nculos de
primera clase. Debemos ver por lo tanto, que cada una de ellas sea consistente
con la otras. Consideremos adem\'{a}s de (\ref{csup1}),
\begin{equation}\label{csup2}
\hat G_{a'} |\psi>=0
\end{equation}
Si multiplicamos (\ref{csup1}) por $\hat G_{a'}$, y (\ref{csup2}) por $\hat
G_a$ y restamos ambas ecuaciones, obtenemos
\begin{equation}\label{csup3}
[\hat G_a, \hat G_{a'}] |\psi>=0
\end{equation}
Luego, esta nueva condici\'{o}n sobre los estados es necesaria por
consistencia. Desear\'{\i}amos que (\ref{csup3}) fuese una consecuencia
directa de (\ref{csup1}), de modo que todas las condiciones de consistencia se
desprendan de la primera, pero esto significa que se requiere que
\begin{equation}\label{cqcs}
[\hat G_a, \hat G_{a'}]=\hat C_{aa'}^{a''} \hat G_{a''}
\end{equation}
Si esto se satisface, entonces (\ref{csup3}) no es una nueva condici\'{o}n
sobre el estado.

A nivel cl\'{a}sico, como los v\'{\i}nculos $G_a$ son de primera clase, se
cumple que el corchete de Poisson entre dos cualquiera de los $G_a$ es una
combinaci\'{o}n lineal de todos ellos. Sin embargo, en el an\'{a}logo
cu\'{a}ntico no necesariamente se cumple (\ref{csup3}) porque en general las
funciones de estructura $C_{aa'}^{a''}$ son operadores que no necesariamente
est\'{a}n a la izquierda de los operadores de v\'{\i}nculo.

Existe una condici\'{o}n similar de consistencia con la ecuaci\'{o}n de
Schr\"odinger. Para que la evoluci\'{o}n preserve las condiciones
(\ref{csup1}) debe satisfacerse
\begin{equation}\label{csuph}
[\hat G_a, \hat H] |\psi>=0,
\end{equation}
lo cual significa que debe cumplirse que
\begin{equation}\label{cqcs2}
[\hat G_a, \hat H]=\hat C_{a0}^{a'} \hat G_{a'}
\end{equation}
para que no surja una nueva condici\'{o}n suplementaria. Nuevamente, a nivel
cl\'{a}sico el Hamiltoniano es de primera clase, por lo tanto el corchete de
Poisson con los v\'{\i}nculos es d\'{e}bilmente nulo (es decir, fuertemente
igual a una combinaci\'{o}n de los v\'{\i}nculos). Sin embargo, de nuevo, esto
no est\'{a} garantizado a nivel cu\'{a}ntico, de manera que las operadores de
las funciones de estructura queden a la izquierda de los operadores de
v\'{\i}nculo. Como vemos, la tarea de lograr una cuantificaci\'{o}n
consistente no es obra de un mero procedimiento met\'{o}dico, o en palabras de
Dirac (1964) refir\'{e}ndose a esta dificultad : ``en general, es necesario un
poco de suerte para obtener una teor\'{\i}a cu\'{a}ntica precisa".

El inter\'{e}s principal de este trabajo es mostrar en ciertos modelos de
dimensi\'{o}n finita con covariancia general, c\'{o}mo deben ordenarse los
operadores de v\'{\i}nculo y de estructura de modo que se satisfagan las
ecuaciones (\ref{cqcs}) y (\ref{cqcs2}).

\subsubsection{El producto interno f\'\i sico.}

~~~~ Analicemos c\'{o}mo deben normalizarse los estados en el m\'{e}todo de
Dirac. Consideremos primero un sistema  de $N$ grados de libertad y con un
\'{u}nico v\'\i nculo $p_k=0$. El producto interno entre dos estados $\psi_1$
y $\psi_2$ del espacio de Hilbert est\'a dado por
\begin{equation}\label{pi1}
<\psi_1\vert \psi_2>=\int{dq^1 ...dq^k ...dq^N \psi_1^*(q^1...q^N)\
\psi_2(q^1...q^N)}.
\end{equation}
Si $\psi_1$ y $\psi_2$ son estados f\'{\i}sicos (es decir que verifican el
v\'\i nculo, y por lo tanto no dependen de $q^k$) la integral diverge. Para
evitar que esto ocurra debe eliminarse la integraci\'on sobre $q^k$, que
corresponde a un grado de libertad no f\'\i sico, introduciendo una
condici\'on de gauge $\chi$ y definiendo el {\it producto interno f\'\i sico}
como
\begin{equation}\label{pi1r}
(\psi_1\vert \psi_2)=\int{dq^1...dq^k...dq^N\ \delta(\chi)\ {\vert [\chi
,p_k]\vert}\ \psi_1^*(q^1...q^N) \ \psi_2(q^1...q^N)},
\end{equation}
donde la condici\'on de gauge $\chi$ da a la coordenada $q^k$, en general,
como funci\'on de las  dem\'as variables, es decir
\begin{equation}\label{cg1}
\chi=0 ~~ \Longleftrightarrow ~~q^k=q^k(q^1,...,q^{k-1},q^{k+1},...q^N)
\end{equation}
El jacobiano asegura que el valor de la integral no dependa de la elecci\'on
de $\chi$: $$\delta (\chi) {\vert [\chi ,p_k]\vert}=\delta  (\chi) \vert
\partial\chi /\partial q^k\vert =\delta (q^k).$$
El producto interno (\ref{pi1r}) es expl\'{\i}citamente igual a
\begin{equation}\label{pi1e}
(\psi_1\vert \psi_2)=\int{dq^1...dq^{k-1} dq^{k+1} ...dq^N
\psi_1^*(q^1...q^{k-1} q^{k+1}...q^n) \ \psi_2(q^1...q^{k-1} q^{k+1}...q^N)}
\end{equation}
para los estados f\'\i sicos, y coincide con el producto escalar en el espacio
reducido, es decir, en el subespacio al cual se restringe el movimiento del
sistema reducido.

Para el caso en que el sistema exhiba v\'{\i}nculos de primera clase tales que
\begin{equation}\label{vinCULOS}
G_a=p_a+{\partial V\over\partial q^a},
\end{equation}
el producto interno f\'{\i}sico puede ser generalizado [\cite{ht92}] como
\begin{equation}\label{pi1gr}
(\psi_1\vert \psi_2)=\int{dq^1...dq^N\ \prod_a \delta(\chi_{_a})\ {\vert
det[\chi ,p_k]\vert}\ \psi_1^*(q^1...q^N) \ \psi_2(q^1...q^N)},
\end{equation}
Para que las condiciones de gauge sean buenas deben ser resolubles para $q^a$,
es decir $\chi_{_a}(q^i)=a_{ab}(q^i)(q^b-f^b)$ con $det~a_{ab}\neq 0$ y $f^b$
independiente de $q^b$.

Para v\'{\i}nculos de forma gen\'{e}rica habr\'{a} que insertar un operador
$\hat \mu(\hat q,\hat p)$ tal que
\begin{equation}\label{pig}
 (\psi_1\vert \psi_2)=<\psi_1\vert \hat\mu\vert \psi_2>,
\end{equation}
con $\hat\mu$ un operador herm\'{\i}tico singular que  elimina la
integraci\'on sobre las variables que son puro gauge.

Denotando las funciones de onda f\'{\i}sicas como
\begin{equation}\nonumber
\psi_1^*(q)=(\psi_1\vert q>
\end{equation} y
\begin{equation}\nonumber
\psi_2(q)=<q\vert \psi_2),
\end{equation}
podemos reescribir el producto interno f\'\i sico en la forma
\begin{equation}\label{pifg}
(\psi_1\vert \psi_2)=\int dq(\psi_1\vert q>\hat\mu<q\vert \psi_2),
\end{equation}
de modo que nos permite definir el operador identidad en el subespacio de
estados f\'\i sicos:
\begin{equation}\label{unomagico}
 {\bf 1}=\int{dq\vert q>\hat\mu<q\vert}.
\end{equation}
Y de manera similar, si los vectores $\vert \Psi_{\alpha})$ forman una base
del subespacio de estados f\'\i sicos, tenemos:
\begin{equation}\label{4.12}
{\bf 1}=\sum_{\alpha} {\vert\Psi_\alpha)(\Psi_\alpha\vert } ;
\end{equation}
los ``bras" y ``kets" curvos  indican  que  el  producto interno debe
interpretarse en el sentido de producto f\'\i sico.

\newpage

\thispagestyle{empty}

 ~\newpage


\chapter{El Formalismo Can\'{o}nico de la Gravedad Cu\'{a}ntica}
\hyphenation{Ge-ne-ral de-sa-rro-lla-do geo-me-tri-ca-men-te}

\bigskip

~~~~{\it La teor\'{\i}a de la Relatividad General y la teor\'{\i}a
Cu\'{a}ntica representan los dos mayores logros de la f\'{\i}sica del
\'{u}ltimo siglo. Ambas teor\'{\i}as son consideradas ``aplicables
universalmente'', es decir que todos los sistemas f\'{\i}sicos deben obedecer
sus principios. Por lo tanto, por esto y por diversas motivaciones adicionales
(ver por ejemplo \cite{ish95}), parece esencial combinar estas dos
teor\'{\i}as en una \'{u}nica teor\'{\i}a consistente. Sin embargo, a pesar de
grandes esfuerzos realizados, a\'{u}n no existe una teor\'{\i}a definitiva.
Esto se debe esencialmente a dificultades de dos tipos: conceptuales y
t\'{e}cnicas.

Las dificultades conceptuales m\'{a}s notorias surgen de diferencias
substanciales en los fundamentos f\'{\i}sicos mismos de cada teor\'{\i}a. Para
comenzar, la mec\'{a}nica cu\'{a}ntica parece reposar excesivamente en
conceptos de la mec\'{a}nica cl\'{a}sica para una teor\'{\i}a que pretende ser
m\'{a}s fundamental, y a pesar de la hegemon\'{\i}a de la interpretaci\'{o}n
de Copenhague, a\'{u}n se discute c\'{o}mo debe interpretarse el formalismo
para eludir ciertos problemas conceptuales que plantea
[\cite{jammer,blok,ballen,popper,sonego,sforza}]. Por otra parte,
independientemente de la interpretaci\'{o}n adoptada para la teor\'{\i}a
cu\'{a}ntica, su formalismo le da un rol privilegiado al tiempo como
``par\'{a}metro de evoluci\'{o}n''; mientras que en relatividad general, la
acci\'{o}n es invariante ante reparametrizaciones, lo cual significa que las
trayectorias del sistema no est\'{a}n parametrizadas por el tiempo sino por un
par\'{a}metro f\'{\i}sicamente irrelevante (de alguna manera, puede decirse
que el tiempo f\'{\i}sico se halla ``escondido'' en el formalismo). Esto
significa que en la pr\'{a}ctica, al carecer de una teor\'{\i}a cu\'{a}ntica
que respete esta invariancia, se hallan enormes dificultades al intentar
cuantificar si no se ha identificado el tiempo f\'{\i}sico de manera
un\'{\i}voca (entre otras propiedades). Esto da lugar al denominado ``problema
del tiempo'' [\cite{ish92,k92}].

Entre los numerosos problemas t\'{e}cnicos que surgen, dos son particularmente
importantes [\cite{ish92,k92}]: La divergencia ultravioleta, que se refiere a
la no renormalizabilidad perturbativa de la gravedad cu\'{a}ntica, lo cual
sugiere que los operadores de la teor\'{\i}a no est\'{a}n bien definidos; el
problema del ordenamiento, que se refiere a la dificultad de realizar como
operadores a los v\'{\i}nculos cl\'{a}sicos de modo que respeten el
\'{a}lgebra.

Estas consideraciones podr\'\i an indicar que su unificaci\'on no
consistir\'\i a ``simplemente'' en tratar de armonizar los conceptos de ambas
teor\'\i as en un esquema coherente. En cambio, sugieren que podemos verlas
como l\'\i mites de una teor\'\i a m\'as general correspondiendo a diferentes
e incompatibles aproximaciones de la misma. Actualmente, se ve como la
candidata m\'{a}s firme para este rol a la teor\'{\i}a de Supercuerdas. La
mayor parte del esfuerzo actual se dirige en esta direcci\'{o}n; aunque las
esperanzas en este camino no son uniformes en la comunidad cient\'{\i}fica,
debido esencialmente a discrepancias que surgen desde el punto de vista de la
f\'{\i}sica relativista en contraposici\'{o}n a la f\'{\i}sica de
part\'{\i}culas: concepciones diferentes de la interacci\'{o}n gravitatoria, y
distinta actitud frente a la necesidad de una formulaci\'{o}n independiente
del fondo [Rovelli (1998,2000)]. Por lo cual, las investigaciones en el
formalismo can\'{o}nico a\'{u}n contin\'{u}an vigorosamente.

En este cap\'{\i}tulo aplicaremos el m\'{e}todo usual de cuantificaci\'{o}n
can\'{o}nica a la relatividad general. Como vimos, un paso previo necesario
ser\'{a} expresarla en forma Hamiltoniana. }

\section{Formulaci\'on Hamiltoniana  de la Relatividad General.}

~~~~ Las ecuaciones de Einstein pueden ser derivadas de la acci\'on de
Einstein-Hilbert
\begin{equation}\label{acchilbert1}
S={1\over 16\pi G}\int d^4x\sqrt{-g}R + S_m.
\end{equation}
\noindent Haciendo las variaciones respecto de $g_{\mu\nu}$, obtenemos
\begin{equation}\label{eceins}
R_{\mu\nu}-{1\over 2}g_{\mu\nu}R = 8\pi G T_{\mu\nu}.
\end{equation}

Esta es una ecuaci\'on para la geometr\'\i a, pero una dada geometr\'\i a
est\'a descripta por toda una clase de m\'etricas equivalentes
(correspondiendo a diferentes sistemas de coordenadas, como lo permite el
principio de covariancia general). Es decir, que al utilizar la m\'etrica
$g_{\mu\nu}$ como variables de campo, estamos utilizando variables
redundantes. En consecuencia, la Relatividad General tendr\'a una
formulaci\'on hamiltoniana con v\'\i nculos.

Para expresarla en forma hamiltoniana, comenzamos con una generalizaci\'on del
tiempo. Cortemos el espacio-tiempo por una hipersuperficie espacial arbitraria
$\Sigma$,
\begin{equation}\label{x}
x^{\alpha}=x^{\alpha}(x^i),
\end{equation}
\noindent donde los \'\i ndices griegos corren de 0 a 3 y los latinos de 1 a
3. Luego, en cada punto de $\Sigma$, tenemos una base formada por tres
vectores tangentes $\xi^{\alpha}_i = \partial_i x^{\alpha}$ y vector normal
unitario $n^{\alpha}$ tales que
\begin{equation}\label{ortono}
{n\cdot\xi_i=0,~~ n^2=-1.}
\end{equation}

Ahora, foliamos el espacio-tiempo deformando en forma continua a $\Sigma$. De
este modo obtenemos la familia de hipersuperficies parametrizadas en $t$,
$x^{\alpha}=x^{\alpha}(x^i,t)$.

Definimos el vector {\it deformaci\'on}:
\begin{equation}\label{vecdef}
N^{\alpha}\equiv \partial_{t}x^{\alpha}(x^i,t)
\end{equation}

\noindent que conecta dos puntos con el mismo r\'otulo $x^i$ en dos
hipersuperficies pr\'oximas. Este vector puede descomponerse en la base
{$n^{\alpha},\xi^{\alpha}_i$}:
\begin{equation}\label{descvec}
N^{\alpha}=N n^{\alpha}+ N^i \xi^{\alpha}_i
\end{equation}

Las componentes $N$ y $N^i$ se denominan funciones de {\it lapso} y {\it
corrimiento}, respectivamente. Su interpretaci\'on f\'\i sica surge de
escribir en la forma 3+1 o ADM, debida a \cite{adm}, la m\'etrica del
espacio-tiempo:
\begin{equation}\label{pitag}
ds^2=-N^2dt^2+g_{ij}(dx^i+N^idt)(dx^j+N^jdt).
\end{equation}

Se puede imaginar el espacio-tiempo foliado por una familia de
hipersuperficies $t=cte$; luego, $N(x)dt$ es el lapso de tiempo propio entre
las hipersuperficies superior e inferior. Por otro lado, la funci\'on de
corrimiento da la correspondencia entre puntos en las dos hipersuperficies;
$x^i + dx^i + N^i dt$ en la hipersuperficie inferior corresponde al punto $x^i
+ dx^i$, $t + dt$ en la superior. Desde este punto de vista, la ecuaci\'on
(\ref{pitag}) es s\'olo una expresi\'on del teorema de Pit\'agoras.

En el marco de la descomposici\'on ADM, es apropiado hacer un an\'alisis del
concepto de curvatura. La curvatura intr\'\i nseca a la 3-geometr\'\i a de una
hipersuperficie espacial puede ser definida y calculada de manera an\'aloga al
caso de la curvatura cuadridimensional. Pero, ahora podemos introducir el
concepto de {\it curvatura extr\'\i nseca} de la 3-geometr\'\i a. Este
concepto carece de significado para una 3-geometr\'\i a concebida s\'olo en
s\'\i ~misma. Su existencia depende de que la 3-geometr\'\i a est\'e embebida
en un espacio-tiempo que la contenga (y estando ambos bien definidos). El
tensor de curvatura extr\'\i nseca $K_{ij}$ de la hipersuperficie mide la
variaci\'on de la normal sobre ella a medida que se traslada sobre la misma
(de modo que $dn_i=-K_{ij}dx^j$). Su expresi\'on es [\cite{grav}]:

\begin{equation}\label{ke}
K_{ij}={1\over{2N}}[N_{i\vert j}+N_{j\vert i}-{\partial g_{ij}\over
\partial t}],
\end{equation}

\noindent donde la barra vertical indica derivaci\'on covariante en la
hipersuperficie espacial.

La acci\'on de Einstein-Hilbert, Ec. (\ref{acchilbert1}), se puede expresar
haciendo uso de la descomposici\'on ADM:
\begin{equation}\label{acchilbert2}
S=\int L~dt=\int
dt~d^3x\sqrt{^{(3)}g}N[K_{ij}K^{ij}-K^2+~^{(3)}R]+t\acute{e}rminos~de ~borde,
\end{equation}

\noindent en la cual suprimimos la acci\'on correspondiente a la materia, para
simplificar el an\'alisis que sigue. Adem\'as $^{(3)}R$ es el escalar de
curvatura tridimensional y por simplicidad tomamos que $16\pi G=1$. A partir
de aqu\'\i  , a menos que se indique lo contrario, $g$ indicar\'a el
determinante de la 3-m\'etrica de la hipersuperficie espacial.

Vemos que en la acci\'on escrita en la forma ADM, no aparecen las derivadas
respecto del tiempo de $N$ y $N^i$, luego sus momentos conjugados son
\begin{equation}\label{vinpri1}
\pi={\delta L\over \delta\dot{N}}=0,
\end{equation}
\begin{equation}\label{vinpri2}
\pi^i={\delta L\over\delta\dot{N^i}}=0.
\end{equation}

Estos son los v\'\i nculos primarios.

El momento can\'onico conjugado a $g_{ij}$ es
\begin{equation}\label{mogij}
\pi^{ij}={\delta L\over{\delta\dot g_{ij}}}=-\sqrt{g}(K^{ij}-g^{ij}K).
\end{equation}

Luego, el Hamiltoniano es
\begin{eqnarray}\label{ham}
\lefteqn{H=\int d^3x (\pi\dot {N}+\pi_i\dot {N^i}+\pi^{ij}{\dot
g}_{ij}-L)}\nonumber \\ &&=\int d^3x (\pi\dot {N}+\pi_i\dot {N^i}+N{\cal
H}+N^i{\cal H}_i),
\end{eqnarray}

\noindent donde $\cal H$ y ${\cal H}_i$ est\'an dados por
\begin{equation}\label{superham}
{\cal H}=G_{ijkl}\pi^{ij}\pi^{kl}-\sqrt{g}~^{(3)}R,
\end{equation}

 y
\begin{equation}\label{supermom}
{\cal H}_i=-2{\pi_i^j}_{\vert j}.
\end{equation}

Estan cantidades se denominan, respectivamente, {\it super-Hamiltoniano} y
{\it supermomentos} de la m\'etrica y en donde
\begin{equation}\label{supermetrica}
G_{ijkl}={1\over{2}}\sqrt{g}(g_{ik}g_{jl}+g_{il}g_{jk}- g_{ij}g_{kl})
\end{equation}

\noindent es la denominada {\it superm\'etrica}.

Finalmente, la acci\'on en forma Hamiltoniana es
\begin{equation}\label{acham}
S=\int dt~d^3x (\pi^{ij} {\dot g}_{ij}- N{\cal H}- N^i {\cal H}_i).
\end{equation}

Comp\'{a}rese esta expresi\'{o}n con la acci\'{o}n Hamiltoniana obtenida en el
cap\'{\i}tulo anterior, Ec. (\ref{pvcv}). Haciendo la variaci\'on  respecto de
$g_{ij}$ y $\pi^{ij}$ obtenemos las ecuaciones de Einstein correspondientes a
$G_{ij}=0$. Haciendo la variaci\'on respecto de $N$ y de $N^i$ se obtienen las
ecuaciones
\begin{equation}\label{vinchamil}
{\cal H}=0,
\end{equation}
\begin{equation}\label{vincmoment}
{\cal H}_i=0,
\end{equation}

\noindent que corresponden, respectivamente, a las ecuaciones restantes
$G_{00}=0$ y $G_{0i}=0$. Estos v\'\i nculos tambi\'en pueden ser obtenidos
imponiendo la condici\'on que la derivada temporal de los v\'\i nculos
primarios se anulen. Esto es esencial para que la estructura de los v\'\i
nculos de nuestro sistema se mantenga durante la evoluci\'on din\'amica. Como
ya vimos, los v\'\i nculos que se obtienen de este modo se denominan
secundarios. Estos v\'\i nculos son, adem\'as, de primera clase; es decir, el
corchete de Poisson entre dos cualesquiera de ellos es nulo en la
hipersuperficie. Esto asegura independencia de la evoluci\'on din\'amica en la
foliaci\'on. Puede calcularse expl\'{\i}citamente que [\cite{hrt,dirac}]:
\begin{eqnarray}\label{pb1}
\lefteqn{~~~~~\{{\mathcal H}(x),{\mathcal H}(x')\}=\left( g^{ij}(x){\mathcal
H}_i(x)+ g^{ij}(x'){\mathcal H}_i(x')\right)\delta_{,j}(x,x'),} \\
&&\{{\mathcal H}_i(x),{\mathcal H}(x')\}={\mathcal
H}(x)\delta_{,i}(x,x'),~~~\label{pb2}\\ &&\{{\mathcal H}_i(x),{\mathcal
H}_j(x')\}={\mathcal H}_i(x)\delta_{,j}(x,x')+ {\mathcal
H}_j(x')\delta_{,i}(x,x'),~~~\label{pb3}
\end{eqnarray}

Estos v\'\i nculos est\'an claramente relacionados con la invariancia de gauge
de la teor\'\i a. El v\'\i nculo (\ref{vinchamil}) aparece como consecuencia
de la invariancia ante reparametrizaciones temporales, y la
Ec.(\ref{vincmoment}) es debida a la invariancia ante reparametrizaciones  de
las coordenadas espaciales en las hipersuperficies. De este modo vemos que el
rol de los v\'\i nculos es reintroducir en la teor\'\i a el principio de
covariancia, que hab\'\i a sido violado al haber elegido la forma particular
(\ref{pitag}) de la m\'etrica.

Debemos se\~nalar un punto importante: en los v\'\i nculos (\ref{vinchamil}) y
(\ref{vincmoment}) no existe referencia alguna a la hipersuperficie $\Sigma$
que posee las cantidades geom\'etricas $g_{ij}$ y $\pi^{ij}$. Esta
hipersuperficie representa un instante de tiempo, por lo tanto el tiempo ha
``desaparecido'' (al menos expl\'\i citamente) del formalismo.

Planteada esta situaci\'on, se podr\'\i a suponer que dada una
hipersuperficie, \'esta queda especificada por las cantidades $g_{ij}$ y
$\pi^{ij}$. Podr\'\i a existir, adem\'as, una transformaci\'on can\'onica
[\cite{k3}]:
\begin{equation}\label{transfcan}
g_{ij}(x),~\pi^{ij}(x) \rightarrow X^A (x), \Pi_A (x), \varrho^r (x), \pi_r (x)
\end{equation}

\noindent que separe las cuatro variables $X^A(x)$, A=0,1,2,3 que especifican
la hipersuperficie de los dos grados de libertad verdaderos del campo
gravitatorio $\varrho^r (x)$, r=1,2. Luego, se pueden hallar los v\'\i nculos
(\ref{vinchamil})-(\ref{vincmoment}) con respecto a los momentos $\Pi_A (x)$ y
reemplazarlos por el conjunto de v\'\i nculos equivalentes
\begin{equation}
H_A(x) \equiv \Pi_A(x)+h_A(x;X,\varrho,\pi)=0.\label{vincequiv}
\end{equation}

Las expresiones $h_A(x)$ representan la densidad de energ\'\i a y el flujo de
energ\'\i a asociados con las variables gravitacionales $\varrho$ y $\pi$ a
trav\'es de la hipersuperficie $X^A (x)$.

En principio, ahora se pueden obtener las leyes din\'amicas a partir de los
v\'\i nculos (\ref{vinchamil})-(\ref{vincmoment}) o de (\ref{vincequiv}). Sin
embargo, no se puede afirmar la existencia de un {\it tiempo global}, es
decir, que se pueda encontrar una variable que siempre crezca a lo largo de
cualquier trayectoria din\'amica, de modo que toda trayectoria intersecte una
hipersuperficie de tiempo constante una sola vez. Esto equivaldr\'\i a a la no
existencia de una transformaci\'on can\'onica (\ref{transfcan}) tal que los
v\'\i nculos originales (\ref{vinchamil})- (\ref{vincmoment}) sean totalmente
equivalentes a los nuevos v\'\i nculos (\ref{vincequiv}). Este problema se
denomina {\it el problema global del tiempo} [\cite{k92}].

La ``desaparici\'on'' del tiempo en este nivel cl\'asico, ser\'a la causante
del problema del tiempo en Gravedad Cu\'antica y de sus problemas de
interpretaci\'on.

\section{Cuantificaci\'on can\'onica.}

~~~~ Procederemos ahora a aplicar el m\'etodo de cuantificaci\'on can\'onica
de sistemas hamiltonianos con v\'\i nculos a la Gravedad
[\cite{dirac,dw67,k73}].

Comenzamos por convertir la m\'etrica $g_{ij}$ y el momento $\pi^{ij}$ en
operadores que satisfacen las reglas de conmutaci\'on:
\begin{equation}\label{conmutad1}
[g_{ij}(x) , g_{kl}(x')]=0~,~~ [\pi^{ij}(x) , \pi^{kl}(x')] = 0,
\end{equation}
\begin{equation}\label{conmutad2}
[g_{ij}(x) , \pi^{kl}(x')] = {i\over{2}}(\delta_i^k
\delta_j^l+\delta_i^l\delta_j^k) \delta (x ,x')
\end{equation}

Podemos ahora adoptar una representaci\'on particular, la representaci\'on de
la m\'etrica (en analog\'\i a con la representaci\'on de posici\'on en
mec\'anica cu\'antica ordinaria). En esta representaci\'on, el funcional de
onda se convierte en un funcional de la 3-m\'etrica, y el momento se reemplaza
por la derivada variacional con respecto a la 3-m\'etrica
\begin{equation}\label{opemom}
\pi^{ij}(x) = -i {\delta\over{\delta g_{ij}(x)}}.
\end{equation}

El paso siguiente es substituir estos operadores en el super-Hamiltoniano
(\ref{superham}) y en el supermomento (\ref{supermom}) e imponer los v\'\i
nculos como restricciones sobre los estados $\Psi$ del sistema:
\begin{equation}\label{ecwdw}
G_{ijkl}(x){\delta^2\Psi\over{\delta g_{ij}(x) \delta g_{kl}(x)}}-
\sqrt{g(x)}R(x)\Psi=0,
\end{equation}

\begin{equation}\label{ecsupermom}
\big[{\delta\Psi\over{\delta g_{ij}(x)}}\big]_{\vert j}=0.
\end{equation}

La ecuaci\'on (\ref{ecsupermom}) (en realidad $3\infty^3$ Ecs.), implica que
el funcional de onda $\Psi$ es invariante ante transformaciones de coordenadas
en la hipersuperficie espacial. Luego, el funcional de onda depende s\'olo de
la geometr\'\i a espacial y no de la m\'etrica particular elegida para
representarla. El dominio del funcional de onda es el {\it superespacio},
espacio abstracto de dimensi\'on infinita en el cual cada punto es una
3-geometr\'\i a [\cite{superesp}].

La versi\'on cu\'antica del v\'\i nculo super-Hamiltoniano (\ref{ecwdw}) es
llamada {\it ecuaci\'on de \break Wheeler-DeWitt}. Es una ecuaci\'on
variacional, y en realidad son $\infty^3$ ecuaciones, una para cada punto de
la hipersuperficie espacial.

Notemos que si en vez de partir de los v\'\i nculos
(\ref{vinchamil})-(\ref{vincmoment}) lo hacemos del v\'\i nculo
(\ref{vincequiv}), obtenemos en lugar de la ecuaci\'on de Wheeler-DeWitt una
ecuaci\'on funcional de Schr\"{o}dinger en las variables hipertemporales $X^A
(x)$ [\cite{k92}]:
\begin{equation}\label{ecscho}
 i {\delta \Psi [X,\varrho] \over \delta X^A (x)}=h_A
 (x;X,\hat{\varrho},\hat{\pi})
\Psi [X,\varrho].
\end{equation}

Esta ecuaci\'on, basada en una cierta elecci\'on de la variable temporal $X^A
(x)$, puede dar una teor\'\i a cu\'antica diferente si se basa en otra
elecci\'on de la variable temporal. Este problema se denomina {\it el problema
de la elecci\'on m\'ultiple}.

La ecuaci\'on de Schr\"{o}dinger autom\'aticamente nos da un producto interno
conservado en la variable temporal seleccionada. En cambio, la ecuaci\'on de
Wheeler-DeWitt, como toda ecuaci\'on diferencial funcional de segundo orden,
presenta problemas cuando se intenta convertir su espacio de soluciones en un
espacio de Hilbert ({\it problema del espacio de Hilbert}).

La consistencia de los v\'\i nculos (\ref{ecwdw})-(\ref{ecsupermom}), o bien
de (\ref{vincequiv}), queda establecida si sus conmutadores no generan nuevos
v\'\i nculos [\cite{dirac,dw67}].  Esto estar\'{a} garantizado toda vez que a
nivel cu\'{a}ntico logre hallarse operadores tales que realicen el \'{a}lgebra
(\ref{pb1}), (\ref{pb2}) y (\ref{pb3}) con los operadores de funciones de
estructura a la izquierda. Esto restringe el posible ordenamiento de los
operadores, pero no lo determina. Vimos que en el caso cl\'asico, los v\'\i
nculos siempre son consistentes, pero la presencia de la m\'{e}trica en
(\ref{pb1}) hace terriblemente dificultoso que esto se satisfaga en el caso
cu\'antico. Esta dificultad, {\it el problema de ordenamiento de los
operadores de v\'{\i}nculo}, es la que estudiaremos en los cap\'{\i}tulos
posteriores.

\section{El tiempo e interpretaciones de Gravedad Cu\'antica}

~~~~ Como hemos mencionado, el problema de ordenamiento, el problema de la
elecci\'on m\'ultiple y el problema del espacio de Hilbert son las tres
mayores dificultades que afronta la Gravedad Cu\'antica. Estas surgen como
consecuencia de no poseer una variable temporal natural a nivel cl\'asico
mientras que la teor\'\i a cu\'antica ordinaria se basa en la existencia de un
tiempo privilegiado. Este contraste es el que dificulta la interpretaci\'on de
la teor\'\i a.

Muchas han sido las propuestas para interpretar la Gravedad Cu\'antica. Pero,
b\'asicamen\-te existen tres maneras de encarar el problema del tiempo
[\cite{k92,ish92,ish95}]:

{\bf I. Marco del Tiempo Interno.} El tiempo est\'a oculto entre las variables
can\'onicas y debe ser identificado antes de la cuantificaci\'on. La
ecuaci\'on en la cual se basa esta interpretaci\'on es la ecuaci\'on de
Schr\"{o}dinger, no la ecuaci\'on de Wheeler-DeWitt. Esta clase de
interpretaciones est\'a expuesta al problema de la m\'ultiple elecci\'on.

{\bf II. Marco de Wheeler-DeWitt.} Los v\'\i nculos se imponen sobre los
estados en la representaci\'on de la m\'etrica dando por resultado la
ecuaci\'on de Wheeler-DeWitt. Se intenta una interpretaci\'on din\'amica de
las soluciones y no se pretende indentificar al tiempo entre las variables
m\'etricas. Esta clase de interpretaciones est\'a expuesta al problema del
espacio de Hilbert.

{\bf III. Marco de la Gravedad Cu\'antica sin Tiempo.} Se basa generalmente
(aunque no necesariamente) en la ecuaci\'on de Wheeler-DeWitt. Se propone que
el tiempo no es necesario para interpretar la Gravedad Cu\'antica e inclusive,
la Mec\'anica Cu\'antica en general. El tiempo puede surgir en situaciones
particulares, pero a\'un no siendo as\'\i, es posible una interpretaci\'on
probabil\'\i stica. Para una exposici\'{o}n no t\'{e}cnica de investigaciones
en esta direcci\'{o}n ver \cite{barbour}.

Esta clasificaci\'on de las interpretaciones es \'util, a\'un cuando los l\'\i
mites entre las tres clases no son precisos.

En esta tesis, trabajaremos dentro del primer marco mencionado, m\'{a}s
espec\'{\i}ficamente en la Interpretaci\'on Interna de Schr\"{o}dinger. En
esta interpretaci\'on, se insiste en que el tiempo debe ser identificado entre
las variables can\'onicas antes de efectuar la cuantificaci\'on. Ya vimos como
obtener la ecuaci\'on funcional de Schr\"{o}dinger que surge de esta visi\'on,
Ec. (\ref{ecscho}).

Dado que el problema de la evoluci\'on funcional sea resuelto, la ecuaci\'on
funcional de Schr\"{o}dinger es autoconsistente y se pueden hallar sus
soluciones $\Psi [X,\varrho]$. Al menos formalmente, la integral funcional
\begin{equation}\label{pischo}
 <\Psi \vert \Psi> \equiv \int D\varrho \vert \Psi[X,\varrho] \vert
^2,
\end{equation}
que no depende en las variables hipertemporales $X^A (x)$. El producto interno
(\ref{pischo}) convierte al espacio ${\cal F}_0$ de soluciones $\Psi
[X,\varrho]$ en un espacio de Hilbert. La estructura del espacio de Hilbert
provee la interpretaci\'on probabil\'\i stica usual del sistema cuantificado.
En particular, $ \vert \Psi[X,\varrho] \vert ^2 D\varrho$ es interpretado como
la probabilidad de hallar los verdaderos grados de libertad gravitacionales
$\varrho^r (x)$ en la celda $D\varrho$ sobre la hipersuperficie $X^A =X^A
(x)$. Adem\'as, el producto interno (\ref{pischo}) permite la construcci\'on
de {\it observables cu\'anticos} bien definidos. Cualquier operador
\begin{equation}\label{op}
\hat {F}=F[X,\hat{\varrho},\hat{\pi}],
\end{equation}
que es autoadjunto en el producto interno (\ref{pischo}) es un observable. Las
reglas usuales de la teor\'\i a cu\'antica nos dan la probabilidad que, en el
estado $\Psi$, el observable $\hat{F}$ tome el valor F permitido por su
espectro, sobre la hipersuperficie $X^A (x)$.

La cuesti\'on b\'asica en este marco interpretativo es c\'omo seleccionar la
variable de tiempo interno. Se han explorado tres opciones:

$\bullet$ {\it Tiempo Intr\'\i nseco.} Se asume que la variable temporal se
construye completamente a partir de la m\'etrica intr\'\i nseca de la
hipersuperficie. \cite{k53} fue uno de los pioneros en hallar un tiempo
intr\'\i nseco en el caso de modelos cosmol\'ogicos homog\'eneos.

$\bullet$ {\it Tiempo Extr\'\i nseco.} Para identificar la hipersuperficie se
necesita, adem\'as de su m\'etrica intr\'\i nseca, su curvatura extr\'\i
nseca; es decir, c\'omo est\'a curvada en el espacio-tiempo que la contiene.
Como ejemplos podemos mencionar el tiempo que se construye a partir de la
curvatura extr\'\i nseca media [\cite{york}], y el que se halla en el caso de
simetr\'\i a cil\'\i ndrica [\cite{k71,k73}].

$\bullet$ {\it Tiempo Material.} El tiempo no se construye a partir de
cantidades geom\'etricas, sino a partir de los campos de materia acoplados a
la gravedad. La introducci\'on de materia facilita el manejo de los v\'\i
nculos que llevan a la ecuaci\'on de Schr\"{o}dinger [\cite{k92}].

En esta interpretaci\'on surge nuevamente el problema global del tiempo a
nivel cl\'asico: puede ocurrir que el sistema de v\'\i nculos
(\ref{vincequiv}) no sea globalmente equivalente al sistema de v\'\i nculos
(\ref{vinchamil})-(\ref{vincmoment}) para cualquier elecci\'on de tiempo
interno. Esto puede ocurrir si no existe una funci\'on de tiempo global en el
espacio de fases tal que cada trayectoria cl\'asica intersecte una sola vez
toda hipersuperficie de tiempo constante. Este problema se ha estudiado en
modelos de sistemas sencillos: un par de osciladores arm\'onicos en un estado
estacionario [\cite{k11,k12}] y en modelos cosmol\'ogicos homog\'eneos. Pero,
es muy poco lo que se sabe acerca de este problema en la teor\'\i a completa.

Tambi\'en aparece el problema de la elecci\'on m\'ultiple: si no hay una
elecci\'on geom\'etrica\-mente natural todas las elecciones de tiempo son
igualmente buenas (o igualmente malas).

Debido a la gran complejidad de la teor\'{\i}a completa, es un procedimiento
habitual el ``congelar'' la mayor\'{\i}a de los grados de libertad de la
teor\'{\i}a (en realidad, infinitos) para quedarse con modelos de
dimensi\'{o}n finita, son los denominados modelos de {\it minisuperespacio}
[\cite{hall}], ejemplos t\'{\i}picos son los modelos usuales de la
cosmolog\'{\i}a. El hecho de trabajar con modelos de dimensi\'{o}n finita, no
s\'{o}lo hace factible el estudio de modelos con las patolog\'{\i}as de la
teor\'{\i}a completa, sino que tambi\'{e}n evita otra dificultad, que es la
divergencia ultravioleta de los operadores al cuantificar la teor\'{\i}a. Por
estas razones, ser\'{a}n modelos de este tipo los que se estudiar\'{a}n en
esta tesis, m\'{a}s espec\'{\i}ficamente modelos con tiempo intr\'{\i}nseco y
extr\'{\i}nseco.

\newpage

\thispagestyle{empty}

 ~\newpage


\hyphenation{for-ma-lis-mo}

\chapter{El formalismo de Becchi-Rouet-Stora-Tyutin}

\bigskip

{\it~~~~ Desde el tratamiento original de Dirac, los sistemas con invariancia
de gauge han sido estudiados extensivamente. Un gran avance se produjo con los
estudios de \cite{fv70} sobre la formulaci\'{o}n covariante de la integral de
camino ante la elecci\'{o}n del gauge para sistemas con \'{a}lgebras que no
cierran, extendiendo as\'{\i} el trabajo pionero de \cite{fp67} para
\'{a}lgebras cerradas. Estos trabajos introdujeron la utilizaci\'{o}n de
variables fermi\'{o}nicas (es decir, anticonmutantes) denominadas
``fantasmas'' que aseguraban la unitariedad de la teor\'{\i}a y la
independencia de la elecci\'{o}n del gauge. Sin embargo, fue el descubrimiento
de la simetr\'{\i}a de Becchi-Rouet-Stora-Tyutin (BRST) [\cite{brs,t,h85}] que
les di\'{o} a los fantasmas un rol prominente. La necesidad de los fantasmas y
la simetr\'{\i}a que revela su importancia fue establecida primero a nivel
cu\'{a}ntico. M\'{a}s tarde se not\'{o} que ten\'{\i}a tambi\'{e}n lugar,
necesariamente y naturalmente, a nivel cl\'{a}sico.

La idea central de la teor\'{\i}a  BRST es reemplazar la simetr\'{\i}a de
gauge original por una supersimetr\'{\i}a global que act\'{u}a en un espacio
de fases extendido apropiadamente. Esta supersimetr\'{\i}a captura la
invariancia de gauge original y conduce a una formulaci\'{o}n m\'{a}s simple
de la teor\'{\i}a.

En este cap\'{\i}tulo desarrollaremos los conceptos fundamentales de la
formulaci\'{o}n BRST, a nivel cl\'{a}sico y cu\'{a}ntico, y mostraremos su
relaci\'{o}n con la cuantificaci\'{o}n can\'{o}nica de Dirac.}

\section{El formalismo BRST cl\'{a}sico}

\bigskip

~~~~Comenzaremos por considerar un conjunto $\{ G_a(q^i, p_i)\}$ de
v\'{\i}nculos de primera clase independientes (sistema irreducible):
\begin{equation}
\{G_a,G_b\}=C_{ab}^c G_c.
\end{equation}

A continuaci\'{o}n, el espacio de fases original de la teor\'{\i}a $(q^i,
p_i)$ es extendido con pares can\'{o}nicamente conjugados de fantasmas
$(\eta^a,{\cal P}_a)$ (uno por cada v\'{\i}nculo) tales que $$\{{\mathcal
P}_a, \eta^b \}=-{\delta}_a^b$$ con $\varepsilon({\mathcal
P}_a)=\varepsilon(\eta^{a})=\varepsilon_a+1$, donde
$\varepsilon_a\doteq\varepsilon(G_a)$. Es decir, son de paridad de
Grassmann\footnote{En el Ap\'{e}ndice A pueden encontrarse las definiciones
b\'{a}sicas y la notaci\'{o}n que involucran al \'{a}lgebra de Grassmann.}
opuesta al v\'{\i}nculo correspondiente (por ej., para v\'{\i}nculos
bos\'{o}nicos se agregar\'{a}n variables fermi\'{o}nicas). Adem\'{a}s, las
nuevas variables cumplen que $\eta^{a*}=\eta^a$ y ${\mathcal
P}_a^*=(-1)^{\varepsilon_a+1} {\mathcal P}_a$. Por otra parte, $(\eta^a,{\cal
P}_a)$ tienen corchete nulo con cualquiera de las variables can\'{o}nicas
originales $(q^i, p_i)$.

Se dota al espacio de fases extendido con una estructura adicional, el
n\'{u}mero de fantasma ({\it ghost}) $$ gh (q^i)= gh
(p_i)=0,~~~~~~~~~~~~~~~~~~~~~gh (\eta^a)=1,~~~~gh({\mathcal P}_a)=-1,$$

El n\'{u}mero de fantasma posee un generador: $$ {\mathcal{G}}\doteq i ~
\eta^a {\mathcal P}_a,~~~~~\varepsilon({\mathcal G})=0,~~~~~ {\mathcal
G}^{\ast}=-{\mathcal G}.$$

En efecto, $\{ \eta^a,{\mathcal G}\}= i \eta^a~~~~\{{\mathcal
 P}_a,{\mathcal G}\}=- i {\mathcal P}_a$ (el coeficiente $i$ est\'{a} puesto
para que los autovalores sean reales en mec\'{a}nica cu\'{a}ntica).

En general, $\{A, {\mathcal G}\} =i ~gh(A) A$.

El formalismo BRST se basa en el siguiente teorema:

\noindent \underline{\bf Teorema 3.1}: en el espacio de fases extendido existe
una cantidad $\Omega=\Omega(q^i,p_j,\eta^a,{\cal P}_b)$, el {\it generador
BRST}, tal que:
\begin{itemize}
  \item  $\Omega^*=\Omega~~~~ gh\Omega=1~~~~\varepsilon(\Omega)=1$
  \item $\Omega=\eta^a G_a+$ (t\'{e}rminos de orden superior en los
  fantasmas)
  \item $\{\Omega,\Omega\}=0$ (no trivial, pues $\Omega$ es fermi\'{o}nico).
\end{itemize}
$\Omega$ es \'{u}nico a menos de transformaciones can\'{o}nicas en el espacio
de fases extendido.

\noindent {\it Demostraci\'{o}n.} Siempre es posible escoger (localmente)
variables $(q,p)$ tales que los v\'{\i}nculos $F_a=0$ son abelianos, es decir,
$\{F_a,F_b\}=0$. En tal caso, $\Omega$ resulta estrictamente $\Omega=\eta^a
F_a$. N\'{o}tese que la propiedad de nilpotencia de $\Omega$
($\{\Omega,\Omega\}=0$), no es mas que otra forma de expresi\'{o}n del
car\'{a}cter abeliano de los v\'{\i}nculos, m\'{a}s adelante veremos que en
casos m\'{a}s generales (con v\'{\i}nculos no abelianos, o \'{a}lgebras
abiertas), de todas formas $\Omega$ captura en la propiedad de nilpotencia
toda el \'{a}lgebra de los v\'{\i}nculos.

Hagamos ahora, una transformaci\'{o}n can\'{o}nica (que conserva
$\{\Omega,\Omega\}=0$) generada por $\eta^b\varepsilon^a_b(q,p){\mathcal P}_a$
con $det \varepsilon^a_b\neq0$, y $\varepsilon^a_b$ es tal que el generador es
real y bos\'{o}nico
($\varepsilon(\varepsilon^a_b)=\varepsilon_a+\varepsilon_b,
~\varepsilon^{a\ast}_b=(-1)^{\varepsilon_a(\varepsilon_b+1)}\varepsilon^a_b$).
Luego,
\begin{equation}
\delta\Omega=\{\Omega,\eta^b\varepsilon^a_b(q,p){\mathcal
P}_a\}=-\{\eta^b\varepsilon^a_b{\mathcal P}_a,
\Omega\}=-\{\eta^b\varepsilon^a_b{\mathcal P}_a,\eta^c
F_c\}=\eta^b\varepsilon^a_b{\mathcal P}_a+\eta\eta\mathcal{P}...+...
\end{equation}
Y obtenemos que
\begin{equation}
\Omega=\eta^b(\delta^a_b+\varepsilon^a_b)F_a+\eta\eta\mathcal{P}...+...
\end{equation}

N\'{o}tese que $G_b=(\delta^a_b+\varepsilon^a_b)F_a$ son funciones de
v\'{\i}nculo que caracterizan la misma superficie de v\'{\i}nculo que las
$F_a$. El resultado general puede obtenerse a partir del infinitesimal por
exponenciaci\'{o}n. Si bien esta demostraci\'{o}n es sencilla, no provee un
m\'{e}todo general para la construcci\'{o}n del generador BRST. \hfill $\Box$

\bigskip
\subsection{La construcci\'{o}n del generador BRST}

~~~~ En esta secci\'{o}n describiremos un m\'{e}todo general recursivo para la
construcci\'{o}n del generador BRST $\Omega$. Su existencia est\'{a}
garantizada a nivel cl\'{a}sico (y como vimos es \'{u}nico a menos de
transformaciones can\'{o}nicas en el espacio de fases extendido).

Sea $\delta$ el operador de Koszul-Tate:
\begin{equation}\label{kt}
\delta q=0=\delta p, ~~~~~\delta\eta=0,~~~~~\delta{\mathcal P}_a=-G_a,
\end{equation}
que satisface $\delta^2$=0. Frente a productos y sumas $\delta$ se comporta
como una diferenciaci\'{o}n impar actuando por derecha:
\begin{equation}
\delta(fg)=f~\delta g+ (-1)^{\varepsilon_g} \delta f ~g.
\end{equation}
Se dice que $\delta$ es un ``diferencial'' (pues $\delta^2$=0).

Por otro lado, desarrollemos $\Omega$ como:
\begin{equation}\label{deso}
\Omega=\sum_{p\geq0}{\Omega}\!\!\!\!\!^{^{^{(p)}}},
~~~~~~~~{\Omega}\!\!\!\!\!^{^{^{(0)}}} =\eta^aG_a,
\end{equation}
donde ${\Omega}\!\!\!\!\!^{^{^{(p)}}}$ tiene la forma gen\'{e}rica:
\begin{equation}
{\Omega}\!\!\!\!\!^{^{^{(p)}}}=\eta^{b_1}...\eta^{b_{p+1}}
{\mathrm{U}\!\!\!\!\!^{^{^{(p)}}}}^{a_1...a_p}_{b_1...b_{p+1}}(q,p) {\mathcal
P}_{a_1}...{\mathcal P}_{a_p}.
\end{equation}

Las funciones $\mathrm{U}\!\!\!\!\!^{^{^{(p)}}}$ se llaman funciones de
estructura. En particular, los v\'{\i}nculos son las funciones de estructura
de orden cero, ${\mathrm{U}\!\!\!\!\!^{^{^{(0)}}}}_a=G_a$, y las de primer
orden, ${\mathrm{U}\!\!\!\!\!^{^{^{(1)}}}}^c_{ab}=-{1\over 2}
(-1)^{\varepsilon_a}C^c_{ba}$.

Luego, puede probarse que $\{\Omega,\Omega\}=0$ es equivalente a que las
$\Omega\!\!\!\!\!^{^{^{(p)}}}$ satisfagan:
\begin{equation}\label{rec}
\delta\Omega\!\!\!\!\!\!\!^{^{^{(p+1)}}}+\mathrm{D}\!\!\!\!\!^{^{^{(p)}}}=0,
\end{equation}
con
\begin{equation}\label{Deq}
\mathrm{D}\!\!\!\!\!^{^{^{(p)}}}={1\over 2}\left[ \sum_{k=0}^p
\{\Omega\!\!\!\!\!^{^{^{(k)}}},\Omega\!\!\!\!\!\!\!^{^{^{(p-k)}}}\}_{orig} +
\sum_{k=0}^{p-1} \{\Omega\!\!\!\!\!\!\!^{^{^{(k+1)}}},
\Omega\!\!\!\!\!\!\!^{^{^{(p-k)}}}\}_{{\mathcal P},\eta}\right],
\end{equation}
donde $\{,\}_{orig}$ es el corchete en el espacio $(q,p)$, mientras que
$\{,\}_{{\mathcal P},\eta}$ es el corchete en el espacio $(\eta, {\mathcal
P})$.

La ecuaci\'{o}n anterior puede ser vista como una herramienta para construir
cada $\Omega\!\!\!\!\!^{^{^{(p)}}}$, partiendo del conocido
$\Omega\!\!\!\!\!^{^{^{(0)}}}$.

\bigskip

Veamos como resulta el generador BRST para algunos casos simples,

\noindent {\bf V\'{\i}nculos abelianos:}

El conjunto m\'{a}s simple posible es el de v\'{\i}nculos abelianos,
\begin{equation}\label{va}
  \{G_a,G_b\}=0.
\end{equation}
En este caso, las funciones de estructura $C^c_{ab}$ se anulan y
$\Omega\!\!\!\!\!^{^{^{(0)}}}$ es nilpotente sin necesidad de t\'{e}rminos
adicionales. Luego, el generador BRST puede ser elegido coincidiendo con
$\Omega\!\!\!\!\!^{^{^{(0)}}}$,
\begin{equation}\label{omegaej}
\Omega=\eta^a  G_a,
\end{equation}
Como ya hab\'{\i}amos visto en la demostraci\'{o}n del teorema,
$\{\Omega,\Omega\}=0$ no es m\'{a}s que la expresi\'{o}n de la condici\'{o}n
abeliana de los v\'{\i}nculos, Ec. (\ref{va}).

\bigskip

\noindent{\bf V\'{\i}nculos que cierran de acuerdo a un grupo}:

El caso de dificultad creciente al caso abeliano, es el dado por un sistema
con transformaciones de gauge que cierran de acuerdo a un grupo. En este caso,
los corchetes de Poisson no se anulan en todo el espacio de fases. En cambio,
se tiene que
\begin{equation}\label{fcv}
  \{G_a,G_b\}=C^c_{ab} G_c,
\end{equation}
donde $C^c_{ab}$ son constantes. La identidad de Jacobi para los corchetes de
Poisson implica la siguiente identidad de Jacobi\footnote{La versi\'on
generalizada de la identidad de Jacobi puede hallarse en el Ap\'{e}ndice A.}
para las constantes de estructura $C^c_{ab}$,
\begin{equation}\label{idjacobi}
C^c_{ab}C^e_{cd} + (-1)^{(\varepsilon_b+\varepsilon_d)\varepsilon_a}
C^c_{bd}C^e_{ca} + (-1)^{(\varepsilon_a+\varepsilon_b)\varepsilon_d}
C^c_{da}C^e_{cb}=0.
\end{equation}

Si $C^c_{ab}\neq 0$, $\Omega\!\!\!\!\!^{^{^{(0)}}}$ no es nilpotente por
s\'{\i} solo. Se necesita agregar el t\'{e}rmino
$\Omega\!\!\!\!\!^{^{^{(1)}}}$. Esto es suficiente para que $\Omega$ sea
nilpotente debido a la identidad de Jacobi (\ref{idjacobi}) y a que las
$C^c_{ab}$ son constantes. Luego, $\Omega$ est\'{a} dado por
\begin{equation}\label{omegaej2}
\Omega=\eta^a  G_a + {1\over 2} (-1)^{\varepsilon_a}  \eta^a \eta^b C_{ab}^c
{\cal P}_c.
\end{equation}
La nilpotencia de $\Omega,~ \{\Omega,\Omega\}=0$, se deduce de (\ref{fcv})-
(\ref{idjacobi}). Y dicha propiedad es de hecho, completamente equivalente a
las ecuaciones de estructura (\ref{fcv})- (\ref{idjacobi}).

\bigskip

En el caso m\'{a}s general de un \'{a}lgebra abierta, la suma
$\Omega\!\!\!\!\!^{^{^{(0)}}}+\Omega\!\!\!\!\!^{^{^{(1)}}}$ puede no ser
nilpotente, y en general se requieren t\'{e}rminos de orden m\'{a}s alto en
$\Omega$. Estos t\'{e}rminos de orden superior caracterizan la estructura
m\'{a}s complicada del \'{a}lgebra de gauge. En verdad, las funciones de
estructura $\mathrm{U}\!\!\!\!\!^{^{^{(2)}}},
\mathrm{U}\!\!\!\!\!^{^{^{(3)}}},...$ podr\'{\i}an ser constru\'{\i}das sin
haber introducido los fantasmas y sus momentos. Esto puede hacerse a partir de
la propiedad de primera clase y explorando sistem\'{a}ticamente las
consecuencias de la identidad de Jacobi.

Sin embargo, es solamente en el espacio de fases extendido que la covariancia
can\'{o}nica de la estructura de un sistema de v\'{\i}nculos de primera clase
es manifiesta. Mas a\'{u}n, es solamente despu\'{e}s de haber introducido los
fantasmas, que la existencia de funciones de estructura de orden superior
revela un contenido algebraico interesante. El uso del formalismo BRST para
estudiar la estructura de \'{a}lgebras abiertas es pues conceptualmente
m\'{a}s claro y mucho m\'{a}s econ\'{o}mico.

\newpage

\noindent {\bf Rango de la teor\'{\i}a:}

Se dice que que un conjunto de v\'{\i}nculos y sus funciones de estructura
asociadas es de rango $r$ si todas las funciones de estructura de orden
estrictamente mayor que $r$ se anulan. Esto significa que el generador BRST
correspondiente contiene a lo sumo $r$ momentos fantasmas ${\cal P}_a$.

Las teor\'{\i}as abelianas son de rango cero. Las teor\'{\i}as basadas en
grupos de gauge verdaderos son de rango uno. Rangos mayores que uno
generalmente ocurren en teor\'{\i}as con \'{a}lgebras abiertas.

El concepto de rango no es intr\'{\i}nseco a la superficie de v\'{\i}nculo,
sino al conjunto de funciones de v\'{\i}nculo que se usan para describirla
(existe ambig\"uedad en la definici\'{o}n de las funciones de estructura, dado
que el conjunto de v\'{\i}nculos puede ser reemplazado por otro equivalente).

\noindent \underline{\bf Teorema 3.2}:  Si todas las funciones de estructura
de orden $k$ se anulan para $r<k\leq 2r+1$, luego todas las funciones de
estructura de orden estrictamente mayor que $2r+1$ pueden ser elegidas nulas,
y el conjunto de funciones de estructura es de rango $r$.

\noindent {\it Demostraci\'{o}n.} Inspeccionando la Ec. (\ref{Deq}), se
observa que si $\Omega\!\!\!\!\!^{^{^{(k)}}}$ se anula (luego, se anula la
funci\'{o}n de estructura de orden $k$ ya que es el coeficiente del polinomio
en los fantasmas $\Omega\!\!\!\!\!^{^{^{(k)}}}$) para $r<k\leq 2r+1$ entonces
se anula $~~\mathrm{D}\!\!\!\!\!\!\!\!^{^{^{(2r+1)}}}$ y por lo tanto
tambi\'{e}n se anula $~~~\Omega\!\!\!\!\!\!\!\!^{^{^{(2r+2)}}}$ y todos los
$~\Omega\!\!\!\!\!^{^{^{(p)}}}$ siguientes. Este resultado puede ser \'{u}til
para la determinaci\'{o}n del rango de un sistema de funciones de estructura.
\hfill $\Box$

\subsection{Observables, evoluci\'{o}n din\'{a}mica y simetr\'{\i}a BRST}

~~~~ La existencia de $\Omega$ con la propiedad $\{\Omega,\Omega\}=0$ en el
espacio de fases extendido, nos permitir\'{a} desarrollar una mec\'{a}nica en
el espacio de las fases extendido.

En la teor\'{\i}a original, los observables son funciones $F(q,p)$ que tiene
corchetes d\'{e}bilmente nulos con los v\'{\i}nculos de primera clase:
\begin{equation}\label{opc}
\{F,G_a\}\approx 0.
\end{equation}
Esto significa que $F$ queda determinada a menos de una combinaci\'{o}n lineal
de los v\'{\i}nculos:
\begin{equation}\label{obsdeq}
F'(q,p)=F(q,p)+k^a(q,p)G_a(q,p)\approx F(q,p)
\end{equation}
y por lo tanto, $F'$ y $F$ deber\'{\i}an considerarse equivalentes.

En el espacio extendido, asignaremos a cada observable $F_0(q,p)$ una cantidad
$F(q,p,\eta,{\cal P})$ tal que:
\begin{equation}\label{equivexte}
F(q,p,\eta,{\cal P})\mid_{_{\eta=0={\cal P}}}=F_0(q,p).
\end{equation}

En general, se dir\'{a} que $F(q,p,\eta,{\cal P})$ es un observable si
\begin{equation}\label{obsbrst}
\{F,\Omega\}=0,~~~~~gh F=0.
\end{equation}

Es decir que $F=F_0(q,p)+\eta^a F_{1a}^b(q,p) {\cal P}_b+$ (al menos cuatro
fantasmas). Obs\'{e}rvese que al orden m\'{a}s bajo en los fantasmas se
recupera la propiedad (\ref{opc}) indicando que la extensi\'{o}n de los
observables es adecuada.

Puede verse que si se tiene una funci\'{o}n $K$ con $gh K=-1$, entonces
$\{K,\Omega\}$ es un observable:
\begin{description}
\item[(i)] $\{\{K,\Omega\},\Omega\}=
(-1)^{\varepsilon_K}\{\Omega,\{K,\Omega\}\}= \{\Omega,\{\Omega,K\}\}=
-(-1)^{\varepsilon_K}\{\Omega,\{K,\Omega\}\}$

$\Longrightarrow \{\{K,\Omega\},\Omega\}=0.$

\item[(ii)] $gh \{K,\Omega\}= gh K + gh \Omega=0.$
\end{description}

Veamos a qu\'{e} observable corresponde en el espacio original. Como $ghK=-1$,
debe ser de la forma: $K=-k^a(q,p){\cal P}_a+\eta^a K_a^{bc}(q,p) {\cal
P}_b{\cal P}_c+...$ Luego, al orden m\'{a}s bajo en los fantasmas es
\begin{equation}\label{kclas}
\{K,\Omega\}\mid_{_{\eta=0={\cal P}}}=k^aG_a.
\end{equation}

Esto significa que la equivalencia apuntada entre los observables del espacio
original, puede trasladarse al espacio extendido como una equivalencia entre
$F$ y $F'=F +\{K,\Omega\}$, con $\varepsilon_K=\varepsilon_F+1$. Es necesario
que la equivalencia sea cerrada ante adici\'{o}n, multiplicaci\'{o}n y
corchete, todas estas propiedades se satisfacen con la ayuda de la propiedad
$\{\Omega,\Omega\}=0$.

En particular, el Hamiltoniano can\'{o}nico es un observable, pues conmuta
d\'{e}bilmente con los v\'{\i}nculos para asegurar consistencia din\'{a}mica.
Luego, existe una extensi\'{o}n BRST del Hamiltoniano que est\'{a} determinada
a menos del t\'{e}rmino $\{K,\Omega\}$, que al orden m\'{a}s bajo en los
fantasmas d\'{a} los t\'{e}rminos que aparecen en el Hamiltoniano extendido.

Podemos dar entonces ecuaciones din\'{a}micas en el espacio de las fases
extendido: para cualquier magnitud $A(q,p,\eta,{\cal P})$ sea o no observable,
su evoluci\'{o}n en el tiempo est\'{a} dada por
\begin{equation}\label{evodinb}
{dA\over dt}=\{A, H\},
\end{equation}
como $gh H=0$ entonces $gh A(t)= gh A(t=0)$. En particular, si los fantasmas
son inicialmente nulos, continuar\'{a}n siendo nulos.

La ambig\"uedad en $H$ ($H$ es equivalente a $H'=H+\{K,\Omega\}$ con
$\varepsilon_K=1$) le agrega a esta ecuaci\'{o}n un t\'{e}rmino
$\{A,\{K,\Omega\}\}=-\{K,\{\Omega,A\}\}-\{\Omega\,\{A,K\}\}$ (donde se ha
usado la identidad de Jacobi en el segundo miembro). Entonces en el caso de un
observable $F$ tenemos:
\begin{equation}\label{evodinobs}
{d'F\over dt}=\{F,H'\}=\{F,H\}-\{\Omega,\{F,K\}\}
\end{equation}
es decir que $d'F\over dt$ est\'{a} en la misma clase de equivalencia que
$dF\over dt$. Adem\'{a}s, a orden cero en los fantasmas esta ecuaci\'{o}n
corresponde a aquella con el Hamiltoniano extendido.

La teor\'{\i}a posee una cantidad conservada: el generador BRST $\Omega$. En
efecto,
\begin{equation}\label{evodin}
{d\Omega\over dt}=\{\Omega, H\}=0
\end{equation}
pues $H$ es un observable. N\'{o}tese adem\'{a}s que en este caso la
ambig\"{u}edad en $H$ no agrega ning\'{u}n t\'{e}rmino debido a la propiedad
$\{\Omega,\Omega\}=0$.

La existencia de una cantidad conservada {\it on-shell} es se\~{n}al de que la
teor\'{\i}a posee una simetr\'{\i}a global. En efecto, la din\'{a}mica puede
obtenerse de la acci\'{o}n
\begin{equation}\label{accionbrst}
S[q,p,\eta,{\cal P}]=\int dt (\dot q^ip_i+\dot\eta^a{\cal P}_a
-H-\{K,\Omega\}).
\end{equation}
Esta acci\'{o}n es invariante frente a la transformaci\'{o}n global (con
par\'{a}metros independientes de $t$) generada por $\Omega$:
\begin{equation}\label{transfom}
\delta f= \{f,\Omega\},
\end{equation}
la cual es nilpotente: $\delta^2 f= \{\{f,\Omega\},\Omega\}=0$ (comp\'{a}rese
con la demostraci\'{o}n de que $\{K,\Omega\}$ es observable). En efecto,
$$\delta H= \{H,\Omega\}=0,$$ $$\delta\{K,\Omega\}=\delta^2K=0,$$ y el
t\'{e}rmino cin\'{e}tico s\'{o}lo cambia por un t\'{e}rmino de superficie pues
la transformaci\'{o}n es can\'{o}nica.

Luego, en el espacio de las fases extendido las simetr\'{\i}as locales (de
gauge) han sido capturadas por una (super) simetr\'{\i}a global generada por
$\Omega$: la simetr\'{\i}a BRST.

De modo que las trayectorias din\'{a}micas del sistema original pueden ser
recuperadas con s\'{o}lo tomar condiciones iniciales adecuadas: fantasmas
inicialmente nulos (ya vimos que en tal caso permanecer\'{a}n siempre nulos) y
$(q,p)$ sobre la superficie de v\'{\i}nculo (recordemos que la evoluci\'{o}n
dejar\'{a} al punto sobre la superficie de v\'{\i}nculo). La din\'{a}mica de
una teor\'{\i}a de gauge puede pensarse como proveniente de la restricci\'{o}n
de las condiciones iniciales de un sistema con un n\'{u}mero mayor de grados
de libertad \underline{no vinculados}, en el cual las simetr\'{\i}as de gauge
han sido reemplazadas por una \'{u}nica simetr\'{\i}a global. En la acci\'{o}n
de partida, los v\'{\i}nculos surg\'{\i}an al variar los multiplicadores de
Lagrange, en cambio la acci\'{o}n BRST no provee ecuaciones de v\'{\i}nculo.

\section{El formalismo BRST cu\'{a}ntico}

~~~~ La clave del formalismo BRST consiste en que agregando m\'{a}s
redundancia (los fantasmas) a un sistema que ya era redundante (porque
ten\'{\i}a v\'{\i}nculos) la descripci\'{o}n se vuelve finalmente m\'{a}s
transparente y, en alg\'{u}n sentido, las dos redundancias se cancelan entre
s\'{\i}.

Para que las redundancias se cancelen entre s\'{\i} se debe imponer una
condici\'{o}n que seleccionar\'{a} un subespacio de estados f\'{\i}sicos. Esta
condici\'{o}n debe ser an\'{a}loga a la demanda de que los estados
f\'{\i}sicos sean aniquilados por los generadores de las transformaciones de
gauge $\hat{G}_a$ en el formalismo sin fantasmas.

A nivel cl\'{a}sico la invariancia de gauge nos lleva a considerar a los
observables $F$ y $F+k^aG_a$ como equivalentes. Esta equivalencia se traduce
en BRST como la equivalencia entre los observables BRST $F$ y
$F+\{K,\Omega\}$. A nivel cu\'{a}ntico, los observables son operadores
lineales que conmutan con $\hat\Omega$: $[\hat F,\hat\Omega]=0$. $\hat F$ y
$\hat F+[\hat K,\hat\Omega]$ deber\'{\i}an considerarse equivalentes.
Realizamos esta equivalencia exigiendo que
\begin{equation}\label{cad}
\hat\Omega\psi_{f\acute{\i}sico}=0,
\end{equation}
donde $\hat{{\Omega}}$, es el operador herm\'{\i}tico asociado del generador
BRST cl\'asico en el producto interno
\begin{equation}\label{pibrst}
\int dq d\eta \psi^*(q,\eta) \varphi(q,\eta),
\end{equation}
donde $\psi,~ \varphi$ son superdensidades de peso ${1\over 2}$.

Adem\'as, la propiedad cl\'asica de nilpotencia tambi\'en debe satisfacerse a
este nivel en la siguiente forma,
$$[\hat{{\Omega}},\hat{{\Omega}}]=2\hat{{\Omega}}^2=0.$$

Recordemos que $[,]$ indica el anticonmutador cuando se toma sobre dos
operadores ambos fermi\'{o}nicos. En particular,
\begin{equation}\label{rccan}
[\hat{\mathcal P}_a,\hat\eta^b]=-i\delta^b_a
\end{equation}
significa
\begin{equation}
\hat{\mathcal
P}_a\hat\eta^b-(-1)^{(\varepsilon_a+1)(\varepsilon_b+1)}\hat\eta^b
\hat{\mathcal P}_a=-i\delta^b_a.
\end{equation}

Si la nilpotencia de $\hat\Omega$ ($\hat{{\Omega}}^2=0$) no puede ser
alcanzada a trav\'{e}s de ning\'{u}n ordenamiento herm\'{\i}tico, se dice que
las teor\'{\i}a presenta anomal\'{\i}as. En este caso la cuantificaci\'{o}n no
ser\'{a} consistente.

La condici\'{o}n que selecciona los estados f\'{\i}sicos tiene las siguientes
propiedades:
\begin{description}
\item[{\it 1.}] Es lineal, por lo tanto define un subespacio.
\item[{\it 2.}] Los observables BRST conmutan con $\hat\Omega$, por lo tanto
cuando act\'{u}an sobre un estado f\'{\i}sico dan por resultado otro estado
f\'{\i}sico.
\item[{\it 3.}] Los observables triviales $[\hat K,\hat\Omega]$ tienen elementos
de matriz nulos entre estados f\'{\i}sicos.
\end{description}

As\'{\i} como existe una relaci\'{o}n de equivalencia entre observables,
tambi\'{e}n se puede establecer una relaci\'{o}n de equivalencia entre estados
f\'{\i}sicos (``libertad de gauge cu\'{a}ntica''). En efecto, la nilpotencia
de $\hat\Omega$ garantiza que si $\chi$ es un estado cu\'{a}ntico cualquiera
entonces $\hat\Omega\chi$ es un estado f\'{\i}sico. Sin embargo,
$\hat\Omega\chi$ deber\'{\i}a ser identificado con cero pues $\hat\Omega\chi$
tiene producto interno nulo con cualquier estado f\'{\i}sico:
\begin{equation}\label{pif}
(\psi_{f\acute{\i}sico},\hat\Omega\chi)=
(\hat\Omega\psi_{f\acute{\i}sico},\chi)=0.
\end{equation}

En particular la aplicaci\'{o}n de $[\hat K,\hat\Omega]$ a un estado
f\'{\i}sico da
\begin{equation}
[\hat K,\hat\Omega]\psi_{f\acute{\i}sico}= \hat\Omega\hat K
\psi_{f\acute{\i}sico},
\end{equation}
es decir, un estado que puede ser identificado con cero.

Por lo tanto, la equivalencia
\begin{equation}
\psi_{f\acute{\i}sico} \sim \psi_{f\acute{\i}sico}+\hat\Omega\chi
\end{equation}
da consistencia a la equivalencia
\begin{equation}\label{equivop}
\hat F \sim \hat F+[\hat K,\hat\Omega]
\end{equation}
(pues la aplicaci\'{o}n de cada uno a un mismo estado f\'{\i}sico da por
resultado estados f\'{\i}sicos equivalentes).

En particular la extensi\'{o}n BRST de los v\'{\i}nculos es

\begin{equation}\label{vincbrst}
\hat G_a= i[\hat{\mathcal P }_a,\hat\Omega].
\end{equation}
Por lo tanto,
\begin{equation}
\hat G_a  \psi_{f\acute{\i}sico}=i
(-1)^{(\varepsilon_a)}\hat\Omega\hat{\mathcal P }_a \psi_{f\acute{\i}sico}
\end{equation}
es un estado que puede ser identificado con cero.

\subsection{Relaci\'{o}n entre la cuantificaci\'{o}n BRST y el m\'{e}todo de
Dirac}

~~~~ Seg\'{u}n hemos visto, el operador BRST que extiende a los operadores de
v\'{\i}nculo es:
\begin{equation}\label{opvext}
\hat G_a(\hat q, \hat p, \hat \eta, \hat{\mathcal P})= i[\hat{ \mathcal
P}_a,\hat\Omega],
\end{equation}
de modo que el operador de Dirac corresponde al t\'{e}rmino sin fantasmas en
el resultado del conmutador.

Para visualizar esto uno puede reordenar en el $\hat\Omega$ herm\'{\i}tico los
$\hat{\mathcal P}$ de modo que queden a la derecha de los $\hat\eta$, teniendo
en cuenta que los (anti)-conmutadores no son en general nulos. Luego de
aplicar este procedimiento resulta:
\begin{equation}\label{omegord}
\hat\Omega=\hat\eta^a\hat G_a(\hat q,\hat p)+
t\acute{e}rminos~con~\hat{\mathcal P}~a~la~derecha.
\end{equation}

Ahora es evidente que:
\begin{equation}
i[\hat{\mathcal P}_a,\hat\Omega]=\hat G_a(\hat q, \hat p) +
~al~menos~dos~fantasmas.
\end{equation}
M\'{a}s a\'{u}n, de $\hat\Omega^2=0$ se obtiene que el operador que multiplica
a $\hat\eta^a$ al cabo del reordenamiento satisface:
\begin{equation}\label{algcua}
[\hat G_a(\hat q, \hat p),\hat G_b(\hat q, \hat p)]=i\hat C_{ab}^c(\hat q,\hat
p)\hat G_c(\hat q, \hat p).
\end{equation}

De modo que $\hat G_a(\hat q, \hat p)$ es el operador que realiza
cu\'{a}nticamente el \'{a}lgebra de los v\'{\i}nculos. Es por lo tanto el
operador que aniquila los estados f\'{\i}sicos de la cuantificaci\'{o}n de
Dirac.

En general $\hat G_a(\hat q, \hat p)$ no va a coincidir con el orden
herm\'{\i}tico de $G_a(\hat q, \hat p)$ sino que tomar\'{a} contribuciones del
reordenamiento de los fantasmas que servir\'{a}n para cancelar anomal\'{\i}as
en el \'{a}lgebra de v\'{\i}nculos. De tal forma que $\hat G_a(\hat q, \hat
p)$ no resultar\'{a} herm\'{\i}tico.

Se asume que es posible hallar el operador $\hat\Omega$ que satisface las
condiciones de nilpotencia y hermiticidad. Sin embargo, a diferencia del caso
cl\'{a}sico, no existe a priori ninguna garant\'{\i}a que esto pueda ser
realizado a partir de una teor\'{\i}a cl\'{a}sica para la cual
$\{\Omega,\Omega\}=0$, puesto que la cuesti\'{o}n del ordenamiento de los
factores resulta crucial. Si la condici\'{o}n $\hat\Omega^2=0$ no se satisface
debido a efectos cu\'{a}nticos, no todos los grados de libertad de gauge
desaparecen del espectro f\'{\i}sico y la teor\'{\i}a cu\'{a}ntica es
an\'{o}mala.


\chapter{Sistemas con tiempo intr\'{\i}nseco}

\bigskip

{\it ~~~~ Como hemos visto, el campo gravitatorio es un sistema con
covariancia general, propiedad que se expresa en la aparici\'{o}n de
v\'{\i}nculos: los {\it supermomentos}, lineales y homogen\'{e}neos en los
momentos del campo), y el {\it super-Hamiltoniano}, cuadr\'{a}tico en los
momentos del campo y que exhibe un t\'{e}rmino ``potencial'' (la curvatura
espacial). La cuantificaci\'{o}n de este tipo de sistemas requiere la
b\'{u}squeda de un ordenamiento para los operadores de v\'{\i}nculo que
preserve el \'{a}lgebra (ausencia de anomal\'{\i}as). Para evitar la
regularizaci\'{o}n de los operadores y poder estudiar el problema en sistemas
m\'{a}s simples pero que preserven las caracter\'{\i}sticas del campo
gravitatorio, es usual congelar la mayor\'{\i}a de los grados de libertad,
para quedarse finalmente con sistemas de dimensi\'{o}n finita (modelos de
``minisuperespacio''). En este esp\'{\i}ritu, \cite{hk90} estudiaron la
cuantificaci\'{o}n de este tipo de sistemas de dimensi\'{o}n finita en el
marco de la cuantificaci\'{o}n de Dirac. El modelo propuesto surge por
analog\'{\i}a con los v\'{\i}nculos presentados por una part\'{\i}cula
relativista en un fondo curvo a la cual se le han agregado grados de libertad
espurios.

En el presente cap\'{\i}tulo, estudiaremos en el marco del formalismo BRST el
sistema propuesto por \cite{hk90}, es decir, un sistema con covariancia
general descripto por $n$ pares de variables can\'{o}nicas $(q^i,p_i)$ sujetas
a $m+1$ v\'{\i}nculos de primera clase, donde $m$ de ellos son lineales y
homog\'{e}neos en los momentos y el restante es una funci\'{o}n cuadr\'{a}tica
de los momentos, con una m\'{e}trica indefinida no degenerada $G^{ij}$,
m\'{a}s un potencial $V$. Hallaremos el generador BRST cl\'{a}sico y su
realizaci\'{o}n cu\'{a}ntica nilpotente en el caso que el potencial $V$ sea
invariante de gauge (es decir, no cambia ante transformaciones generadas por
los v\'{\i}nculos lineales). El producto interno f\'{\i}sico estar\'{a} bien
definido en el caso que el potencial sea definido positivo. Esta propiedad del
potencial significa el sistema bajo estudio no es m\'{a}s que una
part\'{\i}cula relativista y determina que el sistema posea un tiempo
intr\'{\i}nseco [\cite{prd,np00}]. }

\section{El modelo: an\'{a}lisis cl\'{a}sico}

~~~~ Consideremos un sistema Hamiltoniano descripto por $2n$ variables
can\'{o}nicas $(q^i,p_i)$, con $i=1,...,n;$ sujeto a los v\'{\i}nculos {\it
supermomentos}
\begin{equation}\label{vl4}
~~~~~~~~~~~~~~~~~{\cal H}_a=\xi_a^j(q) p_j,~~~~~~~~~~~~ a=1,...,m,
\end{equation}
y al v\'{\i}nculo {\it super-Hamiltoniano}
\begin{equation}\label{vc4}
{\cal H}_o={1\over 2} g^{ij}(q) p_i p_j+\upsilon,
\end{equation}
donde $g^{ij}$ es una m\'{e}trica indefinida no degenerada y $\upsilon$ es un
potencial, en principio general (luego impondremos condiciones seg\'{u}n el
caso).

Supondremos que el sistema de v\'{\i}nculos es de primera clase, es decir que
satisfacen:
\begin{equation}\label{ccgg}
\{{\cal H}_a,{\cal H}_b\}=C_{ab}^{c}(q) {\cal H}_c,
\end{equation}

\begin{equation}\label{cchg}
\{{\cal H}_o,{\cal H}_a\}=C_{oa}^o(q){\cal H}_o + C_{oa}^{b}(q, p){\cal H}_b.
\end{equation}

Veamos qu\'{e} relaciones impone la condici\'{o}n de primera clase.
Reemplazando los v\'{\i}nculos (\ref{vl4}) en (\ref{ccgg}) obtenemos:
\begin{equation}\label{cvlfi}
\xi_b^j\xi_{a,j}^i - \xi_a^j \xi_{b,j}^i= C_{ab}^c \xi_c^i.
\end{equation}

Reemplazando los v\'{\i}nculos (\ref{vl4}) y (\ref{vc4}) en (\ref{cchg}),
tenemos que\footnote{${\cal L}_{\vec \xi_a}$ denota la derivada de Lie
respecto al campo vectorial $\vec\xi_a$, ver por ejemplo
\cite{naka,schutz,grav}.}
\begin{equation}\label{estrlc}
{1\over 2} \xi_a^k g^{ij}_{,k} p_i p_j - g^{ik} p_i \xi_{a,k}^j p_j
+\xi_a^k\upsilon_{,k}= {1\over 2} ({\cal L}_{\vec \xi_a}\bar{\bar{g}})^{ij}
p_i p_j +{\cal L}_{\vec \xi_a}\upsilon =C_{oa}^o(q) {\cal H}_o + C_{oa}^b
{\cal H}_b,
\end{equation}
de donde es inmediato que
\begin{equation}\label{condpot}
C_{oa}^o(q) \upsilon ={\cal L}_{\vec \xi_a}\upsilon.
\end{equation}

De la Ec. (\ref{condpot}), vemos que podemos distinguir dos situaciones
dependiendo de las propiedades que exhiba el potencial $\upsilon$:
\begin{description}
\item[{\it Tipo 1.}] Es invariante de gauge, es decir $\{\upsilon,{\cal H}_a\}
=0={\cal L}_{\vec \xi_a}\upsilon,~\forall a ~~\Longrightarrow C_{oa}^o(q)=0$.
\item[{\it Tipo 2.}] Es definido positivo, $\upsilon>0$, entonces podemos
despejar
\begin{equation}\label{coao}
C_{oa}^o(q) ={\upsilon}^{-1} {\cal L}_{\vec \xi_a}\upsilon.
\end{equation}

\end{description}

Entonces, tenemos que la Ec. (\ref{estrlc}) nos dar\'{a} dos condiciones
diferentes dependiendo si el potencial es de tipo 1 \'{o} 2. Adem\'{a}s
teniendo en cuenta que la funci\'{o}n de estructura $C_{oa}^o(q) $
forzosamente es lineal y homog\'{e}nea en los momentos
\begin{equation}\label{coac}
C_{oa}^b =C_{oa}^{bj}(q) p_j,
\end{equation}
tenemos que:

Si $\upsilon$ es del tipo 1,

\begin{equation}\label{lt1}
{1\over 2} ({\cal L}_{\vec \xi_a} \bar{\bar{g}})^{ij} p_i p_j =C_{oa}^{bj}(q)
\xi_b^ip_ip_j.
\end{equation}

O, si $\upsilon$ es del tipo 2,

\begin{equation}
{1\over 2} ({\cal L}_{\vec \xi_a} \bar{\bar{g}})^{ij} p_i p_j-{1\over 2}
{\upsilon}^{-1}{\cal L}_{\vec \xi_a}\upsilon g^{ij} p_i p_j=C_{oa}^{bj}(q)
\xi_b^ip_ip_j.
\end{equation}

En esta \'{u}ltima ecuaci\'{o}n podemos definir la m\'{e}trica
\begin{equation}\label{escmetrica}
{g'}^{ij}\equiv {\upsilon}^{-1} g^{ij},
\end{equation}
y entonces obtener

\begin{equation}\label{lt2}
{1\over 2} ({\cal L}_{\vec \xi_a} \bar{\bar{g}}')^{ij} p_i p_j
={\upsilon}^{-1} C_{oa}^{bj}(q) \xi_b^ip_ip_j.
\end{equation}

Para despejar las funciones de estructura, debemos especificar ciertos objetos
geom\'{e}tricos en la variedad. Si llamamos $M$ al espacio de
configuraci\'{o}n, los vectores $\{\vec\xi_a\}$ pertenecen al espacio tangente
$TM$. Si son linealmente independientes forman la base de un subespacio de
$TM$, que llamaremos espacio tangente longitudinal $T_{||}M$. Definimos la
base dual en $T_{||}^*M$ como $\{\tilde E^a\}$ donde las 1-formas $\tilde E^a$
son tales que:
\begin{equation}\label{bdual}
\tilde E^a(\vec\xi_b)=\delta_b^a
\end{equation}
Si desarrollamos cada $\tilde E^a$ en la base dual coordenada:
\begin{equation}\label{bdcoord}
\tilde E^a= E^a_j dq^j
\end{equation}
entonces la Ec. (\ref{bdual}) queda $E^a_j\xi_b^j=\delta_b^a$

Luego, usando estas propiedades, podemos despejar de (\ref{cvlfi}), las
funciones de estructura $C_{ab}^c$ como:
\begin{equation}\label{cabc}
C_{ab}^c= E_i^c(\xi_b^j\xi_{a,j}^i - \xi_a^j \xi_{b,j}^i)
\end{equation}

Y seg\'{u}n el potencial sea de tipo 1 \'{o} 2, obtenemos de (\ref{lt1}) o
(\ref{lt2}) respectivamente,

\begin{equation}\label{coabt1}
C_{oa}^{bj}={1\over 2} E^b_i ({\cal L}_{\vec \xi_a}\bar{\bar{g}})^{ij} ~~~~
(tipo~1),
\end{equation}

\begin{equation}\label{coabt2}
C_{oa}^{bj}={1\over 2} \upsilon E^b_i ({\cal L}_{\vec \xi_a}
\bar{\bar{g}}')^{ij} ~~~~ (tipo~2).
\end{equation}

Si observamos la definici\'{o}n (\ref{escmetrica}) y la consideramos al
comparar los resultados (\ref{coabt1}) y (\ref{coabt2}), notamos que lo que
hicimos impl\'{\i}citamente es un {\it cambio de escala} en el
super-Hamiltoniano con el potencial definido positivo. Es decir, pasamos de un
potencial {\it tipo 1} al {\it tipo 2} a trav\'{e}s de la transformaci\'{o}n:

\begin{equation}\label{escham}
{\cal H}_o=g'^{ij}p_ip_j+1  \longrightarrow \upsilon {\cal H}_o= \upsilon
g'^{ij}p_ip_j+ \upsilon
\end{equation}
Y como resultado las funciones de estructura se transforman de modo acorde
\begin{equation}\label{coaoesc}
C_{oa}^o=0  \longrightarrow C_{oa}^o={\upsilon}^{-1} {\cal L}_{\vec \xi_a}
\upsilon
\end{equation}
\begin{equation}\label{coabesc}
C_{oa}^{bj}={1\over 2} E^b_i ({\cal L}_{\vec \xi_a}\bar{\bar{g}})^{ij}
\longrightarrow C_{oa}^{bj}={1\over 2} \upsilon E^b_i ({\cal L}_{\vec
\xi_a}\bar{\bar{g}}')^{ij}
\end{equation}

N\'{o}tese que el cambio de escala en el super-Hamiltoniano (\ref{escham}) es
una transformaci\'{o}n que deja invariante al sistema: los v\'{\i}nculos
${\cal H}_o$ y $\upsilon{\cal H}_o$ con $\upsilon>0$ son equivalentes. Esta
propiedad ser\'{a} explotada al momento de cuantificar el sistema.
Comenzaremos con un super-Hamiltoniano con potencial constante\footnote{Caso
particular de tipo 1, podr\'{\i}amos considerarlo m\'{a}s general pero no
ser\'{\i}a claro como definir el producto interno f\'{\i}sico que
introduciremos m\'{a}s adelante.} y con una transformaci\'{o}n de cambio de
escala pasaremos al caso de potenciales definidos positivos. Esta
transformaci\'{o}n no se realizar\'{a} a nivel de operadores de v\'{\i}nculo
sino al nivel del operador BRST por medio de una transformaci\'{o}n unitaria.

\subsection{El generador BRST cl\'{a}sico}

~~~~ Aplicando el m\'{e}todo recursivo explicado en \S 3.1.1 construiremos el
generador BRST cl\'{a}sico $\Omega$ para el modelo de la secci\'{o}n anterior
en el caso en que el potencial del super-Hamiltoniano es invariante de gauge
(tipo 1).

Los primeros t\'{e}rminos que contribuyen al generado BRST son
\begin{equation}\label{ocero}
~~~~~~~~~~~~~~~~~~~~~\Omega\!\!\!\!\!^{^{^{(0)}}}=\eta^{\alpha}{\cal
H}_{\alpha},~~~~~~\alpha=(o,a)=0,1,...,m
\end{equation}
y
\begin{equation}\label{ouno}
\Omega\!\!\!\!\!^{^{^{(1)}}}={1\over  2}  \eta^\alpha \eta^\beta
C_{\alpha\beta}^\gamma {\cal P}_\gamma~~~~~~
\end{equation}

Luego, para calcular el t\'{e}rmino siguiente debemos usar las Ecs.
(\ref{rec})-(\ref{Deq}),
\begin{equation}\label{rec21}
\delta\Omega\!\!\!\!\!^{^{^{(2)}}}+\mathrm{D}\!\!\!\!\!^{^{^{(1)}}}=0,
\end{equation}
con
\begin{eqnarray}\label{d1}
\lefteqn{\mathrm{D}\!\!\!\!\!^{^{^{(1)}}}={1\over 2}\left[2
\{\Omega\!\!\!\!\!^{^{^{(0)}}},\Omega\!\!\!\!\!^{^{^{(1)}}}\}_{orig} +
\{\Omega\!\!\!\!\!^{^{^{(1)}}},\Omega\!\!\!\!\!^{^{^{(1)}}}\}_{{\mathcal
P},\eta}\right]}\nonumber \\ && =\{\eta^{\alpha}{\cal H}_{\alpha},{1\over 2}
\eta^{\alpha} \eta^{\beta} C_{\alpha\beta}^{\gamma} {\cal P}_{\gamma}
\}_{orig}+{1\over 2}\{ {1\over 2} \eta^{\alpha} \eta^{\beta}
C_{\alpha\beta}^{\gamma} {\cal P}_{\gamma}, {1\over 2} \eta^{\alpha}
\eta^{\beta} C_{\alpha\beta}^{\gamma} {\cal P}_{\gamma} \}_{{\mathcal
P},\eta}.
\end{eqnarray}
Luego,
\begin{equation}\label{d2}
\delta \Omega\!\!\!\!\!^{^{^{(2)}}} =-{1\over 2} \eta^{\alpha} \eta^{\beta}
\eta^{\gamma} \left( \{ {\cal H}_{\alpha},C^{\delta}_{\beta\gamma}\}+
C^{\varepsilon}_{\alpha\beta}C^{\delta}_{\gamma\varepsilon}\right) {\cal
P}_{\delta}.
\end{equation}

Siendo ${\eta^o}^2=0$, en el par\'{e}ntesis s\'{o}lo sobreviven los
t\'{e}rminos lineales en $p$, y los independientes de $p$. Estos \'{u}ltimos
se cancelan, pues as\'{\i} ocurre en una teor\'{\i}a con v\'{\i}nculos
lineales [\cite{mp89,kuns92,fhp93}].

Finalmente queda
\begin{eqnarray}
\lefteqn{\delta \Omega\!\!\!\!\!^{^{^{(2)}}} =-{1\over 2} \eta^o \eta^b \eta^c
\left( \{ {\cal H}_o,C^d_{bc}\}+ C^e_{ob}C^d_{ce}\right) {\cal P}_d~~}
\nonumber \\ &&~-{1\over 2} \eta^a \eta^o \eta^c \left( \{ {\cal H}_a,
C^d_{oc}\}+ C^e_{ao}C^d_{ce}\right) {\cal P}_d\nonumber \\ &&~-{1\over 2}
\eta^a \eta^b \eta^o \left( \{ {\cal H}_a,C^d_{bo}\}+ C^e_{ab}C^d_{oe}\right)
{\cal P}_d
\end{eqnarray}

Redefiniendo \'{\i}ndices mudos y permutando obtenemos,
\begin{equation}
\delta \Omega\!\!\!\!\!^{^{^{(2)}}}=-{1\over 2} \eta^o \eta^b \eta^c \left( \{
{\cal H}_o, C^d_{bc}\}+C^e_{bc}C^d_{oe}-2\{{\cal H}_b,C^d_{oc}\}+2
C^e_{ob}C^d_{ce}\right) {\cal P}_d
\end{equation}
N\'{o}tese que la identidad $\delta^2\Omega\!\!\!\!\!^{^{^{(2)}}}\equiv0$
expresa la identidad de Jacobi:
\begin{equation}
\left( \{ {\cal H}_o,C^d_{bc}\}+C^e_{bc}C^d_{oe}+2\{C^d_{o[c},{\cal H}_{b]}
\}+2 C^e_{o[b}C^d_{c]e}\right){\cal H}_d=0
\end{equation}
Pero como los v\'{\i}nculos lineales son independientes, resulta que
$\Omega\!\!\!\!\!^{^{^{(2)}}}$ es cerrada: $$\delta
\Omega\!\!\!\!\!^{^{^{(2)}}}=0.$$
Por lo tanto escogemos
\begin{equation}\label{o2cerr}
\Omega\!\!\!\!\!^{^{^{(2)}}}=0
\end{equation}
($\Omega\!\!\!\!\!^{^{^{(p)}}}$ siempre est\'{a} determinado a menos de una
cantidad exacta).

Por el teorema 3.2  bastar\'{\i}a ver que $\Omega\!\!\!\!\!^{^{^{(3)}}}=0$,
para probar que la teor\'{\i}a es de rango 1. Calculemos entonces,
\begin{eqnarray}
\lefteqn{\mathrm{D}\!\!\!\!\!^{^{^{(1)}}}={1\over 2}
\{\Omega\!\!\!\!\!^{^{^{(1)}}},\Omega\!\!\!\!\!^{^{^{(1)}}}\}_{orig}}\nonumber
\\ &&={1\over 8}\{ \eta^{\alpha} \eta^{\beta} C_{\alpha\beta}^{\gamma} {\cal
P}_{\gamma}, \eta^{\delta} \eta^{\varepsilon} C_{\delta\varepsilon}^{\sigma}
{\cal P}_{\sigma} \}_{orig} \nonumber \\ &&={1\over 8} \eta^{\alpha}
\eta^{\beta} \eta^{\delta} \eta^{\varepsilon}{\cal P}_{\gamma}{\cal
P}_{\sigma} \{C_{\alpha\beta}^{\gamma},C_{\delta\varepsilon}^{\sigma}\}_{orig}
\end{eqnarray}

S\'{o}lo uno de los $\eta$ puede ser $\eta^o$. Habr\'{a} entonces cuatro
t\'{e}rminos iguales:

\begin{equation}
\delta \Omega\!\!\!\!\!^{^{^{(3)}}}= {1\over 2} \eta^o \eta^b \eta^d  \eta^e
{\cal P}_c {\cal P}_a \{ C_{ob}^c,C_{de}^a \}
\end{equation}

Por lo tanto,
\begin{equation}
0\equiv\delta^2\Omega\!\!\!\!\!^{^{^{(3)}}}= {1\over 2} \eta^o \eta^b \eta^d
\eta^e \left({\cal P}_c (-{\cal H}_a) + {\cal P}_a {\cal H}_c\right) \{
C_{ob}^c,C_{de}^a\}
\end{equation}
\begin{equation}
\Longrightarrow  \{ C_{o[b}^c,C_{de]}^a\} {\cal H}_c - \{
C_{o[b}^a,C_{de]}^c\} {\cal H}_c =0,
\end{equation}

\noindent y como los v\'{\i}nculos lineales son linealmente independientes
\begin{equation}
\{ C_{o[b}^{[c},C_{de]}^{a]}\}=0.
\end{equation}
Por lo tanto,
\begin{equation}
\delta\Omega\!\!\!\!\!^{^{^{(3)}}}=0 \Longrightarrow
\Omega\!\!\!\!\!^{^{^{(3)}}}= 0
\end{equation}

Entonces la teor\'{\i}a es de rango 1. Y el generador BRST cl\'{a}sico para la
teor\'{\i}a es
\begin{eqnarray}
\lefteqn{\Omega=\eta^{\alpha}  {\cal H}_{\alpha}  + {1\over  2}  \eta^{\alpha}
\eta^{\beta} C_{\alpha\beta}^{\gamma} {\cal P}_{\gamma}}\nonumber \\
&&=\eta^o{\cal H}_o+\eta^a{\cal H}_a+ \eta^o \eta^a C_{oa}^b{\cal P}_b+{1\over
2}\eta^a \eta^b C_{ab}^c{\cal P}_c.\label{omegacl4}
\end{eqnarray}

Recordemos entonces que este resultado es v\'{a}lido para sistemas cuyo
v\'{\i}nculo super-Hamiltoniano posea un potencial invariante de gauge
(n\'{o}tese que en (\ref{omegacl4}) no aparece $C_{oa}^o$).

\section{Cuantificaci\'{o}n del sistema.}

~~~~ La cuantificaci\'{o}n del sistema en el marco del formalismo BRST
requiere que el generador BRST cl\'{a}sico, Ec. (\ref{omegacl4}), sea
realizado como un operador herm\'{\i}tico y nilpotente. Para el caso de
solamente un conjunto de v\'{\i}nculos lineales de primera clase este paso ha
sido dado en numerosos trabajos (ver por ej. \cite{fhp93} y referencias
all\'{\i} citadas). Sin embargo, en el caso del modelo presentado, la
presencia de un v\'{\i}nculo cuadr\'{a}tico en los momentos complica
enormemente el hallazgo de la realizaci\'{o}n cu\'{a}ntica del generador BRST.

Debido a la dificultad mencionada, comenzaremos por estudiar en detalle el
caso en el que s\'{o}lo existen v\'{\i}nculos supermomentos y luego
incluiremos el v\'{\i}nculo super-Hamiltoniano. Las transformaciones de
invariancia de la teor\'{\i}a, la reinterpretaci\'{o}n geom\'{e}trica de
ciertos resultados y el tratamiento en pie de igualdad de las variables
can\'{o}nicas originales y los fantasmas en este sistema m\'{a}s simple, nos
dar\'{a}n la clave para hallar el operador BRST en el caso de sistemas que
exhiben adem\'{a}s el v\'{\i}nculo super-Hamiltoniano.

\subsection{V\'{\i}nculos supermomentos}

~~~~ Para un sistema de $m$ v\'{\i}nculos linealmente independientes
\begin{equation}\label{vle}
{\cal H}_a(q^i,p_j)=\xi^k_a(q^i)p_k, ~~~~~~~~~a=1,...,m,
\end{equation}
el problema de encontrar un ordenamiento para los operadores de v\'{\i}nculo
que satisfaga el \'{a}lgebra
\begin{equation}\label{clc4}
[\hat{\cal H}_a,\hat{\cal H}_b]=C_{ab}^c(q) \hat{\cal H}_c
\end{equation}
es resuelto trivialmente por
\begin{equation}\label{vlinq}
\hat{\cal H}_a= f^{1\over 2} \xi^i_a \hat p_i f^{-{1\over 2}},
\end{equation}
donde $f$ es arbitraria. En la cuantificaci\'{o}n de Dirac la funci\'{o}n $f$
puede ser determinada pidiendo que los operadores de v\'{\i}nculo preserven el
car\'{a}cter gem\'{e}trico de la funci\'{o}n de onda [\cite{hk90,mp89,fhp93}].
Dicho car\'{a}cter ge\'{o}metrico est\'{a} determinado por la ley de
transformaci\'{o}n de la funci\'{o}n de onda bajo cambios que dejan invariante
la teor\'{\i}a a nivel cl\'{a}sico: cambios de coordenadas y combinaciones
lineales de los v\'{\i}nculos. La funci\'{o}n de onda deber\'{\i}a
transformarse de manera tal que el producto interno f\'{\i}sico permanezca
invariante.

\bigskip
Por otra parte, en el marco del formalismo BRST, para cuantificar el sistema
extendido se debe realizar el generador BRST para el sistema de primera clase
(\ref{vle}) (resultado de la secci\'{o}n anterior, Ec. (\ref{omegacl4}),
debidamente restringido: poniendo $\eta^o=0$)
\begin{equation}\label{omegacl}
\Omega^{lineal}=\eta^a{\cal H}_a+{1\over 2} \eta^a \eta^b C_{ab}^c {\cal P}_c,
\end{equation}
como un operador herm\'{\i}tico y nilpotente (de esta forma la teor\'{\i}a
estar\'{a} libre de anomal\'{\i}as, cap\'{\i}tulo 3):
\begin{equation}\label{nilq}
[{\hat\Omega},{\hat\Omega}]=2{\hat\Omega}^2=0.
\end{equation}

Antes de llevar a cabo esto, es sumamente importante considerar en pie de
igualdad a las variables can\'{o}nicas originales y fantasmas (esto
simplificar\'{a} la visualizaci\'{o}n del ordenamiento buscado cuando se
incluya el v\'{\i}nculo super-Hamiltoniano). Adoptemos por lo tanto, la
notaci\'{o}n usada en \cite{fhp93}:
\begin{equation}\label{not}
\eta^{c_s}=(q^i, \eta^a),~~~~~~~~~~~{\cal P}_{c_s}=(p_i,{\cal
P}_a),~~~~~~~~~~~~~~~~s=-1,0.
\end{equation}\label{omlin}
Luego, $\Omega^{lineal}$ puede ser escrito como
\begin{equation}\Omega^{lineal}=\sum_{s=-1}^0 \Omega^{c_s}{\cal P}_{c_s},
\end{equation}
donde
\begin{equation}
\Omega^{c_s}\equiv (\eta^a\xi^i_a~,~{1\over 2}\eta^a\eta^b C_{ab}^c).
\end{equation}

El ordenamiento
\begin{equation}\label{ordlin}
\hat{\Omega}^{lineal}=\sum_{s=-1}^0f^{1\over   2}\hat\Omega^{c_s} \hat{\cal
P}_{c_s}f^{-{1\over 2}}
\end{equation}
es nilpotente para cualquier $f(q)$ (es simplemente el resultado cl\'{a}sico
$\{\Omega,\Omega\}=0$) pero $f$ deber\'{\i}a ser elegida de manera tal que
${\hat\Omega}^{lineal}$ resulte herm\'{\i}tico. De esta condici\'{o}n, resulta
que $f$ debe satisfacer
\begin{equation}\label{div}
C^b_{ab}=f^{-1}(f\xi^i_a)_{,i}.
\end{equation}
El operador as\'{\i} obtenido $\hat{\Omega}^{lineal}$ es igual al hallado
simetrizando la ecuaci\'{o}n (\ref{omlin}), es decir
$\hat{\Omega}^{lineal}=\frac{1}{2}(\hat\Omega^{c_s} \hat{\cal P}_{c_s}
+\hat{\cal P}_{c_s}\hat\Omega^{c_s} )$, que es el procedimiento habitual. Esta
realizaci\'{o}n de $\hat{\Omega}^{lineal}$ nos conduce a operadores de
v\'{\i}nculo que coinciden con los hallados mediante el m\'{e}todo
geom\'{e}trico de Dirac:
\begin{equation}\label{vlinqe}
\hat {\cal H}_a= f^{1\over 2} \xi^i_a \hat p_i f^{-{1\over 2}}=\xi^i_a \hat
p_i - {i\over 2} \xi^i_{a,i} + {i\over 2} C^b_{ab}.
\end{equation}

Si bien la ecuaci\'{o}n (\ref{div}) es todo lo que uno necesita para
establecer $\hat{\Omega}^{lineal}$, no define un\'{\i}vocamente a $f$. De
hecho, el lado derecho no cambia si $f$ es multiplicada por una funci\'{o}n
invariante de gauge. La siguiente proposici\'{o}n  har\'{a} m\'{a}s claro el
significado geom\'{e}trico de $f$ en el ecuaci\'{o}n (\ref{div}).

\noindent \underline{{\bf Proposici\'{o}n 4.1}}: Para un conjunto dado de
v\'{\i}nculos de primera clase (\ref{vle}), sea $\tilde\alpha$ el volumen
inducido por los v\'{\i}nculos en el espacio de configuraci\'{o}n original
$M$:
\begin{equation}\label{volnat}
{\tilde\alpha}\equiv {\tilde  E}^1\wedge...\wedge   {\tilde  E}^m \wedge
{\tilde \omega},
\end{equation}
donde $\{\tilde E^a\}$ es la base dual de $\{\vec\xi_a\}$ in $T_{||}M$, el
espacio tangente longitudinal; y ${\tilde \omega}\ =\ \omega(y)\
dy^1\wedge...\wedge dy^{n-m}$ es una $n-m$ forma cerrada, siendo las $y^r$,
$n-m$ funciones que son invariantes ante la acci\'{o}n de las transformaciones
de gauge generadas por los v\'{\i}nculos lineales,\footnote{No denominamos a
estas funciones ``observables'' ya que no ser\'{a}n invariantes ante la
acci\'{o}n del v\'{\i}nculo cuadr\'{a}tico que introduciremos en la
secci\'{o}n siguiente.} es decir, $dy^r(\vec\xi_a)=0~~\forall  r,a$.
$\tilde\alpha$ es el volumen  inducido por los v\'{\i}nculos en la \'{o}rbita,
multiplicado por un volumen (no elegido) en el espacio ``reducido''. Luego,
\begin{equation}\label{prop1}
C_{ab}^b=div_{\tilde\alpha}~\vec\xi_a.
\end{equation}

\noindent {\it Demostraci\'{o}n.} Tomaremos ventaja del hecho que cualquier
base puede ser (localmente) abelianizada. As\'{\i}, probaremos la
proposici\'{o}n para una base abeliana, y luego transformaremos ambos lados de
la ecuaci\'{o}n (\ref{prop1}) mostrando que permanecen iguales para una base
arbitraria de $T_{||}M$.

Sea $\{{\vec\xi}^{\prime}_a\}$ una base abeliana en $T_{||}M$, luego el lado
izquierdo de la ecuaci\'{o}n (\ref{prop1}) es $C^{\prime b}_{ab}=0$. Por otro
lado, por definici\'{o}n la divergencia en el volumen
${\tilde\alpha}^{\prime}$ del campo vectorial ${\vec\xi}^{\prime}_a$ es
escrita en t\'{e}rminos de la derivada exterior de la $(n-1)$-forma
${\tilde\alpha}^{\prime}({\vec\xi}^{\prime}_a)$ [\cite{schutz}]:
\begin{equation}\label{defdiv}
(div_{{\tilde\alpha}^{\prime}}{\vec\xi}^{\prime}_a){\tilde\alpha}^{\prime}
\equiv d[{\tilde\alpha}^{\prime}({\vec\xi}^{\prime}_a)].
\end{equation}
El lado derecho de la ecuaci\'{o}n (\ref{prop1}) es adem\'{a}s nulo porque
${\tilde\alpha}^{\prime}({\vec\xi}^{\prime}_a)$ es cerrada. De hecho, las
formas ${\tilde E}^{\prime a}$ son (localmente) exactas, puesto que una base
abeliana es una base coordenada. Luego, la ecuaci\'{o}n (\ref{prop1}) resulta
verdadera para v\'{\i}nculos abelianos.

Ahora, cambiemos de base
\begin{equation}\label{cbase}
{\vec\xi}_a=A_a^{~b}(q)~{\vec\xi}^{\prime}_b,~~~~~~~~~~~~ {\tilde E}^a
=A^a_{~b}(q)~{\tilde E}^{\prime b}
\end{equation}
siendo ($A^a_{~b}$ la matriz inversa de $A_a^{~b}$). Luego,
\begin{equation}\label{stf}
C^b_{ab}=E^b_i(\xi^j_b \xi^i_{a,j} -\xi^j_a \xi^i_{b,j})=
A^b_{~c}(A^{~c}_{a,j}\xi^j_b - A^{~c}_{b,j}\xi^j_a).
\end{equation}

Por otro lado,
\begin{eqnarray}
\lefteqn{d[{\tilde\alpha}(\vec\xi_a)]=d[det
A^{-1}{\tilde\alpha}^{\prime}(\vec\xi_a)] =d[A_a^{~b}det
A^{-1}{\tilde\alpha}^{\prime}({\vec\xi}^{\prime}_a)]}\nonumber\\
&&=\sum_{b=1}^{m} (-1)^{b-1}(A_a^{~b}~det A^{-1})_{,j}~ dq^j \wedge {\tilde
E}^{\prime 1}\wedge...\nonumber\\ &&~~~...\wedge {\tilde E}^{\prime b-1}
\wedge {\tilde E}^{\prime b+1}\wedge ...\wedge {\tilde E}^{\prime m} \wedge
{\tilde \omega}\nonumber\\ && =(A_a^{~b}~det A^{-1})_{,j}~ A^c_{~b}~ \xi^j_c~
{\tilde E}^{\prime 1}\wedge...\wedge {\tilde E}^{\prime m}\wedge {\tilde
\omega},
\end{eqnarray}
($det A^{-1}\equiv det A^a_{~b}$), debido a que s\'{o}lo la componente
$dq^j({\vec\xi}^{\prime}_b)={\xi}^{\prime j}_b= A^c_{~b}\xi^j_c$ contribuye.

Por lo tanto,
\begin{equation}\label{stf'}
d[{\tilde\alpha}(\vec\xi_a)]=A^b_{~c}(A^{~c}_{a,j}\xi^j_b-A^{~c}_{b,j}
\xi^j_a) {\tilde\alpha}.
\end{equation}

De este modo, las ecuaciones (\ref{stf}) y (\ref{stf'}) nos dicen ambos lados
de la ecuaci\'{o}n (\ref{div}) tienen el mismo valor independientemente de
cual sea la base de $T_{||}M$. Luego, la proposici\'{o}n resulta demostrada
para cualquier conjunto de v\'{\i}nculos de primera clase lineales y
homog\'{e}neos en los momentos. \hfill  $\Box$

El resultado de la proposici\'{o}n significa en la ecuaci\'{o}n (\ref{div}),
$f$ puede ser vista como la componente de $\tilde\alpha$ en la base coordenada
$\{dq^i\}$:
\begin{equation}
{\tilde\alpha}= f~dq^1\wedge...\wedge dq^n.
\end{equation}

En el marco del formalismo BRST, una redefinici\'{o}n de los v\'{\i}nculos
como la de la ecuaci\'{o}n (\ref{cbase}) es vista como el cambio de variables
$\eta^a \rightarrow \eta^{\prime a}=\eta^b A_b^{~a}(q)$. Puesto que la
funci\'{o}n de onda BRST se comporta como una superdensidad de peso 1/2 en el
espacio ($q,~\eta$) (de modo de preservar el producto interno invariante),
puede concluirse que los factores $f^{1 \over 2}$ y $f^{-{1 \over 2}}$ en la
ecuaci\'{o}n (\ref{ordlin}) son exactamente lo que se necesita para que
${\hat{\Omega}^{lineal}}\psi$ se comporte de la misma manera que $\psi$ bajo
tal cambio (de hecho, $f \rightarrow f^{\prime}=det A~f$). Esta propiedad de
$f$ deber\'{\i}a tenerse en cuenta al momento de cuantificar un sistema con un
v\'{\i}nculo cuadr\'{a}tico, ya que podr\'{\i}a facilitar la b\'{u}squeda del
operador $\hat\Omega$.

\bigskip

\subsection{V\'{\i}nculo super-Hamiltoniano}

~~~~ Como ya se ha adelantado, consideraremos ahora la inclusi\'{o}n de un
v\'{\i}nculo cuadr\'{a}tico con un potencial que no se anula. Esta propiedad
permitir\'{a} factorizar el potencial, y reemplazar al v\'{\i}nculo
cuadr\'{a}tico por uno equivalente pero con un potencial constante (o como
mencionamos antes, un caso particular de tipo 1: invariante de gauge).
Comencemos, por lo tanto por considerar un v\'{\i}nculo Hamiltoniano ${\cal
H}_o(q^i,p_j)$:
\begin{equation}\label{vce}
{\cal H}_o(q^k,p_j)={1\over 2} g^{ij}(q^k) p_i p_j+\lambda,
~~~~~~\lambda=constante,
\end{equation}
siendo $g^{ij}$ una m\'{e}trica indefinida no degenerada. M\'{a}s adelante, se
introducir\'{a} un potencial que no se anula $V = \lambda \vartheta(q)$.

De esta forma tenemos el sistema completo Ecs. (\ref{vl4})-(\ref{vc4}), junto
con las condiciones de primera clase, Ecs. (\ref{ccgg})-(\ref{cchg}).
Seg\'{u}n demostramos en \S 4.1.1, en este caso el generador BRST es, Ec.
(\ref{omegacl4}),

\begin{equation}\label{omegacl2}
\Omega=\eta^o {\cal H}_o +  \eta^a  {\cal H}_a + \eta^o \eta^a c_{oa}^b {\cal
P}_b + {1\over 2} \eta^a \eta^b  C_{ab}^c  {\cal P}_c \equiv
\Omega^{cuad}+\Omega^{lineal},
\end{equation}
donde $\Omega^{lineal}$ es el mismo de (\ref{omlin}), y $\Omega^{cuad}$ es

\begin{equation}\label{omcuad}
\Omega^{cuad}={1\over  2}\sum_{r,s=-1}^0  {\cal  P}_{a_r} \Omega^{a_rb_s}
{\cal P}_{b_s}+ \eta^o\lambda,
\end{equation}

con
\begin{equation}
\Omega^{a_rb_s} \equiv\pmatrix{ \eta^o g^{ij} &\eta^o\eta^a c_{oa}^{bi}\cr ~&
~\cr \eta^o\eta^b c_{ob}^{aj} &0\cr}.
\end{equation}

\medskip

El sistema se cuantifica convirtiendo al generador BRST en el operador
herm\'{\i}tico y nilpotente $\hat\Omega$:
\begin{equation}\label{nilpo}
[\hat{{\Omega}},\hat{{\Omega}}]=[\hat\Omega^{cuad},
\hat\Omega^{cuad}]+2[\hat\Omega^{cuad},\hat\Omega^{lineal}]+
[\hat\Omega^{lineal},\hat\Omega^{lineal}]=0.
\end{equation}

El t\'{e}rmino $[\hat\Omega^{cuad},\hat\Omega^{cuad}]$ se anula trivialmente
ya que ${\eta^o}^2=0$ (n\'{o}tese que $\Omega$ no depende de ${\cal P}_o$). El
\'{u}ltimo t\'{e}rmino se anula porque $\hat\Omega^{lineal}$ ya es nilpotente.
Luego, s\'{o}lo debe hallarse un ordenamiento para $\hat\Omega^{cuad}$ que
satisfaga $[\hat\Omega^{cuad},\hat\Omega^{lineal}]=0$. La estructura de
$\hat{\Omega}^{lineal}$ sugiere fuertemente el siguiente ordenamiento
herm\'{\i}tico para $\hat{\Omega}^{cuad}$:
\begin{equation}\label{ordcuad}
\hat{\Omega}^{cuad}={1\over 2} \sum_{r,s=-1}^0 f^{-{1\over  2}}\hat{\cal
P}_{a_r} f \hat\Omega^{a_rb_s} \hat{\cal P}_{b_s}f^{-{1\over
2}}+\hat\eta^o\lambda.
\end{equation}

Efectivamente, por c\'{a}lculo directo puede probarse que $\hat\Omega$ es
nilpotente. Esta demostraci\'{o}n, que es central en esta tesis, puede
hallarse en el Ap\'{e}ndice B.

\subsection{Operadores de v\'{\i}nculo de Dirac}

~~~~ Ahora identificaremos los operadores de v\'{\i}nculo de Dirac. Hemos
visto en la secci\'{o}n \S 3.2.1 que dichos operadores pueden ser
identificados f\'{a}cilmente si se escribe de manera apropiada el operador
$\hat\Omega$:
\begin{eqnarray}\label{omegatot}
\lefteqn{~~ \hat\Omega={1\over 2} \sum_{r,s=-1}^0 f^{-{1\over  2}}\hat{\cal
P}_{a_r} f \hat\Omega^{a_rb_s} \hat{\cal P}_{b_s}f^{-{1\over
2}}+\hat\eta^o\lambda+\sum_{s=-1}^0 \hat\Omega^{c_s} \hat{\cal
P}_{c_s}}\nonumber
\\ &&= {\hat\eta}^o {1\over 2}  f^{-{1\over 2}} \hat p_i  g^{ij} f \hat p_j
f^{-{1\over 2}} + {1\over 2} f^{-{1\over 2}}\hat{\cal P}_a\hat\eta^o
\hat\eta^c c_{oc}^{aj} f \hat p_j f^{-{1\over 2}}+\nonumber \\ &&+{1\over 2}
f^{-{1\over 2}}\hat\eta^o \hat\eta^c \hat p_i c_{oc}^{ai} f \hat{\cal P}_a
f^{-{1\over 2}} + \hat\eta^o\lambda + \hat \eta^a f^{1\over 2} \xi^i_a \hat
p_i f^{-{1\over 2}} + {1\over 2} \hat \eta^a \hat \eta^b C_{ab}^c \hat {\cal
P}_c.
\end{eqnarray}

Los operadores de v\'{\i}nculo del m\'{e}todo de Dirac son los coeficientes de
los operadores fantasmas en el operador BRST escrito en el ordenamiento
$\hat\eta - \hat{\cal P}$, es decir que todos lo momentos fantasmas son
llevados a la derecha de las variables fantasmas conjugadas usando
repetidamente las relaciones de anticonmutaci\'{o}n. Obtenemos de este modo:
\begin{eqnarray}\label{omeord'}
\lefteqn{\hat\Omega={\hat\eta}^o({1\over 2}  f^{-{1\over 2}} \hat p_i  g^{ij}
f \hat p_j f^{-{1\over 2}} + {i\over 2} f^{1\over 2} c_{oa}^{aj}\hat p_j
f^{-{1\over 2}}+ \lambda) + \hat \eta^a f^{1\over 2} \xi^i_a \hat p_i
f^{-{1\over 2}}+}\nonumber\\ &&+ {1\over 2} \hat \eta^o \hat \eta^a (f^{1\over
2} c_{oa}^{bj}\hat p_j f^{-{1\over 2}} + f^{-{1\over 2}} \hat p_j c_{oa}^{bj}
f^{1\over 2} ) \hat {\cal P}_b + {1\over 2} \hat \eta^a \hat \eta^b C_{ab}^c
\hat {\cal P}_c.
\end{eqnarray}

Luego, el coeficiente de ${\hat\eta}^o$ en la ecuaci\'{o}n (\ref{omeord'}) es
el operador de v\'{\i}nculo super-Hamiltoniano $\hat {\cal H}_o$ y los
coeficientes de ${\hat\eta}^a$ son los operadores de v\'{\i}nculos
supermomentos ${\hat {\cal H}}_a$. El segundo t\'{e}rmino en $\hat{\cal H}_o$
es la contribuci\'{o}n de los fantasmas al operador de v\'{\i}nculo.
N\'{o}tese que la hermiticidad de $\hat\Omega$ no implica que los operadores
de v\'{\i}nculo sean herm\'{\i}ticos. Claramente esta propiedad no es
necesaria porque el \'{u}nico autovalor de inter\'{e}s es el cero. En cambio
el t\'{e}rmino no herm\'{\i}tico garantiza que el \'{a}lgebra de v\'{\i}nculos
cierre.

\bigskip

\subsubsection{Cambio de escala del super-Hamiltoniano}

~~~~ Hasta ahora hemos tratado con un potencial constante (tipo 1). La
introducci\'{o}n de un potencial definido positivo (tipo 2) $\lambda
\vartheta(q)$ en el v\'{\i}nculo super-Hamiltoniano, puede hacerse al nivel
del operador BRST por medio de la transformaci\'{o}n unitaria
\begin{equation}\label{tu}
\hat \Omega  \rightarrow  e^{i\hat  C}~\hat\Omega~ e^{-i\hat C},
\end{equation}
que provee un nuevo operador BRST herm\'{\i}tico y nilpotente. Elijamos, por
lo tanto,
\begin{equation}\label{de}
\hat C={1\over 2}[\hat\eta^o~ ln~  \vartheta(q)~\hat{\cal
P}_o-\hat{\cal P}_o~ ln~ \vartheta(q)~\hat\eta^o],~~~~~~~\vartheta(q)>0.
\end{equation}
As\'{\i} obtenemos,
\begin{eqnarray}\label{omegaesc}
\lefteqn{\hat\Omega={\hat\eta}^o({1\over 2}{\vartheta}^{1\over 2} f^{-{1\over
2}}\hat p_i g^{ij}f \hat p_j f^{-{1\over 2}}{\vartheta}^{1\over 2}+ {i\over
2}{\vartheta}^{1\over 2}  f^{1\over 2} c_{oa}^{aj}\hat p_j f^{-{1\over 2}}
{\vartheta}^{1\over 2} +\lambda {\vartheta}) + \hat\eta^a
{\vartheta}^{-{1\over 2}}f^{1\over 2} \xi^i_a \hat p_i f^{-{1\over
2}}{\vartheta}^{1\over 2}}\nonumber\\ &&+\hat\eta^o\hat\eta^a \xi^i_a (ln
{\vartheta})_{,i}\hat {\cal P}_o + {1\over 2} \hat \eta^o \hat \eta^a
{\vartheta}^{{1\over 2}} (f^{1\over 2} c_{oa}^{bj} \hat p_j f^{-{1\over 2}} +
f^{-{1\over 2}} \hat p_j c_{oa}^{bj} f^{1\over 2} ){\vartheta}^{1\over 2} \hat
{\cal P}_b + {1\over 2} \hat \eta^a \hat \eta^b C_{ab}^c \hat {\cal P}_c.
\end{eqnarray}

El operador $\hat \Omega$ resultante corresponde a un v\'{\i}nculo
cuadr\'{a}tico con el cambio de escala $H\ =\ \vartheta\ {\cal H}_o$ (luego,
$C^{bj}_{oa}\ =\ \vartheta\ c^{bj}_{oa}$). Una vez m\'{a}s, los nuevos
operadores de v\'{\i}nculo pueden leerse de la ecuaci\'{o}n (\ref{omegaesc}):
\begin{equation}\label{vcuadq'v}
\hat H={1\over 2}{\vartheta}^{1\over 2} f^{-{1\over 2}}\hat p_i g^{ij}f \hat
p_j f^{-{1\over 2}}{\vartheta}^{1\over 2}+ {i\over 2}{\vartheta}^{-{1\over 2}}
f^{1\over 2} C_{oa}^{aj}\hat p_j f^{-{1\over 2}} {\vartheta}^{1\over 2}
+\lambda {\vartheta},
\end{equation}
\begin{equation}\label{vlinq'}
\hat {\cal H}_a={\vartheta}^{-{1\over 2}}f^{1\over 2} \xi^i_a\hat p_i
f^{-{1\over 2}}{\vartheta}^{1\over 2},
\end{equation}
con las correspondientes funciones de estructura,
\begin{equation}
\hat C_{oa}^o=\xi^i_a (ln ~{\vartheta})_{,i},
\end{equation}
\begin{equation}
\hat C_{oa}^b = {1\over 2} \left({\vartheta}^{-{1\over 2}} f^{1\over 2}
C_{oa}^{bj} \hat p_j f^{-{1\over 2}} \vartheta^{1\over  2} + \vartheta^{1\over
2} f^{-{1\over 2}} \hat p_j C_{oa}^{bj} f^{1\over 2} {\vartheta}^{-{1\over 2}}
\right),
\end{equation}
\begin{equation}
\hat C_{ab}^c=C_{ab}^c.
\end{equation}
Todos los operadores est\'{a}n apropiadamente ordenados de manera que
satisfacen el \'{a}lgebra de v\'{\i}nculos a nivel cu\'{a}ntico libre de
anomal\'{\i}as,
\begin{equation}\label{ccq'}
[\hat  H,\hat  {\cal H}_a]=\hat C_{oa}^o \hat H  +  \hat C_{oa}^b(q,p) \hat
{\cal H}_b,
\end{equation}
\begin{equation}\label{clq'}
[\hat {\cal H}_a,\hat {\cal H}_b]=\hat C_{ab}^c(q) \hat {\cal H}_c.
\end{equation}

\goodbreak

El  resultado  (\ref{vcuadq'v}) dice que el operador asociado con el
v\'{\i}nculo de primera clase $H={1\over   2}  G^{ij}(q)p_ip_j+  V(q)$,  con
$V(q)>0~~\forall q$, no es el Laplaciano para la m\'{e}trica $G^{ij}$ m\'{a}s
$V$, sino
\begin{equation}\label{vcuadG}
\hat H={1\over 2} V^{1\over 2} f^{-{1\over 2}} \hat  p_i  V^{-1} G^{ij} f \hat
p_j f^{-{1\over 2}} V^{1\over 2}+{i\over 2} V^{-{1\over 2}}  f^{1\over 2}
C_{oa}^{aj}\hat p_j f^{-{1\over 2}} V^{1\over 2} +V,
\end{equation}
ya que la m\'{e}trica en el t\'{e}rmino cin\'{e}tico debe ser $g_{ij}=V
G_{ij}$. Por simplicidad usamos un potencial definido positivo, pero debe
notarse que, en general, lo que se requiere es que no se anule. Como se ve, el
formalismo BRST provee contribuciones de los fantasmas a los operadores de
v\'{\i}nculo, que son necesarias para satisfacer el \'{a}lgebra y preservar el
car\'{a}cter geom\'{e}trico de la funci\'{o}n de onda. La contribuci\'{o}n de
los fantasmas al operador de v\'{\i}nculo cuadr\'{a}tico es el segundo
t\'{e}rmino en la ecuaci\'{o}n (\ref{vcuadG}) y lo analizaremos luego. Los
operadores de v\'{\i}nculo lineales adquieren dos t\'{e}rminos
antiherm\'{\i}ticos asociados con las trazas de las funciones de estructura:
\begin{equation}\label{varlinqe}
\hat {\cal H}_a= V^{-{1\over 2}} f^{1\over 2} \xi^i_a \hat p_i f^{-{1\over 2}}
V^{1\over 2}=\xi^i_a \hat p_i - {i\over 2} \xi^i_{a,i} + {i\over 2} C^b_{ab} +
{i\over 2} C^o_{ao},
\end{equation}
donde ${i\over 2} C^o_{ao} = -{i\over 2} \xi^i_a (\ln V)_{,i}$.  De acuerdo
con lo que vimos al principio este cap\'{\i}tulo, la restricci\'{o}n sobre el
potencial puede ser relajada, debido a que los resultados no cambian si
$\lambda$, en vez de ser constante, es una funci\'{o}n $\lambda(y)$ invariante
en \'{o}rbitas asociadas v\'{\i}nculos lineales (tipo 1). El potencial
estar\'{\i}a entonces s\'{o}lo restringido a factorizar como $V = \vartheta(q)
\lambda(y)$, $\vartheta(q)> 0$.\footnote{Esta factorizaci\'{o}n permite la
existencia de un ``gauge f\'{\i}sico'' globalmente bien definido en
\cite{hk90}: $\Omega (q)$ en esa cita puede ser tomado como $\ln \vartheta(q)$
y  ${i\over 2} C^o_{ao}$ es  el all\'{\i} denominado ``cociclo''.}  Sin
embargo, potenciales que no son definidos positivos hacen menos evidente la
manera de construir el producto interno f\'{\i}sico en el espacio de Hilbert
f\'{\i}sico [\cite{bf}]. Luego, el t\'{e}rmino cin\'{e}tico en el
super-Hamiltoniano y los supermomentos son sensibles a la existencia del
potencial.

\subsection{Invariancias y producto interno f\'{\i}sico}

~~~~ El operador de v\'{\i}nculo Hamiltoniano (\ref{vcuadG}) difiere del
hallado en \cite{hk90}, donde es agregado un t\'{e}rmino de curvatura para
retener la invariancia de la teor\'{\i}a ante cambio de escala. De un modo
diferente, aqu\'{\i} se consigue la invariancia ante cambios de escala gracias
al papel jugado por el potencial en los operadores de v\'{\i}nculo. De hecho,
el papel jugado por los factores $f^{{\pm}{1\over 2}}$, $V^{{\pm}{1\over 2}}$
es claro si se presta atenci\'{o}n a las transformaciones que deber\'{\i}an
dejar invariante la teor\'{\i}a; estas transformaciones son (i) cambios de
coordenadas, (ii) combinaciones de v\'{\i}nculos supermomentos, Ec.
(\ref{cbase}), y (iii) cambio de escala del v\'{\i}nculo super-Hamiltoniano
($H \rightarrow e^{\Theta}~H$). El producto interno f\'{\i}sico de las
funciones onda de Dirac,
\begin{equation}\label{prodesc}
(\varphi_1,\varphi_2)=\int dq\ \big[\prod^{m+1} \delta(\chi)\big]\ J\ \varphi^*_1(q)\
\varphi_2(q),
\end{equation}
(donde $J$ es el determinante de Faddeev-Popov y $\chi$ representa las $m+1$
condiciones de gauge) debe ser invariante bajo cualquiera de estas
transformaciones. De acuerdo al cambio del determinante de Faddeev-Popov bajo
(ii) y (iii), el producto interno permanecer\'{a} invariante si la funci\'{o}n
de onda de Dirac cambia de acuerdo a
\begin{equation}\label{unmedio}
\varphi \rightarrow \varphi'=(det A)^{1\over 2} e^{-{\Theta\over 2}} \varphi.
\end{equation}

De esta manera, los factores $f^{{\pm}{1\over 2}}$, $V^{{\pm}{1\over 2}}$ en
los operadores de v\'{\i}nculo son justamente lo que se necesita para que
$\hat {\cal H}_a\varphi,~\hat H\varphi,$ y ${\hat C}_{oa}^b\varphi$
transformen como $\varphi$, preservando de esta manera el car\'{a}cter
geom\'{e}trico de la funci\'{o}n de onda de Dirac.

En las Ecs. (\ref{prodesc})-(\ref{unmedio}) la funci\'{o}n de onda es una
densidad de peso ${1\over 2}$. Si se prefiere ver a la funci\'{o}n de onda
como invariante bajo las transformaciones relevantes (i)-(iii), se
deber\'{\i}a realizar la transformaci\'{o}n
\begin{eqnarray}
\lefteqn{~~~~~~\varphi ~ \rightarrow \phi=f^{-{1\over 2}}V^{+{1\over
2}}\varphi, }\label{fun} \\ &&\hat{\cal O} ~ \rightarrow f^{-{1\over
2}}V^{+{1\over 2}}\hat{\cal O} f^{+{1\over 2}}V^{-{1\over 2}}. \label{opi}
\end{eqnarray}
El producto interno f\'{\i}sico correspondiente resulta de la integraci\'{o}n
del invariante $\phi^*_1\phi_2$ en el volumen invariante $V^{-1}~J~[\prod
\delta(\chi)]~{\tilde\alpha}$.
\begin{equation}\label{peinv}
(\phi_1,\phi_2)=(\varphi_1,\varphi_2)=\int \tilde\alpha\ V^{-1} \ J\
\big[\prod \delta(\chi)\big]\ \phi^*_1\ \phi_2.
\end{equation}

\subsubsection{Reducci\'{o}n del sistema}

~~~~ Puesto que el producto interno (\ref{prodesc}) o (\ref{peinv}) es
invariante ante la transformaci\'{o}n (ii), se puede elegir la base coordenada
abeliana $\vec\xi_a^{\prime}={\partial/ \partial Q^a}$ ($G^{\prime}_a=P_a$).
De este modo, el volumen se expresa
\begin{equation}\label{volab}
{\tilde\alpha}^{\prime}=dQ^1 \wedge ... \wedge  dQ^m \wedge \omega(y) dy^1
\wedge ... dy^{n-m}.
\end{equation}

Luego, las ecuaciones de v\'{\i}nculo lineales para la funci\'{o}n de onda de
Dirac $\phi$ son
\begin{equation}\label{linfacil}
{\partial\phi\over\partial Q^a} = 0.
\end{equation}
Estas ecuaciones simplifican notablemente la ecuaci\'{o}n de v\'{\i}nculo
correspondiente al super-Hamiltoniano que, cuando se escribe en la base
coordenada $\{dQ^a, dy^r\}$ [luego $f' = \omega(y)$], se reduce a
\begin{equation}\label{casilaplace}
\left(-{1\over  2}  V  {\partial\over\partial  Q^a}\ V^{-1}  G^{ar}
{\partial\over\partial y^r} -{1\over 2}
V{\omega(y)}^{-1}{\partial\over\partial y^r}\ {\omega(y)}V^{-1} G^{rs}
{\partial\over\partial y^s} + {1\over 2} C^{ar}_{oa} {\partial\over \partial
y^r} + V\right)\ \phi=0.
\end{equation}
El potencial puede ser factorizado. Luego, teniendo en cuenta las ecuaciones
(\ref{cchg}) y (\ref{coac}), se obtiene que $V^{-1}C^{br}_{oa}=
c^{br}_{oa}=\partial g^{br}/\partial Q^a$ y $V^{-1}G^{rs}=g^{rs}=g^{rs}(y)$.
Entonces,
\begin{equation}\label{laplac}
\left(-{1\over 2} \omega(y)^{-1} {\partial\over\partial y^r}\ \omega(y)
g^{rs}(y) {\partial\over\partial y^s} + 1\right)\ \phi =  0.
\end{equation}
Por lo tanto, la contribuci\'{o}n de los fantasmas al super-Hamiltoniano
permite el surgimiento del ``Laplaciano" en t\'{e}rminos de las variables
reducidas $\{y^r\}$.  Para obtener el Laplaciano verdadero para la m\'{e}trica
reducida invariante ante cambio de escala $g_{rs}(y)$, se deber\'{\i}a elegir
$\omega(y)$ como $\vert \det (g_{rs})\vert^{1/2}$. Es claro que el formalismo
BRST no puede dar un valor para $\omega(y)$, porque $\hat\Omega$ es
herm\'{\i}tico y nilpotente para cualquier $\omega(y)$.\footnote{Si, por
ejemplo, el sistema tiene solamente un v\'{\i}nculo cuadr\'{a}tico ${\cal H}$,
luego $\hat\Omega = \hat\eta \hat {\cal H}$ ser\'{\i}a herm\'{\i}tico y
nilpotente para cualquier ordenamiento herm\'{\i}tico de $\hat {\cal H}$.}

Para analizar la relaci\'{o}n entre la cuantificaci\'{o}n en el marco del
formalismo de Dirac y la cuantificaci\'{o}n en el espacio de fases reducido,
elijamos las funciones de fijado de gauge $\chi^a=Q^a,~ \{\chi^o,P_a\}=0$.
Luego, en la ecuaci\'{o}n (\ref{peinv}) se integra en las variables $Q^a$
usando el volumen $\tilde{\alpha}^{\prime}$ y se obtiene
\begin{equation}\label{pered}
(\phi_1,\phi_2)=\int \tilde\omega\ V^{-1} \ J_o\ \delta(\chi^o)\
\phi^*_1[q^i(Q^a=0,y^r)]\ \phi_2[q^i(Q^a=0,y^r)].
\end{equation}

Se puede definir una densidad de peso 1/2 ante cambios de las $y^r$:
\footnote{La funci\'{o}n $\omega^{-1} f'$ juega el papel de $\mu$ en
\cite{fhp93} y el de ${\cal M}$ en \cite{bk93}. De hecho, si elegimos usar la
base Abelianizada ${\vec\xi}^{\prime}_a$ en $T_{||}M$ que es una base
coordenada: ${\vec\xi}^{\prime}_a={\partial/ \partial Q^a}$
($G^{\prime}_a=P_a$), ${\tilde E}^{\prime a}=dQ^a$.  Luego, $\omega^{-1} f'$
es el Jacobiano del cambio de coordenadas $(Q^a,y^r) \rightarrow q^i$.}
\begin{eqnarray}\label{densred}
\lefteqn{\varphi_R(y)=\omega (y)^{1\over  2}\   V[q^i(Q^a=0,y^r)]^{-{1\over
2}}\ \phi[q^i(Q^a=0,y^r)]}\nonumber\\ &&~~~=\omega (y)^{1\over 2}\ f'
[q^i(Q^a=0,y^r)]^{-{1\over 2}}\ \varphi[q^i(Q^a=0,y^r)].
\end{eqnarray}

Luego,
\begin{equation}\label{peredens}
(\phi_1,\phi_2)=(\varphi_{R_1},\varphi_{R_2})=\int dy \ J_o\ \delta(\chi^o)
\varphi^*_{R_1}(y) \varphi_{R_2}(y),
\end{equation}
y $\varphi_R$ es funci\'{o}n de onda de Dirac en un espacio ``reducido" donde
s\'{o}lo subsiste el v\'{\i}nculo cuadr\'{a}tico: $\varphi_R$ est\'{a}
vinculada por la ecuaci\'{o}n (\ref{laplac}) que es satisfecha por $\phi
[q^i(Q^a=0,y^r)]$.

\section{Ep\'{\i}logo: Principio de m\'{\i}nima acci\'{o}n de Jacobi}

~~~~ El modo en que el potencial entra en el ordenamiento de los operadores de
v\'{\i}nculo puede parecer artificial (especialmente en el t\'{e}rmino
cin\'{e}tico del v\'{\i}nculo cuadr\'{a}tico). Sin embargo, vi\'{e}ndolo con
cuidado el papel que desempe\~{n}a resulta completamente natural. Con este fin
exponemos a continuaci\'{o}n un argumento a nivel cl\'{a}sico sumamente
elegante. Los sistemas con covariancia general son invariantes ante cambios
del par\'{a}metro en la acci\'{o}n funcional. Esto significa que el
par\'{a}metro es f\'{\i}sicamente irrelevante: no es el tiempo. El tiempo
puede estar escondido entre las variables din\'{a}micas y, como resultado de
ello, el Hamiltoniano est\'{a} restringido a anularse [\cite{koxf}]. En este
caso el tiempo podr\'{\i}a ser identificado como una funci\'{o}n $t(q,p)$ en
el espacio de fases, que crece mon\'{o}tonamente sobre todas las trayectorias
din\'{a}micas. Puesto que el $H$ aqu\'{\i} estudiado es equivalente a un
super-Hamiltoniano con un potencial constante, los sistemas a los cuales
pueden aplicarse los resultados hallados son aquellos semejantes al caso de
una part\'{\i}cula relativista en un espacio tiempo curvo. Luego, el tiempo se
halla escondido en el espacio de configuraci\'{o}n: es decir se trata de un
{\it tiempo intr\'{\i}nseco} (recordar lo expuesto en \S 2.3). Esto significa
que la trayectoria en el espacio de configuraci\'{o}n contiene toda la
informaci\'{o}n sobre el sistema. En mec\'{a}nica cl\'{a}sica, el principio de
Jacobi [\cite{lanczos,goldstein}] es el principio variacional para obtener las
trayectorias en el espacio de configuraci\'{o}n, para una energ\'{\i}a fija
$E$, sin informaci\'{o}n alguna sobre la evoluci\'{o}n del sistema en el
par\'{a}metro de la acci\'{o}n funcional. Los caminos son obtenidos cuando se
var\'{\i}a la funcional
\begin{equation}\label{jac}
I=\int_{q'}^{q''} \sqrt {2 |E-V|G_{ij}dq^i dq^j}.
\end{equation}

En nuestro caso, la energ\'{\i}a es cero (por la condici\'{o}n de
v\'{\i}nculo), y la ecuaci\'{o}n (\ref{jac}) tiene la forma de la acci\'{o}n
funcional de una part\'{\i}cula relativista de masa unidad en un fondo curvo
con m\'{e}trica $2 V G_{ij}=2g_{ij}$. Los caminos son geod\'{e}sicas de esta
m\'{e}trica $2 g_{ij}$, en vez de $G_{ij}$. Cuando se fija el gauge como $Q^a
= 0$, el principio de Jacobi se reduce a la variaci\'{o}n de la funcional
\begin{equation}\label{jac2}
I_R = \int_{y'}^{y''} \sqrt {2 V G_{rs} dy^r dy^s} ,
\end{equation}
dando de esta manera un sustento cl\'{a}sico a la ecuaci\'{o}n de v\'{\i}nculo
(\ref{laplac}), donde el Laplaciano es el asociado con la m\'{e}trica
invariante ante cambio de escala $2 V G_{rs} = 2 g_{rs}(y)$ que aparece en la
ecuaci\'{o}n (\ref{jac2}).

\newpage


\chapter{Sistemas con tiempo extr\'{\i}nseco}

\bigskip

{\it ~~~~ Al intentar cuantificar un sistema como la relatividad general,
vimos que una de sus caracter\'{\i}sticas m\'{a}s complicadas es el problema
del tiempo (\S 2.3). En mec\'{a}nica cu\'{a}ntica, el tiempo es un
par\'{a}metro absoluto; no est\'{a} en pie de igualdad con las otras
coordenadas que luego se convierten en operadores y observables. En
relatividad general, en cambio, el ``tiempo'' es una mera etiqueta de una
hipersuperficie espacial, y las cantidades f\'{\i}sicamente relevantes son
independientes de dichas etiquetas: ellas son invariantes ante difeomorfismos.
La relatividad general es un ejemplo de un sistema parametrizado (un sistema
cuya acci\'{o}n es invariante ante cambios del par\'{a}metro de
integraci\'{o}n). Tal tipo de sistemas puede ser obtenido a partir de una
acci\'{o}n que no posee invariancia de reparametrizaci\'{o}n, elevando al
tiempo al rango de variable din\'{a}mica. De este modo, los grados de libertad
originales y el tiempo son funciones de cierto par\'{a}metro f\'{\i}sicamente
irrelevante. El tiempo puede ser variado independientemente de los otros
grados de libertad cuando se agrega un v\'{\i}nculo con su correspondiente
multiplicador de Lagrange. En este proceso, surge una caracter\'{\i}sica
especial: el Hamiltoniano est\'{a} restringido a anularse.

La mayor\'{\i}a de los esfuerzos dirigidos a cuantificar la relatividad
general (o ciertos modelos de minisuperespacio) enfatizan la analog\'{\i}a con
la part\'{\i}cula relativista. Es justamente el tipo de tratamiento que
realizamos en el cap\'{\i}tulo anterior . Esto se debe a que ambos sistemas
poseen v\'{\i}nculos Hamiltonianos ${\cal H}_o$ que son hiperb\'{o}licos en
los momentos. Si el papel de la masa al cuadrado es interpretado por un
potencial definido positivo, luego la analog\'{\i}a es completa en el sentido
que el tiempo est\'{a} escondido en el espacio de configuraci\'{o}n
[\cite{prd,hk90}]. M\'{a}s a\'{u}n, un potencial definido positivo garantiza
que la componente temporal del momento nunca es nula sobre la superficie de
v\'{\i}nculo. De este modo, el corchete de Poisson $\{q^o, {\cal H}_o \}$
nunca se anula, lo cual indica que $q^o$ evoluciona mon\'{o}tonamente sobre
cualquier trayectoria din\'{a}mica; esta es la propiedad esencial del tiempo.
En este caso, hemos mostrado c\'{o}mo obtener un ordenamiento consistente de
los operadores de v\'{\i}nculo a partir del formalismo BRST [\cite{prd}].

Desafortunadamente, la analog\'{\i}a no puede ser considerada muy seriamente
puesto que el potencial en relatividad general es la curvatura espacial (no
necesariamente definida positiva). Esto significa que el tiempo en relatividad
general debe ser sugerido por otro modelo mec\'{a}nico. En este cap\'{\i}tulo
construiremos un modelo con tiempo extr\'{\i}nseco cuyo potencial no est\'{a}
restringido a ser definido positivo, pero es invariante de gauge (tipo 1) lo
cual nos permitir\'{a} utilizar los resultados ya obtenidos como punto de
partida para su cuantificaci\'{o}n en un caso m\'{a}s general
[\cite{prd99,np00}].}

\section{El modelo: escondiendo el tiempo}

~~~~ Con el objetivo de hallar un modelo mec\'{a}nico mejor para la
relatividad general, comencemos con un sistema que posee $n$ grados de
libertad genuinos con un Hamiltoniano $h={1\over
2}g^{\mu\nu}p_{\mu}p_{\nu}+v(q^{\mu})$ cuya m\'{e}trica $g^{\mu\nu}$ es
definida positiva.

La din\'{a}mica del sistema no cambiar\'{a} si se agrega una funci\'{o}n del
tiempo al Hamiltoniano, digamos $-{t^2\over 2}$. La acci\'{o}n para dicho
sistema es
\begin{equation}
S=\int \left[p_{\mu}{dq^{\mu}\over dt} - h(q^{\mu},p_{\mu})+ {t\over
2}^2\right] dt,~~~~~ {\mu}=1,...,n
\end{equation}

El sistema puede ser parametrizado si la variable de integraci\'{o}n $t$ es
vista como una variable can\'{o}nica cuyo momento conjugado es (menos) el
Hamiltoniano. Esta \'{u}ltima condici\'{o}n entra en la acci\'{o}n como un
v\'{\i}nculo ${\cal H}_o=p_t+h-{t\over 2}^2$, y la acci\'{o}n queda expresada
como

\begin{equation}\label{act}
S[q^{\mu},p_{\mu},t,p_t,N^o]=\int\left[p_t{dt\over d{\tau}}+
p_{\mu}{dq^{\mu}\over d{\tau}} - N^o \left( p_t+h(q^{\mu},p_{\mu})-{t\over
2}^2 \right)\right] d\tau
\end{equation}

\noindent donde $N$ es multiplicador de Lagrange.

Hasta aqu\'{\i} el v\'{\i}nculo es parab\'{o}lico en los momentos. Sin
embargo, si realizamos la transformaci\'{o}n can\'{o}nica,
\begin{equation}\label{tc}
q^0=p_t ,~~~~~~~~~~~~~~~  p_0=-t
\end{equation}
convierte al v\'{\i}nculo ${\cal H}_o$ en una funci\'{o}n hiperb\'{o}lica de
los momentos [\cite{bf}]:
\begin{equation}\label{ham5}
{\cal H}_o=q^0+h-{1\over 2}p_0^2=-{1\over 2}p_0^2+{1\over 2} g^{\mu\nu}
p_{\mu}p_{\nu}+v(q^{\mu})+q^0={1\over 2} {\cal G}^{rs} p_r p_s+{\cal V}(q^r),
\end{equation}
con $r,s=0,1,...,n$. Las componentes de la m\'{e}trica son ${\cal G}^{00}=-1$,
${\cal G}^{0\nu}=0$, ${\cal G}^{\mu\nu}=g^{\mu\nu}$ y el potencial es ${\cal
V}(q^i)=v(q^{\mu})+q^0$. Luego, ${\cal G}^{rs}$ es una m\'{e}trica
Lorentziana, como lo es la superm\'{e}trica en el formalismo de
Arnowitt-Deser-Misner (ADM) en relatividad general. El v\'{\i}nculo
(\ref{ham5}) describe un sistema parametrizado con tiempo extr\'{\i}nseco
(escondido en el espacio de fases), cuyo potencial no es definido positivo.

Para una analog\'{\i}a m\'{a}s completa con relatividad general, los
v\'{\i}nculos supermomentos pueden ser introducidos agregando $m$ grados de
libertad $q^a$. Su car\'{a}cter espurio es expresado por $m$ v\'{\i}nculos
lineales y homog\'{e}neos en los momentos ${\cal H}_a\equiv\xi_a^r p_r$, donde
$\vec\xi_a$ son $m$ campos vectoriales tangentes a las curvas coordenadas
asociadas con las coordenadas $q^a$. Estos $m$ v\'{\i}nculos ${\cal H}_a$
pueden adem\'{a}s ser combinados linealmente
\begin{equation}\label{combl}
{\cal H}_a \rightarrow {\cal H}_{a'}= A_{a'}^{~a}(q)~{\cal
H}_a,~~~~~~~~~~~~{\rm det}~A\ne 0,
\end{equation}
y obtener un conjunto equivalente de v\'{\i}nculos supermomentos lineales y
hom\'{o}geneos. El conjunto $({\cal H}_o, {\cal H}_a)$ es de primera clase.

Finalmente la din\'{a}mica del sistema se obtiene variando la acci\'{o}n
\begin{equation}
S[q^i,p_i,N^o ,N^a]=\int \left[p_i{dq^i\over d{\tau}} - N^o {\cal H}_o - N^a
{\cal H}_a\right] d\tau, ~~~~~~~i=0,1,...,n+m
\end{equation}
donde $N^a$ son los multiplicadores de Lagrange correspondientes a los
v\'{\i}nculos ${\cal H}_a$.

Tal sistema satisface las siguientes condiciones (que pueden ser le\'{\i}das
del v\'{\i}nculo Hamiltoniano, Ec. (\ref{ham5})):
\begin{eqnarray}
\lefteqn{ ~~~~~ {\partial {\cal G}^{ij}\over \partial q^0}\approx
0}\label{conda}\\ &&{\partial {\cal V}\over \partial q^0} = 1\label{condb}
\end{eqnarray}
El s\'{\i}mbolo ``$\approx$'' significa ``igualdad d\'{e}bil'' (la igualdad
est\'{a} restringida a la subvariedad definida por los v\'{\i}nculos ${\cal
H}_a\approx 0$) y reemplaza la igualdad ordinaria debido a que la m\'{e}trica
posee un sector no f\'{\i}sico que puede depender de $q^0$.

La parametrizaci\'{o}n del sistema, que es a\'{u}n visible en la ecuaci\'{o}n
(\ref{ham5}) debido a la forma espacial del potencial y las componentes de la
m\'{e}trica, puede ser enmascarada por medio de una transformaci\'{o}n general
de coordenadas. Sin embargo, las propiedades geom\'{e}tricas distintivas del
sistema, expresadas por las ecuaciones (\ref{conda}),(\ref{condb}), pueden ser
escritas en forma geom\'{e}trica (es decir, independiente de las coordenadas)
usando derivadas de Lie. De este modo, las ecuaciones (\ref{ham5}) y
(\ref{conda}), (\ref{condb}) indican que existe un vector de Killing temporal
d\'{e}bilmente {\it unitario} que satisface
\begin{eqnarray}
\lefteqn{ ~~~~~ {\cal L}_{\vec\xi _0}{\bf {\cal G}}\approx 0,}
\label{condga}\\ && {\cal L}_{\vec\xi _0}{\cal V} = 1. \label{condgb}
\end{eqnarray}

A diferencia de otros tratamientos donde un v\'{\i}nculo hiperb\'{o}lico como
(\ref{ham5}) es comparado con el de una part\'{\i}cula relativista, y el
par\'{a}metro del vector de Killing es visto como el tiempo [\cite{koxf}], en
nuestro enfoque el tiempo ($t=-p_0$) es la variable din\'{a}mica {\it
conjugada} al par\'{a}metro del vector de Killing [\cite{prd99}].

\section{El generador BRST cl\'{a}sico y cu\'{a}ntico}

~~~~ Si bien este modelo se ha obtenido a trav\'{e}s de un proceso de
parametrizaci\'{o}n, el resultado final desde el punto de vista de los
v\'{\i}nculos y las funciones de estructura, debido a que el potencial es
invariante de gauge (tipo 1), es enteramente equivalente al del cap\'{\i}tulo
anterior [\cite{np00}]. El hecho de poseer un tiempo de caracter\'{\i}sticas
diferentes al caso anterior s\'{o}lo influir\'{a} en la forma en la cual se
define el producto interno f\'{\i}sico pero no en el generador BRST. Esto es
as\'{\i}, ya que al aplicar el formalismo BRST lo \'{u}nico que interesa es el
conjunto de v\'{\i}nculos (de primera clase) en s\'{\i} y sus funciones de
estructura asociadas.

Es decir que podemos explotar esta equivalencia a nivel de v\'{\i}nculos para
obtener sin necesidad de c\'{a}lculo adicional el generador BRST a nivel
cl\'{a}sico y cu\'{a}ntico.

La invariancia de gauge del potencial se corresponde con el tipo 1 descripto
en \S 4.1, en ese caso vimos que la funci\'{o}n de estructura $C_{oa}^o(q)$ se
anula, y el \'{a}lgebra resultante es la (\ref{ccgg})-(\ref{cchg}) con $
C_{oa}^o=0$.

Es decir que el generador BRST cl\'{a}sico ser\'{a}, Ec. (\ref{omegacl4}),
\begin{eqnarray}
\Omega=\eta^o{\cal H}_o+\eta^a{\cal H}_a+ \eta^o \eta^a C_{oa}^b{\cal
P}_b+{1\over 2}\eta^a \eta^b C_{ab}^c{\cal P}_c.
\end{eqnarray}

Por lo tanto, tambi\'{e}n tenemos que el operador BRST herm\'{\i}tico y
nilpotente, es entonces, Ec. (\ref{omeord'}),
\begin{eqnarray}
\hat\Omega={\hat\eta}^o{\cal H}_o+ \hat \eta^a {\cal H}_a+ {1\over 2} \hat
\eta^o \hat \eta^a (f^{1\over 2} c_{oa}^{bj}\hat p_j f^{-{1\over 2}} +
f^{-{1\over 2}} \hat p_j c_{oa}^{bj} f^{1\over 2} ) \hat {\cal P}_b + {1\over
2} \hat \eta^a \hat \eta^b C_{ab}^c \hat {\cal P}_c.
\end{eqnarray}
con
\begin{equation}\label{vcuad}
\hat{\cal H}_o={1\over 2} f^{-{1\over 2}}\hat p_i {\cal G}^{ij}f \hat p_j
f^{-{1\over 2}}+ {i\over 2} f^{1\over 2} c_{oa}^{aj}\hat p_j f^{-{1\over 2}} +
{\cal V}
\end{equation}
y
\begin{equation}\label{vlin}
\hat {\cal H}_a= f^{1\over 2} \xi^i_a \hat p_i f^{-{1\over 2}},
\end{equation}
donde la funci\'{o}n $f=f(q)$ satisface, en forma an\'{a}loga al caso
anterior,
\begin{equation}\label{div5}
C^b_{ab}=f^{-1}(f\xi^i_a)_{,i}={\rm div}_{\tilde\alpha}~\vec\xi_a,
\end{equation}
con la excepci\'{o}n que ahora $\tilde\alpha$ es el volumen
${\tilde\alpha}\equiv f~dq^0\wedge...\wedge dq^{n+m}$). \footnote{N\'{o}tese
que la dimensi\'{o}n del volumen se ha incrementado por la inclusi\'{o}n de
``t'' al parametrizar. Ahora tenemos que la $(n+m+1)$-forma $\tilde\alpha$ que
resuelve la ecuaci\'{o}n (5.15) es un volumen en el espacio de
configuraci\'{o}n ${\cal M}$: ${\tilde\alpha}\equiv {\tilde
E}^1\wedge...\wedge {\tilde E}^m \wedge {\tilde \omega}$, donde $\{\tilde
E^a\}$ es la base dual de $\{\vec\xi_a\}$ en $T^*_{||}{\cal M}$ (el espacio
tangente ``longitudinal'', y ${\tilde \omega}\ =\ \omega(y)\
dy^0\wedge...\wedge dy^{n}$ es una $n$ forma cerrada donde las $y^r$ son $n+1$
funciones que son invariantes ante las transformaciones de gauge generadas por
los v\'{\i}nculos lineales ($dy^r(\vec\xi_a)=0,~~\forall r,a$). $\tilde\alpha$
es el volumen inducido por los v\'{\i}nculos en la \'{o}rbita, multiplicado
por un volumen (no elegido) en el espacio ``reducido".}

Los operadores de v\'{\i}nculo de Dirac (\ref{vcuad})-(\ref{vlin}) fueron
obtenidos del generador BRST cu\'{a}ntico, el objeto central del m\'{e}todo.

En el cap\'{\i}tulo anterior, comenzamos con una m\'{e}trica
pseudo-Riemanniana y un potencial constante (una part\'{\i}cula relativista en
un espacio curvo). Dicho sistema ten\'{\i}a la propiedad $c^0_{0a}=0$, que
facilitaba la b\'{u}squeda del generador BRST cu\'{a}ntico. Despu\'{e}s de
eso, un potencial m\'{a}s general (pero definido positivo) era introducido a
trav\'{e}s de una transformaci\'on unitaria del generador BRST, y $c^0_{0a}$
surg\'{\i}a como no nula. Este procedimiento nos di\'{o} los operadores de
v\'{\i}nculo invariantes ante cambio de escala del v\'{\i}nculo
super-Hamiltoniano (sin tener que recurrir a un t\'{e}rmino de curvatura).

En el caso presente, el sistema no es una part\'{\i}cula relativista: el
potencial no es constante (ni tampoco definido positivo), y el  tiempo no
est\'{a} escondido entre las coordenadas. Sin embargo, $c^0_{0a}$ es a\'{u}n
nula debido a la invariancia de gauge del potencial. Una vez m\'{a}s, la
invariancia ante cambio de escala del v\'{\i}nculo superhamiltoniano ser\'{a}
introducida a trav\'{e}s de una transformaci\'{o}n unitaria en el espacio
extendido.

\subsection{Cambio de escala del super-Hamiltoniano}

~~~~ El cambio de escala del v\'{\i}nculo super-Hamiltoniano
\begin{equation}\label{escc}
{\cal H}_o \rightarrow  H = F ~{\cal H}_o,~~~~~~~~~~~~~F>0
\end{equation}
(luego ${\cal G}^{ij} \rightarrow G^{ij}= F~ {\cal G}^{ij},~ {\cal V}
\rightarrow V= F~{\cal V}$) relaja las propiedades geom\'{e}tricas de
$\vec\xi_0$:
\begin{eqnarray}
|\vec\xi_0|= 1~~ &&\rightarrow ~~|\vec\xi_0|= F^{-{1\over 2}}
\label{condg'0}\\ {\cal L}_{\vec\xi _0}{\bf {\cal G}}\approx 0 ~~&&\rightarrow
~~ {\cal L}_{\vec\xi_0} {\bf G} \approx C~ {\bf G}\label{condga'}\\ {\cal
L}_{\vec\xi _0}{\cal V}=1~~ &&\rightarrow ~~{\cal L}_{\vec\xi _0}V= C~
V+|\vec\xi_0|^{-2}\label{condgb'}
\end{eqnarray}
con $C(q)=\vec\xi_0 ({\rm ln} F)=-2\vec\xi_0 ({\rm ln} |\vec\xi_0|)$. As\'{\i}
$\vec\xi_0$ se convierte en un vector de Killing d\'{e}bilmente conforme no
unitario.

A nivel cu\'{a}ntico, la operaci\'{o}n de cambio de escala correspondiente
puede ser llevada a cabo realizando la transformaci\'{o}n unitaria del
operador BRST
\begin{equation}\label{tu5}
\hat \Omega  \rightarrow  e^{i\hat  M}~\hat\Omega~ e^{-i\hat M},
\end{equation}
con
\begin{equation}
\hat M=[\hat{\cal P}_o~{\rm ln} |\vec\xi_0|~\hat\eta^o -\hat\eta^o ~{\rm ln}
|\vec\xi_0|~\hat{\cal P}_o],~~~~~~~|\vec\xi_0|>0
\end{equation}
dando un nuevo operador BRST herm\'{\i}tico y nilpotente,
\begin{eqnarray}\label{omegaesc5}
\hat\Omega=&&{\hat\eta}^o\left({1\over 2} |\vec\xi_0|^{-1} f^{-{1\over 2}}\hat
p_i {\cal G}^{ij}f \hat p_j f^{-{1\over 2}} |\vec\xi_0|^{-1} + {i\over 2}
|\vec\xi_0|^{-1} f^{1\over 2} c_{oa}^{aj}\hat p_j f^{-{1\over 2}}
|\vec\xi_0|^{-1}+ |\vec\xi_0|^{-2}{\cal V}\right)\nonumber\\ &&+ \hat\eta^a
|\vec\xi_0| f^{1\over 2} \xi^i_a \hat p_i f^{-{1\over 2}}|\vec\xi_0|^{-1}-2
\hat\eta^o\hat\eta^a \xi^i_a ({\rm ln} |\vec\xi_0|)_{,i}\hat {\cal
P}_o\nonumber\\ && + {1\over 2} \hat \eta^o \hat \eta^a |\vec\xi_0|^{-1}
(f^{1\over 2} c_{oa}^{bj} \hat p_j f^{-{1\over 2}} + f^{-{1\over 2}} \hat p_j
c_{oa}^{bj} f^{1\over 2})|\vec\xi_0|^{-1} \hat {\cal P}_b + {1\over 2} \hat
\eta^a \hat\eta^b C_{ab}^c \hat {\cal P}_c,
\end{eqnarray}

que corresponde a operadores de v\'{\i}nculo que satisfacen la invariancia
ante cambio de escala (de modo que $G^{ij}=|\vec\xi_0|^{-2}~{\cal G}^{ij}$,
$V=|\vec\xi_0|^{-2}~{\cal V}$, y $C_{oa}^{bj}=|\vec\xi_0|^{-2}~c_{oa}^{bj}$)

\begin{equation}\label{vcuadq'v5}
\hat H={1\over 2}|\vec\xi_0|^{-1} f^{-{1\over 2}} \hat p_i |\vec\xi_0|^2
G^{ij}f \hat p_j f^{-{1\over 2}}|\vec\xi_0|^{-1}+ {i\over 2}|\vec\xi_0|
f^{1\over 2} C_{oa}^{aj}\hat p_j f^{-{1\over 2}} |\vec\xi_0|^{-1} + V ,
\end{equation}
\begin{equation}\label{vlinq5'}
\hat {\cal H}_a=|\vec\xi_0| f^{1\over 2} \xi^i_a\hat p_i f^{-{1\over
2}}|\vec\xi_0|^{-1},
\end{equation}
con las correspondientes funciones de estructura,
\begin{equation}
\hat C_{oa}^o=-2 \xi^i_a ({\rm ln}|\vec\xi_0|)_{,i},
\end{equation}
\begin{equation}
\hat  C_{oa}^b = {1\over 2} \left( |\vec\xi_0| f^{1\over 2} C_{oa}^{bj}  \hat
p_j  f^{-{1\over  2}} |\vec\xi_0|^{-1} + |\vec\xi_0|^{-1} f^{-{1\over 2}} \hat
p_j C_{oa}^{bj} f^{1\over 2} |\vec\xi_0| \right),
\end{equation}
\begin{equation}\label{ef}
\hat C_{ab}^c=C_{ab}^c,
\end{equation}
tales que preservan el \'{a}lgebra a nivel cu\'{a}ntico,
\begin{equation}\label{ccq'5}
[\hat  H,\hat  {\cal H}_a]=\hat C_{oa}^o \hat H  +  \hat C_{oa}^b(q,p) \hat
{\cal H}_b,
\end{equation}
\begin{equation}\label{clq'5}
[\hat {\cal H}_a,\hat {\cal H}_b]=\hat C_{ab}^c(q) \hat {\cal H}_c.
\end{equation}

\subsection{Producto interno f\'{\i}sico}

~~~~ El proceso de cuantificaci\'{o}n no est\'{a} completo sin un producto
interno f\'{\i}sico donde los grados de libertad espurios son congelados por
medio condiciones de fijado de gauge. Adem\'{a}s de las $m$ condiciones de
gauge $\chi^a$ relacionadas con los grados de libertad espurios, se puede
tratar la reparametrizaci\'{o}n temporal, que est\'{a} asociada con la
inclusi\'{o}n del tiempo entre las variables din\'{a}micas. Esta es una tarea
f\'{a}cil, siempre y cuando se siga el proceso de parametrizaci\'{o}n expuesto
al principio de este cap\'{\i}tulo. Al nivel de la ecuaci\'{o}n (\ref{act}),
es evidente que se puede insertar una funci\'{o}n delta, $\delta(t-t_0)$
($\{t-t_0,{\cal H}_o\}=1)$, para regularizar el producto interno, que
significa tomar el producto interno a un tiempo dado $t_0$,
\begin{equation}
(\varphi_1,\varphi_2)_{t_0}=\int dt dq\ \left[\prod^m \delta(\chi)\right]\ J\
\delta(t-t_0)\ \varphi^*_1(t,q^{\gamma})\ \varphi_2(t,q^{\gamma}),
\end{equation}
donde $\gamma=1, ..., n+m$ y $J$ es el determinante de Faddeev-Popov asociado
con los v\'{\i}nculos lineales. Luego, a trav\'{e}s de una transformaci\'{o}n
can\'{o}nica, el tiempo es asociado con los momentos, Ec. (\ref{tc}); luego,
cambiando a esta representaci\'{o}n (transformando Fourier la funci\'{o}n de
onda) se obtiene el producto interno:
\begin{equation}\label{prodescreg}
(\varphi_1,\varphi_2)={1\over 2\pi}\int dq\ dq^0\ dq'^0\ \left[\prod^m
\delta(\chi)\right]\ J\ e^{-it_0(q^0-q'^0)}\varphi^*_1(q^0,q^{\gamma})\
\varphi_2(q'^0,q^{\gamma}).
\end{equation}

Cuando se realiza un cambio de escala del v\'{\i}nculo Hamiltoniano,
ecuaci\'{o}n (\ref{escc}), el producto interno f\'{\i}sico debe permanecer
invariante. De acuerdo al comportamiento de la funci\'{o}n de onda ante cambio
de escala, lo cual es evidente en la estructura de los operadores de
v\'{\i}nculo (ve\'{a}se ecuaci\'{o}n (\ref{wf}) m\'{a}s abajo), la norma
(originalmente unitaria) de $\vec\xi_0$ debe aparecer en el producto interno
f\'{\i}sico
\begin{equation} \label{pescrege}
(\varphi_1,\varphi_2)={1\over 2\pi}\int dq\ dq^0 \left[\prod^m
\delta(\chi)\right]\ J\ \varphi^*_1(q^0,q^{\gamma}) \ |\vec\xi_0|^{-1} \int
dq'^0\ e^{-it_0(q^0-q'^0)} |\vec\xi_0|^{-1}\varphi_2(q'^0,q^{\gamma}).
\end{equation}
Deber\'{\i}a notarse que la integraci\'{o}n en la coordenada $q'^0$ es
evaluada a lo largo de las l\'{\i}neas del campo vectorial $\vec\xi_0$.

El ordenamiento obtenido, ecuaciones (\ref{vcuadq'v5})-(\ref{ef}), satisface
como antes, las propiedades de invariancia impuestas a la teor\'{\i}a: (i)
cambios de coordenadas, (ii) combinaciones de v\'{\i}nculos supermomentos [Ec.
(\ref{combl})], y (iii) cambio de escala del super-Hamiltoniano [Ec.
(\ref{escc})]. El producto interno f\'{\i}sico invariante de gauge,
ecuaci\'{o}n (\ref{pescrege}), debe ser invariante ante cualquiera de estas
transformaciones. De acuerdo al cambio del determinante de Faddeev-Popov bajo
(ii) y (iii), el producto interno permanecer\'{a} invariante si la funci\'{o}n
de onda de Dirac cambia de acuerdo con
\begin{equation}\label{wf}
\varphi \rightarrow \varphi'=({\rm det} A)^{1\over 2} |\vec\xi_0|~ \varphi.
\end{equation}
Luego, los factores $f^{{\pm}{1\over 2}}$, $|\vec\xi_0|^{{\pm}1}$ en los
operadores de v\'{\i}nculo son justo los necesarios para que $\hat {\cal
H}_a\varphi,~\hat H\varphi,$ y ${\hat C}_{oa}^b\varphi$ transformen como
$\varphi$, preservando as\'{\i} el car\'{a}cter geom\'{e}trico de la
funci\'{o}n de onda de Dirac [\cite{prd99}]. N\'{o}tese que el rol jugado por
el potencial definido positivo en el caso con tiempo intr\'{\i}nseco, ahora es
tomado por el m\'{o}dulo del vector de Killing conforme temporal de la
teor\'{\i}a.

\bigskip

En lo concerniente a la extensi\'{o}n del tratamiento aqu\'{\i} expuesto al
caso de relatividad general, se ha mostrado en [\cite{koxf}] que en verdad
existe un vector de Killing conforme temporal en el superespacio del
formalismo ADM. Pero, la cuesti\'{o}n de si satisface la propiedad
(\ref{condgb'}) permanece abierta.

Es importante mencionar tambi\'{e}n que si bien la idea de un tiempo
extr\'{\i}nseco no es nueva en relatividad general [\cite{york}], su uso en el
problema de cuantificaci\'{o}n es muy poco frecuente [\cite{k71,kt2}]. Como se
ha mostrado, el modelo mec\'{a}nico presentado aqu\'{\i} puede conducir a una
mejor comprensi\'{o}n de c\'{o}mo lograr su implementaci\'{o}n.

\newpage


\chapter{Sistemas con dos super-Hamiltonianos}

\bigskip

{\it ~~~~ Los peculiares v\'{\i}nculos de las teor\'{\i}as con covariancia
general han sido estudiados, especialmente en lo concerniente a su
cuantificaci\'{o}n, utilizando ciertos modelos de dimensi\'{o}n finita
[\cite{hk90}; Barvinsky (1990,1996); Ferraro \& Sforza (1997,1999, 2000a)].
Sin embargo, en todos esos casos los modelos considerados contienen un solo
v\'{\i}nculo super-Hamlitoniano, a pesar que la relatividad general contiene
una infinidad de v\'{\i}nculos super-Hamiltonianos (uno por cada punto del
espacio) con un \'{a}lgebra no trivial entre ellos y los supermomentos:
\begin{eqnarray}\label{pb6}
\lefteqn{~~~~~\{{\mathcal H}(x),{\mathcal H}(x')\}={\mathcal
H}^a(x)\delta_{,a}(x,x')+ {\mathcal H}^a(x')\delta_{,a}(x,x'),} \\
&&\{{\mathcal H}_a(x),{\mathcal H}(x')\}={\mathcal
H}(x)\delta_{,a}(x,x'),\label{pb62}\\ &&\{{\mathcal H}_a(x),{\mathcal
H}_b(x')\}={\mathcal H}_b(x)\delta_{,a}(x,x')+ {\mathcal
H}_a(x')\delta_{,b}(x,x'),\label{pb63}
\end{eqnarray}
que aseguran que el sistema evolucione de modo consistente satisfaciendo
\begin{equation}
{\mathcal H}=0,\ \ \ \ \ {\mathcal H}_a=0.
\end{equation}
La cuantificaci\'{o}n de dicho sistema requiere la b\'{u}squeda de un
ordenamiento de factores de modo que los operadores de v\'{\i}nculo preserven
el \'{a}lgebra. Modelos con varios v\'{\i}nculos Hamiltonianos han sido
considerados, pero en todos los casos se trata de un conjunto de
part\'{\i}culas relativistas en un fondo plano y sin v\'{\i}nculos lineales
[\cite{lusanna,lusanna2}]. Recientemente, \cite{mrt99} propusieron un modelo
con tres v\'{\i}nculos, $H_1$, $H_2$ y $G$ en un fondo plano, con un
\'{a}lgebra cerrada que tiene la estructura
\begin{equation}\label{algmon}
\{H_1, H_2\} \sim G,~~~~~ \{H_i, G\} \sim H_i,
\end{equation}
que imita parcialmente a (\ref{pb6})-(\ref{pb63}) (al contener un s\'{o}lo
v\'{\i}nculo lineal, no est\'{a} presente la \'{u}ltima relaci\'{o}n). Este ha
sido un gran paso, aunque lamentablemente, los v\'{\i}nculos propuestos son
tan simples que no presentan problemas de ordenamiento, por lo cual no se
realiza aporte alguno en la comprensi\'{o}n de uno de los problemas
fundamentales de la cuantificaci\'{o}n de este tipo de sistemas.

En este cap\'{\i}tulo, desarrollaremos y cuantificaremos un modelo de
dimensi\'{o}n finita en un fondo curvo que imita completamente el \'{a}lgebra
(\ref{pb6})-(\ref{pb63}) (pero en forma cerrada) y que presenta problemas de
ordenamiento no triviales [\cite{prd00}]. }

\section{Definici\'{o}n del modelo}

~~~~ Consideremos un sistema descripto por $4n$ coordenadas can\'{o}nicas
$(q^i,p_i)$ con $i=(i_1,i_2)$ donde $i_1=1,...,n$ e $i_2=n+1,...,2n$, sometido
a dos v\'{\i}nculos super-Hamiltonianos

\begin{equation}\label{h1}
{\cal H}_1={1\over 2}g^{i_1j_1}(q^{k_1})p_{i_1}p_{j_1}+{\upsilon}_1(q^{k_2}),
\end{equation}
\begin{equation}\label{h2}
{\cal H}_2={1\over 2}g^{i_2j_2}(q^{k_2})p_{i_2}p_{j_2}+{\upsilon}_2(q^{k_1}).
\end{equation}

\medskip

La m\'{e}trica del super-Hamiltoniano ${\cal H}_1$ depende solamente de las
coordenadas $q^{i_1}$ y la m\'{e}trica del super-Hamiltoniano  ${\cal H}_2$
depende solamente de las coordenadas $q^{i_2}$, ambas son indefinidas y no
degeneradas. N\'{o}tese que los potenciales, en cambio tienen una dependencia
funcional tal que ${\upsilon}_1={\upsilon}_1(q^{i_2})$, es s\'{o}lo
funci\'{o}n de las coordenadas $q^{i_2}$ y
${\upsilon}_2={\upsilon}_2(q^{i_1})$ lo es s\'{o}lo de $q^{i_1}$. El sistema
est\'{a} restringido adem\'{a}s por $m$ v\'{\i}nculos supermomentos
linealmente independientes
\begin{equation}\label{g}
{\cal H}_a=\xi_a^ip_i,  ~~~~~~a=3,...,m+2; ~~i=(i_1,i_2)
\end{equation}
donde
\begin{equation}\label{proplin}
\xi_a^{i_1}=\xi_a^{i_1}(q^{k_1}),~~ \xi_a^{i_2}=\xi_a^{i_2}(q^{k_2}).
\end{equation}

La forma en que las m\'{e}tricas, los potenciales y los vectores $\vec \xi_a$
dependen de las coordenadas $q^{i_1}$ y $q^{i_2}$ tiene por objetivo la
obtenci\'{o}n de un \'{a}lgebra de v\'{\i}nculos que imite al \'{a}lgebra
(\ref{pb6})-(\ref{pb63}) de la relatividad general.

Es importante destacar que este sistema representa a dos part\'{\i}culas con
interacci\'{o}n, y no es separable en el sentido que no es posible describir
el sistema como si fuesen dos part\'{\i}culas libres (separando la
descripci\'{o}n de ellas en cada subespacio $(q^{i_1}, p_{i_1})$ y $(q^{i_2},
p_{i_2})$). Esto lo hace muy interesante ya que hasta la fecha y hasta donde
sabemos, no exist\'{\i}an modelos en espacio-tiempo curvo con interacci\'{o}n
(toda vez que se introduc\'{\i}a m\'{a}s de un super-Hamiltoniano, finalmente
se terminaba describiendo un sistema de part\'{\i}culas libres [\cite{k86}]).

El sistema de v\'{\i}nculos (\ref{h1}), (\ref{h2}) y (\ref{g}) ser\'{a} de
primera clase siempre que satisfaga las relaciones
\begin{equation}\label{chh}
\{{\cal H}_1 ,{\cal H}_2 \}=c_{12}^a{\cal H}_a
\end{equation}
\begin{equation}\label{ch1g}
\{{\cal H}_1,{\cal H}_a\}=c_{1a}^1 {\cal H}_1
\end{equation}
\begin{equation}\label{ch2g}
\{{\cal H}_2,{\cal H}_a\}=c_{2a}^2 {\cal H}_2
\end{equation}
\begin{equation}\label{cgg}
\{ {\cal H}_a,{\cal H}_b \}=C_{ab}^c {\cal H}_c
\end{equation}

Para comenzar, nos restringiremos a \'{a}lgebras cerradas (las funciones de
estructura son constantes). M\'{a}s adelante, a partir de cambios de escala en
los super-Hamiltonianos podremos generalizar los resultados al caso de un
\'{a}lgebra abierta (aunque la estructura del \'{a}lgebra
(\ref{chh})-(\ref{cgg}) resultar\'{a} modificada).

Las combinaciones de v\'{\i}nculos que se obtienen como resultado del corchete
son tales que respetan rigurosamente el \'{a}lgebra que encontramos en \S 2.1
para la relatividad general. Veamos qu\'{e} condiciones imponen estas
ecuaciones. Si reemplazamos expl\'{\i}citamente los v\'{\i}nculos en
(\ref{chh}) obtenemos
\begin{equation}\label{chhcond}
 g^{i_2k_2} {\upsilon}_{1,k_2}p_{i_2} -g^{i_1k_1}
{\upsilon}_{2,k_1}p_{i_1}=c_{12}^a( \xi_a^{i_1}p_{i_1}+\xi_a^{i_2}p_{i_2}),
\end{equation}
es decir\footnote{N\'{o}tese que las ecuaciones (6.15) y (6.16) no obligan a
$c_{12}^a$  a ser constantes; bastar\'{\i}a que se cumpla:
$c_{12}^a=~^{(1)}c_{12}^a(q^{i_1})-~^{(2)}c_{12}^a(q^{i_2})$ con
$~^{(1)}c_{12}^a\xi_a^{i_1}=0$ y  $~^{(2)}c_{12}^a\xi_a^{i_2}=0$.}
\begin{equation}\label{vcg61}
c_{12}^a\xi_a^{i_2}=g^{i_2k_2} {\upsilon}_{1,k_2}
\end{equation}
y
\begin{equation}\label{vcg62}
c_{12}^a\xi_a^{i_1}=-g^{i_1k_1} {\upsilon}_{2,k_1}
\end{equation}

Reemplazando los v\'{\i}nculos en (\ref{ch1g}), obtenemos
\begin{equation}\label{ch1gcond}
{1\over 2} (g^{i_1j_1}_{,k_1}\xi_a^{k_1}
-2g^{i_1k_1}\xi_{a,k_1}^{j_1})p_{i_1}p_{j_1}+
{\upsilon}_{1,k_2}\xi_a^{k_2}=c_{1a}^1(g^{i_1j_1}p_{i_1}p_{j_1}+ {\upsilon}_1)
\end{equation}
es decir,
\begin{equation}\label{cond611}
g^{i_1j_1}_{,k_1}\xi_a^{k_1}-2g^{i_1k_1}\xi_{a,k_1}^{j_1}=c_{1a}^1g^{i_1j_1}
\end{equation}
y
\begin{equation}\label{cond612}
\xi_a^{k_2}{\upsilon}_{1,k_2}=c_{1a}^1 {\upsilon}_1
\end{equation}
Expresando estas condiciones en forma geom\'{e}trica,
\begin{eqnarray}\label{lie1}
&&{\cal L}_{\vec\xi_a}\bar{\bar{g_1}}=c_{1a}^1\bar{\bar{g_1}} \\ &&{\cal
L}_{\vec\xi_a} {\upsilon}_1=c_{1a}^1 {\upsilon}_1
\end{eqnarray}

En forma totalmente an\'{a}loga, al reemplazar los v\'{\i}nculos en
(\ref{ch2g}), obtenemos
\begin{equation}\label{ch2gcond}
{1\over 2}(g^{i_2j_2}_{,k_2}\xi_a^{k_2}
-2g^{i_2k_2}\xi_{a,k_2}^{j_2})p_{i_2}p_{j_2}+
{\upsilon}_{2,k_1}\xi_a^{k_1}=c_{2a}^2(g^{i_2j_2}p_{i_2}p_{j_2}+ {\upsilon}_2)
\end{equation}
de donde
\begin{equation}\label{cond621}
g^{i_2j_2}_{,k_2}\xi_a^{k_2}-2g^{i_2k_2}\xi_{a,k_2}^{j_2}=c_{2a}^2g^{i_2j_2}
\end{equation}
y
\begin{equation}\label{cond622}
\xi_a^{k_1}{\upsilon}_{2,k_1}=c_{2a}^2 {\upsilon}_2
\end{equation}
O en forma geom\'{e}trica,
\begin{eqnarray}\label{lie2}
&&{\cal L}_{\vec\xi_a}\bar{\bar{g_2}}=c_{2a}^2\bar{\bar{g_2}} \\ &&{\cal
L}_{\vec\xi_a} {\upsilon}_2=c_{2a}^2 {\upsilon}_2
\end{eqnarray}

Vemos entonces que la condici\'{o}n de primera clase entre los
super-Hamiltonianos y los supermomentos corresponde a que estos \'{u}ltimos
sean vectores de Killing conformes de las m\'{e}tricas y los potenciales (lo
cual es natural, pues la definici\'{o}n de la derivada de Lie est\'{a}
asociada al corchete).

Finalmente si reemplazamos los supermomentos en (\ref{cgg}), vemos
que\footnote{La Ec. (6.27) no obliga a que $C_{ab}^c$ sean constantes, sino a
que se descompongan seg\'{u}n:
$C_{ab}^c=~^{(1)}C_{ab}^c(q^{i_1})-~^{(2)}C_{ab}^c(q^{i_2})$ con
$~^{(1)}C_{ab}^c\xi_c^{i_1}=0$ y  $~^{(2)}C_{ab}^c\xi_c^{i_2}=0$.}
\begin{equation}\label{cggcond}
(\xi_a^{i_1}\xi_{b,i_1}^{j_1}-\xi_b^{i_1}\xi_{a,i_1}^{j_1})p_{j_1}+
(\xi_a^{i_2}\xi_{b,i_2}^{j_2}-\xi_b^{i_2}\xi_{a,i_2}^{j_2})p_{j_2}= C_{ab}^c (
\xi_c^{i_1}p_{i_1}+\xi_c^{i_2}p_{i_2}).
\end{equation}

\section{El generador BRST cl\'{a}sico y cu\'{a}ntico}

~~~~ Debido a que el \'{a}lgebra de v\'{\i}nculos cierra de acuerdo a un grupo
(funciones de estructura constantes) el sistema es de rango 1 (como vimos en
\S 3.1.1) y el generador BRST en el espacio de fases debidamente extendido es
entonces
\begin{equation}\label{omegaclm}
\Omega=\eta^1 {\cal H}_1 +\eta^2 {\cal H}_2 + \eta^a  {\cal H}_a + \eta^1
\eta^2 c_{12}^a {\cal P}_a + \eta^1 \eta^a c_{1a}^1 {\cal P}_1 +  \eta^2
\eta^a c_{2a}^2 {\cal P}_2 + {1\over  2} \eta^a  \eta^b C_{ab}^c {\cal P}_c.
\end{equation}

Para hallar el operador $\hat\Omega$ herm\'{\i}tico y nilpotente es \'{u}til
escribir, como hicimos previamente, a $\Omega$ con los momentos originales y
los fantasmas en pie de igualdad. $\Omega$ es adem\'{a}s la suma de un
t\'{e}rmino cuadr\'{a}tico en los momentos
\begin{equation}
{\Omega}^{cuad}=\eta^1 {\cal H}_1 +\eta^2 {\cal H}_2\equiv{1\over 2}
\sum_{r,s=-1}^0 \Omega^{a_rb_s}{\cal P}_{a_r} {\cal P}_{b_s}
+\eta^1\upsilon_1+\eta^2\upsilon_2
\end{equation}
m\'{a}s otro lineal en los momentos
\begin{equation}\label{omlin6}
\Omega^{lineal}= \eta^a  {\cal H}_a + \eta^1 \eta^2 c_{12}^a {\cal P}_a +
\eta^1 \eta^a c_{1a}^1 {\cal P}_1 +  \eta^2 \eta^a c_{2a}^2 {\cal P}_2 +
{1\over  2} \eta^a  \eta^b C_{ab}^c {\cal P}_c\equiv\sum_{s=-1}^0
\Omega^{c_s}{\cal P}_{c_s}
\end{equation}

En general, obtener el operador BRST herm\'{\i}tico nilpotente a partir de su
contraparte cl\'{a}sica es un paso muy dif\'{\i}cil (e inclusive puede no ser
realizable pues a nivel cu\'{a}ntico su mera existencia no est\'{a}
garantizada). Sin embargo, en este caso particular, a pesar que el sistema
presenta dos v\'{\i}nculos super-Hamiltonianos, el \'{a}lgebra es
extremadamente simple al ser cerrada. Notemos adem\'{a}s que, como antes,
podemos comenzar por ordenar la parte lineal de manera que
\begin{equation}\label{ordlin6}
\hat{\Omega}^{lineal}=\sum_{s=-1}^0f^{1\over 2}\hat\Omega^{c_s} \hat{\cal
P}_{c_s}f^{-{1\over 2}}
\end{equation}
sea herm\'{\i}tico (pero en este caso no es nilpotente por s\'{\i} solo) y
donde $f=f(q^i)$ (depende de todas las coordenadas).

La condici\'{o}n de hermiticidad exige que $f$ satisfaga (recordar la
Ec.(\ref{div}))
\begin{equation}\label{div6}
C^{\beta}_{a\beta}=f^{-1}(f\xi^i_a)_{,i},
\end{equation}
con $\beta=(A,a)=1,2,3,...,m+2$. Es decir que ahora
$C^{\beta}_{a\beta}=C^b_{ab}+c^1_{a1}+c^2_{a2}$.

Es f\'{a}cil ver adem\'{a}s, que de este sector lineal podemos identificar al
operador de v\'{\i}nculo de los supermomentos como
\begin{equation}\label{gc}
\hat {\cal H}_a=f^{1\over 2} \xi^i_a \hat p_i f^{-{1\over 2}}.
\end{equation}

La parte cuadr\'{a}tica restante s\'{o}lo presenta momentos can\'{o}nicos (a
diferencia de los casos anteriores en los que exist\'{\i}a un t\'{e}rmino con
una funci\'{o}n de estructura lineal multiplicada por un momento fantasma).
Considerando los resultados previos (cap\'{\i}tulos 4 y 5) proponemos el
ordenamiento herm\'{\i}tico:
\begin{equation}\label{ordcuad6}
\hat{\Omega}^{cuad}={1\over 2} \sum_{r,s=-1}^0 f^{-{1\over  2}}\hat{\cal
P}_{a_r} f \hat\Omega^{a_rb_s} \hat{\cal P}_{b_s}f^{-{1\over 2}}+\hat
\eta^1\upsilon_1+\hat \eta^2\upsilon_2.
\end{equation}
Es decir que los operadores de v\'{\i}nculo de los super-Hamiltonianos
ser\'{a}n
\begin{equation}\label{h1c}
\hat{\cal H}_1=\frac{1}{2}  f^{-{1\over 2}} \hat p_{i_1} g^{i_1j_1} f \hat
p_{j_1} f^{-{1\over 2}}+ {\upsilon}_1,
\end{equation}
\begin{equation}\label{h2c}
\hat{\cal H}_2=\frac{1}{2}  f^{-{1\over 2}} \hat p_{i_2} g^{i_2j_2} f \hat
p_{j_2} f^{-{1\over 2}}+ {\upsilon}_2.
\end{equation}

Finalmente, el operador BRST herm\'{\i}tico propuesto, en el ordenamiento
$\hat \eta-\hat {\cal P}$, es entonces:
\begin{equation}\label{omeord'm}
\hat\Omega={\hat\eta}^1 \hat{\cal H}_1+ {\hat\eta}^2 \hat{\cal H}_2+ \hat
\eta^a \hat {\cal H}_a+\hat \eta^1 \hat\eta^2 c_{12}^a \hat {\cal P}_a
+\hat\eta^1 \hat\eta^a c_{1a}^1 \hat {\cal P}_1+ \hat\eta^2  \hat\eta^a
c_{2a}^2 \hat{\cal P}_2 + {1\over 2} \hat \eta^a \hat \eta^b C_{ab}^c \hat
{\cal P}_c.
\end{equation}

Debemos demostrar ahora que este ordenamiento de $\hat \Omega$ es nilpotente.
Es razonable esperar que surja alguna condici\'{o}n adicional sobre $f(q^i)$,
que manifieste la presencia de dos v\'{\i}nculos super-Hamiltonianos.
Estudiaremos esto en la pr\'{o}xima secci\'{o}n.

\subsection{Demostraci\'{o}n de $\hat\Omega^2=0$}

~~~~ La demostraci\'{o}n de la nilpotencia de $\hat\Omega$ la realizaremos por
c\'{a}lculo expl\'{\i}cito. Varios t\'{e}rminos en $[\hat\Omega,\hat\Omega]$
son nulos trivialmente debido a que las funciones de estructura son constantes
o a que aparecen fantasmas al cuadrado:
\begin{equation}\label{n1}
[{\hat\eta}^1 \hat{\cal H}_1+{\hat\eta}^2 \hat{\cal H}_2, \hat \eta^1
\hat\eta^2 c_{12}^b \hat {\cal P}_b+{1\over 2} \hat \eta^a \hat \eta^b
C_{ab}^c \hat {\cal P}_c]=0,
\end{equation}
\begin{equation}\label{n2}
[{\hat\eta}^1 \hat{\cal H}_1, \hat\eta^2 \hat\eta^a c_{2a}^2 \hat {\cal
P}_2]=0,
\end{equation}
\begin{equation}\label{n3}
[{\hat\eta}^2 \hat{\cal H}_2, \hat\eta^1 \hat\eta^a c_{1a}^1 \hat {\cal
P}_1]=0,
\end{equation}
\begin{equation}\label{n4}
[{\hat\eta}^a\hat{\cal H}_a,\hat\eta^1 \hat\eta^b c_{1b}^1 \hat {\cal
P}_1+\hat\eta^2 \hat\eta^a c_{2a}^2 \hat {\cal P}_2]=0.
\end{equation}
Tambi\'{e}n es nulo por construcci\'{o}n
\begin{equation}\label{nl}
[{\hat\eta}^a\hat{\cal H}_a+{1\over 2} \hat \eta^a \hat \eta^b C_{ab}^c \hat
{\cal P}_c, {\hat\eta}^d \hat{\cal H}_d+{1\over 2} \hat \eta^d \hat \eta^e
C_{de}^f \hat {\cal P}_f]=0.
\end{equation}

Por lo tanto, debemos calcular los restantes t\'{e}rminos no nulos en
\begin{eqnarray}\label{nilp}
\lefteqn{\!\!\!\!\!\!\!\![\hat \Omega,\hat \Omega]= 2[{\hat\eta}^1 \hat{\cal
H}_1, {\hat\eta}^2 \hat{\cal H}_2] +2[{\hat\eta}^1 \hat{\cal H}_1,
{\hat\eta}^a \hat{\cal H}_a]+ 2[{\hat\eta}^2 \hat{\cal H}_2, {\hat\eta}^a
\hat{\cal H}_a]} \nonumber \\ &&+2[{\hat\eta}^1\hat{\cal H}_1,\hat\eta^1
\hat\eta^a c_{1a}^1 \hat {\cal P}_1]+ 2[{\hat\eta}^2\hat{\cal H}_2,\hat\eta^2
\hat\eta^a c_{2a}^2 \hat {\cal P}_2] + 2[{\hat\eta}^a \hat{\cal H}_a,\hat
\eta^1 \hat\eta^2 c_{12}^b \hat {\cal P}_b]\nonumber \\ &&+2[\hat \eta^1
\hat\eta^2 c_{12}^a \hat {\cal P}_a, \hat\eta^1 \hat\eta^b c_{1b}^1 \hat {\cal
P}_1+\hat\eta^2 \hat\eta^b c_{2b}^2 \hat {\cal P}_2+{1\over 2} \hat \eta^b
\hat \eta^c C_{bc}^d \hat {\cal P}_d].
\end{eqnarray}

Cada uno de ellos contribuye con
\begin{equation}\label{c11}
[{\hat\eta}^1\hat{\cal H}_1,\hat\eta^1 \hat\eta^a c_{1a}^1 \hat {\cal P}_1]=
-i \hat\eta^1 \hat\eta^a c_{1a}^1\hat{\cal H}_1,
\end{equation}
\begin{equation}\label{c22}
[{\hat\eta}^2\hat{\cal H}_2,\hat\eta^2 \hat\eta^a c_{2a}^2 \hat {\cal P}_2]=
-i \hat\eta^2 \hat\eta^a c_{2a}^2\hat{\cal H}_2,
\end{equation}
\begin{equation}\label{c2}
[{\hat\eta}^a \hat{\cal H}_a,\hat \eta^1 \hat\eta^2 c_{12}^b \hat {\cal
P}_b]=-i\hat \eta^1 \hat\eta^2 c_{12}^b \hat {\cal H}_b,
\end{equation}
\begin{eqnarray}\label{c3}
\lefteqn{\!\!\!\!\!\!\!\!\!\!\!\!\!\!\!\!\!\![\hat \eta^1 \hat\eta^2 c_{12}^a
\hat {\cal P}_a, \hat\eta^1 \hat\eta^b c_{1b}^1 \hat {\cal P}_1+\hat\eta^2
\hat\eta^b c_{2b}^2 \hat {\cal P}_2+{1\over 2} \hat \eta^b \hat \eta^c
C_{bc}^d \hat {\cal P}_d]=\ \ \ \ \ \ \ \ }\nonumber \\ && \ \ \ \ \ \ \ \ \ \
\ \ \ \ \ \ \ \ \ \ \ \ \ \ \ \ \ \ \ =i\hat \eta^1 \hat\eta^2\hat \eta^a
(c_{12}^c (c_{1a}^1+c_{2a}^2)-c_{12}^b C_{ba}^c)\hat {\cal P}_c.
\end{eqnarray}

Este \'{u}ltimo t\'{e}rmino es efectivamente nulo ya que es la expresi\'{o}n
de la identidad de Jacobi:
\begin{eqnarray}\label{idj}
\lefteqn{\{\{{\cal H}_1,{\cal H}_2\},{\cal H}_a\}+\{\{{\cal H}_2,{\cal
H}_a\},{\cal H}_1\}+\{\{{\cal H}_a,{\cal H}_1\},{\cal H}_2\}=0}\nonumber \\
&&\Longrightarrow c_{12}^b C_{ba}^c- c_{12}^c (c_{1a}^1+c_{2a}^2)=0.
\end{eqnarray}

Finalmente, obtenemos que
\begin{eqnarray}\label{nilpfin}
\lefteqn{\!\!\!\!\!\!\!\![\hat \Omega,\hat \Omega]= 2 {\hat\eta}^1
{\hat\eta}^2([\hat{\cal H}_1, \hat{\cal H}_2]-i c_{12}^b \hat {\cal H}_b)}
\nonumber \\ && +2{\hat\eta}^1 {\hat\eta}^a([ \hat{\cal H}_1, \hat{\cal
H}_a]-i  c_{1a}^1\hat{\cal H}_1) \nonumber \\ && + 2{\hat\eta}^2
{\hat\eta}^a([ \hat{\cal H}_2, \hat{\cal H}_a] -i c_{2a}^2\hat{\cal H}_2)
\end{eqnarray}

Es decir que debemos mostrar expl\'{\i}citamente que el ordenamiento propuesto
para $\hat{\cal H}_1,~\hat{\cal H}_2$ y $\hat{\cal H}_A$ satisfacen las
condiciones de primera clase a nivel de conmutadores:
\begin{equation}\label{q12}
[\hat{\cal H}_1,\hat{\cal H}_2]=\!\!\!\!\!^{^{^{_?}}}~~ i c_{12}^a \hat {\cal
H}_a,
\end{equation}
\begin{equation}\label{q1a}
\ \ \ \ \ \ \ \ \ \ \ \ \ \ \ \ \ \ \ \ \ \ \ \ [\hat{\cal H}_A,\hat{\cal
H}_a]=\!\!\!\!\!^{^{^{_?}}}~~ i c_{Aa}^A \hat {\cal H}_A~~~~~~~~~~~~A=1,2.
\end{equation}

Calculemos entonces:
\begin{eqnarray}\label{ccord12}
\lefteqn{\!\!\!\!\!\!\!\!\!\!\!\!\! [\hat{\cal H}_1,\hat{\cal
H}_2]=[\frac{1}{2} f^{-{1\over 2}} \hat p_{i_1} g^{i_1j_1} f \hat p_{j_1}
f^{-{1\over 2}}+ {\upsilon}_1,\frac{1}{2} f^{-{1\over 2}} \hat p_{i_2}
g^{i_2j_2} f \hat p_{j_2} f^{-{1\over 2}}+ {\upsilon}_2]}\nonumber \\
&&=f^{1\over 2}  [\frac{1}{2} f^{-1} \hat p_{i_1} g^{i_1j_1} f \hat p_{j_1}+
{\upsilon}_1,\frac{1}{2} f^{-1} \hat p_{i_2} g^{i_2j_2} f \hat p_{j_2}+
{\upsilon}_2] f^{-{1\over 2}}\nonumber
\\ &&=f^{1\over 2}  [\frac{1}{2}g^{i_1j_1} \hat p_{i_1} \hat p_{j_1} -{i\over
2} f^{-1} (g^{i_1j_1} f)_{,i_1} \hat p_{j_1} + {\upsilon}_1,\nonumber \\ && \
\ \ \ \ \ \ \ \ \ \ \ \ \ \ \ \ \ \ \ \ \ \ \ \ \ \ \ \ \   ,\frac{1}{2}
g^{i_2j_2} \hat p_{i_2} \hat p_{j_2}-{i\over 2} f^{-1} (g^{i_2j_2} f)_{,i_2}
\hat p_{j_2}+ {\upsilon}_2] f^{-{1\over 2}}
\end{eqnarray}

Para satisfacer el \'{a}lgebra deben anularse necesariamente los t\'{e}rminos
cuadr\'{a}ticos en los momentos que emergen del c\'{a}lculo del corchete
\begin{equation}\label{h1h2c2}
[-{i\over 2} f^{-1} (g^{i_1j_1} f)_{,i_1} \hat p_{j_1},\frac{1}{2} g^{i_2j_2}
\hat p_{i_2} \hat p_{j_2}]=[-{i\over 2} f^{-1} (g^{i_2j_2} f)_{,i_2} \hat
p_{j_2},\frac{1}{2}g^{i_1j_1} \hat p_{i_1} \hat p_{j_1}]=0.
\end{equation}

Vemos entonces que si pedimos como condici\'{o}n adicional sobre $f(q^i)$ que
pueda ser factorizable seg\'{u}n
\begin{equation}\label{condfvol}
f(q^i)=f_1(q^{i_1})f_2(q^{i_2}),
\end{equation}
entonces efectivamente los conmutadores en (\ref{h1h2c2}) se anular\'{a}n.
N\'{o}tese que la condici\'{o}n (\ref{condfvol}) sobre $f$ es elegible en la
condici\'{o}n previa (\ref{div6}).

Como consecuencia de esta propiedad de $f$ tambi\'{e}n se anular\'{a}
\begin{equation}\label{h1h2c1}
[-{i\over 2} f^{-1} (g^{i_1j_1} f)_{,i_1} \hat p_{j_1},-{i\over 2} f^{-1}
(g^{i_2j_2} f)_{,i_2} \hat p_{j_2}]=0.
\end{equation}

Quedan entonces los t\'{e}rminos:
\begin{equation}\label{fvc1}
[-{i\over 2} f_1^{-1} (g^{i_1j_1} f_1)_{,i_1} \hat
p_{j_1},\upsilon_2]=-{1\over 2} f_1^{-1} (g^{i_1j_1} f_1)_{,i_1}
\upsilon_{2,j_1},
\end{equation}
\begin{equation}\label{fvc2}
[{i\over 2} f_2^{-1} (g^{i_2j_2} f_2)_{,i_2} \hat p_{j_2},\upsilon_1]={1\over
2} f_2^{-1} (g^{i_2j_2} f_2)_{,i_2} \upsilon_{1,j_2},
\end{equation}
y
\begin{equation}\label{g1g2c}
f^{1\over 2}[\frac{1}{2}g^{i_1j_1} \hat p_{i_1} \hat p_{j_1}+
{\upsilon}_1,\frac{1}{2}g^{i_2j_2} \hat p_{i_2} \hat p_{j_2}+
{\upsilon}_2]f^{-{1\over 2}}=i c_{12}^af^{1\over 2} \hat{\cal H}_a f^{-{1\over
2}}-g^{i_1j_1} \upsilon_{2,i_1j_1}+g^{i_2j_2} \upsilon_{1,i_2j_2}.
\end{equation}
Donde, en el miembro derecho se utiliz\'{o} la relaci\'{o}n (\ref{chhcond})
del \'{a}lgebra cl\'{a}sica para poder reconstruir $\hat {\cal H}_a$.

Finalmente deber\'{\i}amos mostrar que es posible anular la suma de
t\'{e}rminos ``sobrantes'' en las Ecs. (\ref{fvc1}), (\ref{fvc2}) y
(\ref{g1g2c}):
\begin{equation}\label{ecfinal}
-{1\over 2} f_1^{-1} (g^{i_1j_1} f_1)_{,i_1} \upsilon_{2,j_1}-g^{i_1j_1}
\upsilon_{2,i_1j_1}+{1\over 2} f_2^{-1} (g^{i_2j_2} f_2)_{,i_2}
\upsilon_{1,j_2}+g^{i_2j_2} \upsilon_{1,i_2j_2}=0.
\end{equation}

Es decir que deber\'{a} cumplirse:
\begin{equation}\label{cond6}
-f_1^{-1} (g^{i_1j_1} \upsilon_{2,j_1}f_1)_{,i_1}+ f_2^{-1} (g^{i_2j_2}
\upsilon_{1,j_2} f_2)_{,i_2}=0.
\end{equation}
Como consecuencia de reemplazar las identidades (\ref{vcg61}) y (\ref{vcg62})
en (\ref{cond6}), se obtiene
\begin{equation}\label{cond6'}
f_1^{-1} (c_{12}^a\xi_a^{i_1}f_1)_{,i_1}+ f_2^{-1} (
c_{12}^a\xi_a^{i_2}f_2)_{,i_2}=0.
\end{equation}
Luego, deber\'{\i}a satisfacerse que
\begin{equation}\label{condvol}
f_1^{-1} (c_{12}^a\xi_a^{i_1}f_1)_{,i_1}+f_2^{-1} (
c_{12}^a\xi_a^{i_2}f_2)_{,i_2}=c_{12}^a  f^{-1} (\xi_a^{i}f)_{,i}=c_{12}^a(
C_{ab}^b+c^1_{a1}+c^2_{a2})=0.
\end{equation}

Lo que efectivamente se cumple como consecuencia de la identidad de Jacobi:
para hacerlo expl\'{\i}cito solo debe tomarse la traza en (\ref{idj}).

Nos resta a\'{u}n evaluar (recordar que $A=1,2$)
\begin{eqnarray}\label{ccqlin1}
\lefteqn{\!\!\!\! [\hat{\cal H}_A,\hat{\cal H}_a]=[\frac{1}{2} f^{-{1\over 2}}
\hat p_{i_A} g^{i_Aj_A} f \hat p_{j_A} f^{-{1\over 2}}+ {\upsilon}_A,f^{1\over
2} \xi^k_a \hat p_k f^{-{1\over 2}}]=} \nonumber \\ && \ \ \ \ \
=\frac{1}{2}f^{-{1\over 2}}( \hat p_{i_A} g^{i_Aj_A} f \hat p_{j_A}\xi^k_a
\hat p_k - \xi^k_a \hat p_k \hat p_{i_A} g^{i_Aj_A} f \hat p_{j_A}
)f^{-{1\over 2}}+i \xi^k_a \upsilon_{A,k}\nonumber \\ && \ \ \ \ \
=\frac{1}{2}f^{-{1\over 2}}\left(i \hat p_{i_A}
(g^{i_Aj_A}_{,k_A}\xi_a^{k_A}-2g^{i_Ak_A}\xi_{a,k_A}^{j_A })\hat p_{j_A}+ f
g^{i_Aj_A} (( \xi_a^k (ln f)_{,k})_{,i_A}+ \xi_{a,i_Ak}^k )\hat p_{j_A}
\right)f^{-{1\over 2}}\nonumber \\ && \ \ \ \ \ \ \ +i \xi^k_a \upsilon_{A,k}
\end{eqnarray}
donde, usando las relaciones (\ref{cond611})-(\ref{cond612}) o
(\ref{cond621})-(\ref{cond622}), obtenemos
\begin{eqnarray}\label{ccqlin2}
\lefteqn{\!\!\!\!\!\!\!\!\!\!\!\! [\hat{\cal H}_A,\hat{\cal H}_a]=i c_{Aa}^A
\frac{1}{2}f^{-{1\over 2}}\hat p_{i_A}g^{i_Aj_A}f\hat p_{j_A}f^{-{1\over 2}}
+i \xi^k_a \upsilon_{A,k}} \nonumber \\ && \ \ +\frac{1}{2} f^{-{1\over
2}}\left( f g^{i_Aj_A} (( \xi_a^k (ln f)_{,k})_{,i_A}+ \xi_{a,i_Ak}^k )\hat
p_{j_A} \right)f^{-{1\over 2}}
\end{eqnarray}
Los dos primeros t\'{e}rminos en (\ref{ccqlin2}) son justamente $i
c_{Aa}^{(A)} \hat{\cal H}_{(A)}$ (no existe suma sobre el \'{\i}ndice $A$).
Nos resta ver si el \'{u}ltimo t\'{e}rmino se anula.

En efecto,
\begin{equation}\label{anula}
( \xi_a^k (ln f)_{,k})_{,i_A}+ \xi_{a,i_Ak}^k= ( \xi_a^k (ln f)_{,k}+
\xi_{a,k}^k)_{,i_A}= (f^{-1}(\xi_a^k f)_{,k})_{,i_A}=C_{a\beta,i_A}^{\beta}=0.
\end{equation}

Lo cual completa la demostraci\'{o}n de la nilpotencia de $\hat\Omega$.
N\'{o}tese que la \'{u}nica condici\'{o}n adicional sobre $f$ para la
nilpotencia es que $f(q^i)=f_1(q^{i_1})f_2(q^{i_2})$. \hfill $\Box$

\subsection{Transformaci\'{o}n unitaria y operadores de v\'{\i}nculo}

~~~~ En los sistemas estudiados previamente (cap\'{\i}tulos 4 y 5), vimos que
era posible generalizar la forma de los operadores de v\'{\i}nculo y los
operadores de funciones de estructura para el caso en que se realizara un
cambio de escala del \'{u}nico v\'{\i}nculo super-Hamiltoniano, de ma\-nera
que el sistema respetara dicha transformaci\'{o}n de invariancia. Dichos
cambios de escala estaban relacionados o bien con un potencial definido
positivo (casos con tiempo intr\'{\i}nseco) o con el m\'{o}dulo de un vector
de Killing conforme (casos con tiempo extr\'{\i}nseco). El  sistema
presentemente bajo estudio, posee dos v\'{\i}nculos super-Hamiltonianos, por
lo tanto los operadores de v\'{\i}nculo y de funciones de estructura
deber\'{a}n respetar los dos cambios de escala posibles. Otro punto a tener en
cuenta es que los potenciales de los super-Hamiltonianos, hasta ahora no
fueron obligados a ser definidos positivos ni a poseer un vector de Killing
conforme como suced\'{\i}a previamente, sino que s\'{o}lo fueron condicionados
a depender de ciertos sectores (excluyentes entre s\'{\i}) de las coordenadas.
Sin embargo, podemos relajar esta condici\'{o}n mediante cambios de escala de
los super-Hamiltonianos (si bien al definir el producto interno f\'{\i}sico
ser\'{a} necesario alguna propiedad adicional sobre ellos, secci\'{o}n \S
6.2.3).

La manera de realizar a nivel cu\'{a}ntico el cambio de escala de los
super-Hamiltonianos ${\cal H}_1 \rightarrow e^{F_1(q^i)}{\cal H}_1, ~~ {\cal
H}_2 \rightarrow e^{F_2(q^i)}{\cal H}_2$ es mediante la transformaci\'{o}n
unitaria

\begin{equation}
\hat \Omega  \rightarrow  e^{i\hat  C}~\hat\Omega~ e^{-i\hat C},\label{tu6}
\end{equation}
donde
\begin{eqnarray}\label{de6}
\lefteqn{\hat C={1\over 2}[\hat\eta^1 ~ F_1(q^i) ~ \hat{\cal P}_1-\hat{\cal
P}_1 ~ F_1(q^i) ~ \hat\eta^1 + \hat\eta^2 ~ F_2(q^j) ~ \hat{\cal
P}_2-\hat{\cal P}_2 ~ F_2(q^j) ~ \hat\eta^2] }\nonumber \\ &&=\hat\eta^1 ~
F_1(q^i) ~ \hat{\cal P}_1+  \hat\eta^2 ~ F_2(q^j) ~ \hat{\cal P}_2 +
\frac{i}{2}F_1(q^i)+\frac{i}{2} F_2(q^i) \nonumber \\ &&=-\hat{\cal P}_1 ~
F_1(q^i) ~ \hat\eta^1 -\hat{\cal P}_2 ~ F_2(q^j) ~ \hat\eta^2 -
\frac{i}{2}F_1(q^i) - \frac{i}{2}F_2(q^i) ,~~~~~~F_{1,2}(q)>0.
\end{eqnarray}

que dar\'{a} un nuevo operador BRST herm\'{\i}tico y nilpotente.

Ser\'{a} \'{u}til usar las siguientes identidades ($A=1,2$)
\begin{equation}
e^{i\hat\eta^A ~ F_A(q^i) ~ \hat{\cal P}_A} = 1+ i \hat\eta^A ~(
e^{F_A(q^i)}-1) ~ \hat{\cal P}_A,
\end{equation}
\begin{equation}
e^{i \hat{\cal P}_A~ F_A(q^i) ~\hat\eta^A } = 1+ i  \hat{\cal P}_A ~(
e^{F_A(q^i)}-1) ~\hat\eta^A,
\end{equation}
para calcular expl\'{\i}citamente la Ec. (\ref{tu6}), de donde obtenemos
\begin{eqnarray}\label{omegaescf}
\lefteqn{\hat\Omega={\hat\eta}^1 e^{\frac{F_1-F_2}{2}}\hat {\cal H}_1
e^{\frac{F_1+F_2}{2}}+{\hat\eta}^1{\hat\eta}^2 ie^{\frac{F_1+F_2}{2}} [\hat
{\cal H}_1,e^{-F_2}] e^{\frac{F_1+F_2}{2}}\hat{\cal P}_2}\nonumber\\ &&~~+
{\hat\eta}^2 e^{\frac{F_2-F_1}{2}}\hat {\cal H}_2 e^{\frac{F_1+F_2}{2}}
+{\hat\eta}^2{\hat\eta}^1 ie^{\frac{F_1+F_2}{2}}[\hat {\cal H}_2,e^{-F_1}]
e^{\frac{F_1+F_2}{2}}\hat{\cal P}_1 \nonumber\\
&&~~+\frac{1}{2}{\hat\eta}^1{\hat\eta}^2 e^{F_1+F_2}\hat c_{12}^a \hat{\cal
P}_a + {\hat\eta}^a e^{-\frac{F_1+F_2}{2}}\hat {\cal H}_a
e^{\frac{F_1+F_2}{2}} \nonumber \\&&~~+ {\hat\eta}^1{\hat\eta}^a \left(\hat
c_{1a}^1-ie^{\frac{F_1-F_2}{2}} [\hat {\cal H}_a,e^{-F_1}]
e^{\frac{F_1+F_2}{2}}\right)\hat{\cal P}_1 \nonumber\\&&~~
+{\hat\eta}^2{\hat\eta}^a \left(\hat c_{1a}^1-ie^{\frac{-F_1+F_2}{2}}[\hat
{\cal H}_a,e^{-F_2}] e^{\frac{F_1+F_2}{2}}\right)\hat{\cal P}_2+\frac{1}{2}
{\hat\eta}^a{\hat\eta}^b C_{ab}^c\hat{\cal P}_c
\end{eqnarray}


Podemos identificar los operadores de v\'{\i}nculo y las funciones de
estructura en la Ec. (\ref{omegaescf}) como
\begin{equation}\label{vcuadq'v1}
\hat H_1={1\over 2} e^{\frac{F_1-F_2}{2}}f_1^{-{1\over 2}}\hat p_{i_1}
g^{i_1j_1}f_1 \hat p_{j_1} f_1^{-{1\over
2}}e^{\frac{F_1+F_2}{2}}+e^{F_1}{\upsilon}_1,
\end{equation}

\begin{equation}\label{vcuadq'v2}
\hat H_1={1\over 2} e^{\frac{F_2-F_1}{2}}f_2^{-{1\over 2}}\hat p_{i_2}
g^{i_2j_2}f_2 \hat p_{j_2} f_2^{-{1\over
2}}e^{\frac{F_1+F_2}{2}}+e^{F_2}{\upsilon}_2,
\end{equation}

\begin{equation}\label{vlinq6'}
\hat {\cal H}_a= e^{-\frac{F_1+F_2}{2}}f^{1\over 2} \xi^i_a\hat p_i
f^{-{1\over 2}}e^{\frac{F_1+F_2}{2}}.
\end{equation}

Deber\'{\i}a notarse que los operadores de v\'{\i}nculo
(\ref{vcuadq'v1})-(\ref{vcuadq'v2}) corresponden a los v\'{\i}nculos
super-Hamiltonianos con cambio de escala $H_A=e^{F_A} {\cal H}_A,~(
G^{i_Aj_A}= e^{F_A} g^{i_Aj_A},~ V_A= e^{F_A}{\upsilon}_A$), con $(A,B=1,2)$:

\begin{equation}\label{vcuadq'v61}
\hat H_1={1\over 2} e^{\frac{F_1-F_2}{2}}f_1^{-{1\over 2}}\hat p_{i_1}
e^{-F_1} G^{i_1j_1}f_1 \hat p_{j_1} f_1^{-{1\over
2}}e^{\frac{F_1+F_2}{2}}+V_1,
\end{equation}

\begin{equation}\label{vcuadq'v62}
\hat H_2={1\over 2} e^{\frac{F_2-F_1}{2}}f_2^{-{1\over 2}}\hat p_{i_2}
e^{-F_2} G^{i_2j_2}f_2 \hat p_{j_2} f_2^{-{1\over
2}}e^{\frac{F_1+F_2}{2}}+V_2,
\end{equation}

con las correspondientes funciones de estructura,
\begin{eqnarray}\label{newssf}
\lefteqn{\hat  C_{AB}^B= i e^{\frac{F_1+F_2}{2}}[\hat{\cal H}_A,e^{-F_B}]
e^{\frac{F_1+F_2}{2}}} \nonumber\\ &&~~=\frac{1}{2}e^{\frac{F_A}{2}}
f^{-{1\over2}}[\hat p_{i_A} f g^{i_Aj_A} F_{B,j_A} +  F_{B,j_A} f
g^{i_Aj_A}\hat p_{j_A}] f^{-{1\over 2}} e^{\frac{F_A}{2}},
\end{eqnarray}
\begin{equation}
\hat C_{12}^a= e^{\frac{F_1+F_2}{2}}\hat c_{12}^a,
\end{equation}
\begin{equation}
\hat C_{Aa}^A=\hat c_{Aa}^A+\xi^i_a (F_A)_{,i},
\end{equation}
\begin{equation}
\hat C_{ab}^c=\hat C_{ab}^c,
\end{equation}
todos los operadores y las funciones de estructura est\'{a}n ordenados de modo
que satisfacen
\begin{equation}\label{ccqf}
[\hat  H_1,\hat  H_2]=\hat C_{12}^1(q,p) \hat H_1  +  \hat
C_{12}^2(q,p) \hat H_2+\hat C_{12}^a\hat{\cal H}_a,
\end{equation}
\begin{equation}\label{ccqfnw}
[\hat  H_A,\hat  {\cal H}_a]=\hat C_{Aa}^{(A)} \hat H_{(A)},
\end{equation}
\begin{equation}\label{clqf}
[\hat {\cal H}_a,\hat {\cal H}_b]=\hat C_{ab}^c \hat {\cal H}_c,
\end{equation}
(no existe suma sobre el \'{\i}ndice $A$). Es decir, el \'{a}lgebra est\'{a}
libre de anomal\'{\i}as a nivel cu\'{a}ntico. Debe notarse que como resultado
de los cambios de escala realizados en los super-Hamiltonianos, en el
resultado del primer conmutador aparecen ahora todos los operadores de
v\'{\i}nculo, y las funciones de estructura ya no son m\'{a}s constantes (las
\'{u}nicas que permanecen constantes son las asociadas a los conmutadores
entre supermomentos).

\subsection{Producto interno f\'{\i}sico}

~~~~ Para completar el proceso de cuantificaci\'{o}n debemos definir un
producto interno f\'{\i}sico. Veamos una vez m\'{a}s el rol de las
transformaciones que deber\'{\i}an dejar invariante la teor\'{\i}a; estas
transformaciones son (i) cambios de coordenadas, (ii) combinaciones de
v\'{\i}nculos supermomentos, y (iii) cambios de escala de los v\'{\i}nculos
super-Hamiltonianos (${\cal H}_A \rightarrow e^{F_A}~{\cal H}_A$). El producto
interno f\'{\i}sico de las funciones onda de Dirac,
\begin{equation}\label{prodesc6}
(\varphi_1,\varphi_2)=\int dq\ \big[\prod^{m+2} \delta(\chi)\big]\ J\
\varphi^*_1(q)\ \varphi_2(q),
\end{equation}
(donde $J$ es el determinante de Faddeev-Popov y $\chi$ representa las $m+2$
condiciones de gauge) debe ser invariante bajo cualquiera de estas
transformaciones. De acuerdo al cambio del determinante de Faddeev-Popov bajo
(ii) y (iii), el producto interno permanecer\'{a} invariante si la funci\'{o}n
de onda de Dirac cambia de acuerdo a
\begin{equation}\label{unmedio6}
\varphi \rightarrow \varphi'=(det A)^{1\over 2} e^{-{F_1+F_2\over 2}} \varphi.
\end{equation}

De esta manera, los factores $f^{{\pm}{1\over 2}}$, $e^{{\pm}{F_A\over 2}}$ en
los operadores de v\'{\i}nculo son justamente lo que se necesita para que
$\hat {\cal H}_a\varphi,~\hat H_A\varphi,$ y ${\hat C}_{\alpha\beta}^{\gamma}
\varphi$ transformen como $\varphi$ (las letras griegas pueden tomar cualquier
valor entre $1$ y $m+2$), preservando de esta manera el car\'{a}cter
geom\'{e}trico de la funci\'{o}n de onda de Dirac.

En los casos estudiados previamente, los factores de escala estaban asociados
o con un potencial definido positivo (tiempo intr\'{\i}nseco) o con el
m\'{o}dulo de un vector de Killing conforme (tiempo extr\'{\i}nseco). Hasta
aqu\'{\i} no fue necesario hacer suposiciones adicionales sobre los
potenciales presentes en los v\'{\i}nculos super-Hamiltonianos. Sin embargo,
para poder definir adecuadamente el producto interno f\'{\i}sico, es necesario
fijar adecuadamente las condiciones de gauge para eliminar la integraci\'{o}n
sobre las variables espurias. Ante\-riormente, esto se hizo fijando las $m$
coordenadas a los v\'{\i}nculos supermomentos y teniendo en cuenta el tiempo
f\'{\i}sico que conten\'{\i}a la teor\'{\i}a, que se correspond\'{\i}a con
ciertas caracter\'{\i}sticas del v\'{\i}nculo super-Hamiltoniano. De este
modo, dependiendo de si el sistema ten\'{\i}a un tiempo intr\'{\i}nseco o
extr\'{\i}nseco, se proced\'{\i}a seg\'{u}n correspondiera a fijar una
condici\'{o}n de gauge adicional. Sin embargo, ahora tenemos, adem\'{a}s de
los $m$ v\'{\i}nculos supermomentos (sobre los que sabemos c\'{o}mo fijar la
libertad de gauge asociada con ellos), {\it dos} v\'{\i}nculos
super-Hamiltonianos, lo cual implica que debemos tener {\it dos} condiciones
de gauge adicionales. Podemos pensar que una condici\'{o}n de gauge se
determinar\'{a} con el tipo de tiempo que posea el sistema; si por ejemplo,
los v\'{\i}nculos super-Hamiltonianos poseen potenciales definidos positivos,
entonces el tiempo ser\'{a} intr\'{\i}nseco y los factores que cambian de
escala a los super-Hamiltonianos ser\'{a}n los potenciales: $F_A= ln V_A$ y el
producto interno f\'{\i}sico y la condici\'on de gauge an\'{a}logas al caso ya
estudiado. En cambio, si el tiempo es extr\'{\i}nseco, los factores de cambio
de escala estar\'{a}n asociados con el m\'{o}dulo de los vectores de Killing
conformes de cada super-Hamiltoniano: $F_A= ln |\vec\xi_A|^{-2}$ y el producto
interno estar\'{a} definido an\'{a}logamente al caso estudiado en esas
circunstancias.

Sin embargo, a\'{u}n nos falta una condici\'{o}n de gauge. Dicha condici\'{o}n
deber\'{\i}a de alguna forma forzar la equivalencia entre ambos v\'{\i}nculos
super-Hamiltonianos, es decir, cada variable espuria asociada a estos
v\'{\i}nculos puede ser relacionada con un tiempo, por lo tanto, la
condici\'{o}n que parece razonable es que esos tiempos sean el mismo. Es decir
que el tipo de condici\'{o}n que proponemos es:
\begin{equation}\label{condgaugtemp}
\chi=\delta(t_1-t_2).
\end{equation}
Esta elecci\'{o}n de gauge puede ser justificada m\'{a}s rigurosamente dentro
del marco del formalismo de tiempo m\'{u}ltiple [\cite{lusanna,lusanna2}], en
el cual es usual este tipo de condici\'{o}n.

De esta forma, en el caso presente podemos tratar tanto el caso de tiempo
intr\'{\i}nseco como el de extr\'{\i}nseco. Por supuesto, para ser
consistentes con la condici\'{o}n de gauge (\ref{condgaugtemp}), ambos
super-Hamiltonianos contendr\'{a}n el mismo tipo de tiempo. Finalmente, el
producto interno quedar\'{a} completamente regularizado mediante la
inclusi\'{o}n de (\ref{condgaugtemp}), junto con las ya conocidas $m+1$
condiciones de gauge restantes.


\newpage

\thispagestyle{empty}

 ~\newpage


\fancyhead{} \fancyhead[LE]{\bf \thepage} \fancyhead[RE]{\sl Conclusiones}
\fancyhead[RO]{\bf \thepage}\fancyhead[LO]{}

\addcontentsline{toc}{chapter}{Conclusiones}

\noindent \Huge{\bf Conclusiones}

\thispagestyle{empty}

\bigskip

\normalsize

En esta tesis se estudi\'{o} la cuantificaci\'{o}n de sistemas con covariancia
general mediante la aplicaci\'{o}n del formalismo can\'{o}nico y del
formalismo BRST a modelos de dimensi\'{o}n finita que emulan la estructura de
v\'{\i}nculos de la relatividad general. Uno de los aspectos de mayor
inter\'{e}s fue encontrar los ordenamientos adecuados de los operadores de
v\'{\i}nculo de manera que respeten el \'{a}lgebra a nivel cu\'{a}ntico (es
decir que no se generen nuevos v\'{\i}nculos). La aplicaci\'{o}n del
formalismo BRST en la obtenci\'{o}n de dichos ordenamientos consistentes para
los operadores de v\'{\i}nculo de Dirac, ha dado resultados sumamente
novedosos con respecto a los tratamientos usuales.

El punto de partida ha sido el estudio de sistemas que contienen varios
v\'{\i}nculos supermomentos y un \'{u}nico v\'{\i}nculo super-Hamiltoniano
cuyo potencial es invariante de gauge. La principal gu\'{\i}a en la
b\'{u}squeda del ordenamiento adecuado del operador BRST (aquel con el cual
resulta herm\'{\i}tico y nilpotente) fueron las transformaciones de
invariancia de la teor\'{\i}a: (i) transformaci\'{o}n general de coordenadas,
(ii) combinaci\'{o}n lineal de los supermomentos y (iii) cambios de escala en
el super-Hamiltoniano. Si bien este tipo de invariancias hab\'{\i}an sido
tenidas en cuenta previamente (tanto desde el punto de vista de la
cuantificaci\'{o}n can\'{o}nica de Dirac, como por ejemplo en \cite{k86}, o
adem\'{a}s mediante el formalismo BRST en \cite{fhp93,b96,mp89}) en este
trabajo se explot\'{o} estas propiedades en la b\'{u}squeda de un ordenamiento
adecuado para la parte cuadr\'{a}tica del generador BRST (asociada con el
v\'{\i}nculo super-Hamiltoniano). En especial, el tipo de invariancia (ii)
llev\'{o} a reconocer que la contribuci\'{o}n no herm\'{\i}tica de los
fantasmas en los operadores de v\'{\i}nculo supermomentos puede asociarse al
volumen natural inducido por estos v\'{\i}nculos en las \'{o}rbitas. Se
mostr\'{o} que este volumen juega un papel fundamental al ordenar el sector
cuadr\'{a}tico del operador BRST nilpotente. El tipo de invariancia (iii) --
cambio de escala del v\'{\i}nculo super-Hamiltoniano -- tambi\'{e}n tiene un
rol importante en la determinaci\'{o}n del ordenamiento, ya que dicha
invariancia debe ser respetada por la teor\'{\i}a a nivel cu\'{a}ntico (la
cual se expresa mediante una transformaci\'{o}n unitaria de la carga BRST). Si
bien la introduci\'{o}n de esta propiedad es independiente del factor de
escala en s\'{\i}, su interpretaci\'{o}n f\'{\i}sica surge a partir de la
definici\'{o}n del producto interno f\'{\i}sico, lo cual requiere la
identificaci\'{o}n previa del tiempo f\'{\i}sico en la teor\'{\i}a. En el caso
de tiempo intr\'{\i}nseco, el factor de escala es el potencial definido
positivo. Debido a esto el potencial modifica el t\'{e}rmino cin\'{e}tico del
super-Hamiltoniano, esta caracter\'{\i}stica que parece poco natural es
justificada elegantemente mediante la formulaci\'{o}n de Jacobi del principio
de m\'{\i}nima acci\'{o}n. Una consecuencia totalmente novedosa es que con
este tratamiento no se requiere la inclusi\'{o}n de un t\'{e}rmino de
curvatura para lograr la invariancia ante cambios de escala, lo que
constitu\'{\i}a un procedimiento habitual [ver por ej. \cite{k86}]. Una
dificultad seria que surge al intentar extrapolar estos resultados a la
relatividad general es que el potencial en el super-Hamiltoniano en ese caso
es la curvatura espacial, que evidentemente no est\'{a} restringida a ser
definida positiva.

Esto induce la b\'{u}squeda de un nuevo modelo mec\'{a}nico que represente de
una manera m\'{a}s acabada a la relatividad general. Esta motivaci\'{o}n
llev\'{o} al desarrollo de un modelo  con un tiempo extr\'{\i}nseco mediante
un proceso de parametrizaci\'{o}n. En este caso, los resultados a nivel de
ordenamiento de los operadores de v\'{\i}nculo resultan similares, salvo que
el rol jugado por el potencial definido positivo en el caso con tiempo
intr\'{\i}nseco, ahora es tomado por el m\'{o}dulo del vector de Killing
conforme de la teor\'{\i}a. Fue necesario adem\'{a}s una nueva definici\'{o}n
del producto interno f\'{\i}sico de acuerdo al nuevo tiempo que contiene la
teor\'{\i}a. Es decir, la inserci\'{o}n que regulariza el producto interno
difiere del caso previo, ya que al estar el tiempo asociado con un momento (en
vez de una coordenada) involucra un operador integral que realiza una
transformaci\'{o}n de Fourier. Resulta llamativo que a pesar de que la idea de
un tiempo extr\'{\i}nseco en relatividad general no es reciente [\cite{york}],
no es usual que se lo utilice en el problema de la cuantificaci\'{o}n (entre
los pocos ejemplos existentes pueden mencionarse a \cite{k71,kt2}). El modelo
que hemos desarrollado podr\'{\i}a ser de ayuda para comprender como
implementar una cuantificaci\'{o}n basada en este esquema.

Finalmente, una caracter\'{\i}stica no expresada en ninguno de los sistemas
hasta ahora mencionados, es el hecho que la relatividad general contiene una
infinidad de v\'{\i}nculos super-Hamiltonianos (uno por cada punto del
espacio) con un \'{a}lgebra de v\'{\i}nculos no trivial entre ellos y los
supermomentos. Hasta la fecha, los \'{u}nicos modelos en dimensi\'{o}n finita
con covariancia general con varios v\'{\i}nculos super-Hamiltonianos se
formulaban en un fondo plano y no presentaban problemas de ordenamiento
[\cite{mrt99,lusanna,lusanna2}]. Consideramos pues, sumamente importante,
haber logrado resolver un modelo que permitiese estudiar el problema de
ordenamiento con m\'{a}s de un v\'{\i}nculo super-Hamiltoniano en
espacio-tiempos curvos. La aplicaci\'{o}n de los procedimientos desarrollados
previamente a este caso result\'{o} natural, y s\'{o}lo condicion\'{o} las
caracter\'{\i}sticas del volumen natural de la teor\'{\i}a a factorizar en
sectores excluyentes de las coordenadas. Mediante la transformaci\'{o}n de
cambio de escala de los super-Hamiltonianos se hallaron los operadores de
v\'{\i}nculo que respetan dicha invariancia mediante una transformaci\'{o}n
unitaria apropiada (procedimiento con el cual el \'{a}lgebra perdi\'{o} su
car\'{a}cter de cerrada). Los cambios de escala en este modelo pueden estar
asociados tanto a un tiempo intr\'{\i}nseco como a uno extr\'{\i}nseco,
dependiendo de las caracter\'{\i}sticas del potencial. La regularizaci\'{o}n
del producto interno requiere hallar una condici\'{o}n de gauge extra, que
resulta en la inserci\'{o}n de una funci\'{o}n delta que hace que el producto
interno f\'{\i}sico est\'{e} evaluado a un \'{u}nico tiempo.

De este modo, el problema de ordenamiento de los operadores de v\'{\i}nculo de
teor\'{\i}as con covariancia general ha sido estudiado en un amplio espectro
de modelos de dimensi\'{o}n finita, con resultados originales que pueden ser
un buen indicio de c\'{o}mo atacar el problema en la relatividad general.

\newpage

\thispagestyle{empty}

 ~\newpage


\appendix
\renewcommand{\theequation}{\thechapter.\arabic{equation}}

\fancyhead{} \fancyhead[LE]{\bf \thepage} \fancyhead[RE]{\sl Ap\'{e}ndice A}
\fancyhead[RO]{\bf \thepage} \fancyhead[LO]{\sl \'{A}lgebras de Grassmann}

\thispagestyle{empty}

\chapter{\'{A}lgebras de Grassmann}

\normalsize

~~~~ Un \'{a}lgebra de Grassmann $G_m$ tiene $m$ generadores $\eta^a$ que
satisfacen
\begin{equation}\label{cgras}
\eta^a\eta^b+\eta^b\eta^a=0 ~~~~~~~~~~({\mathrm en~ particular}~{\eta^a}^2=0).
\end{equation}
Decimos que las $\eta^a$ son {\it variables fermi\'{o}nicas} (anticonmutan).

$G_m$ es un espacio lineal de $2^n$ dimensiones, ya que cualquier elemento de
$G_m$ debe poder escribirse como combinaci\'{o}n lineal (a coeficientes
constantes) de: $1, \eta^1,...,\eta^m, \eta^1\eta^2, ...,$
$\eta^1\eta^2...\eta^m$, debido a que no puede repetirse un mismo generador en
ning\'{u}n monomio. Un elemento general de $G_m$ ser\'{a}
\begin{equation}\label{eg}
f=f_0+f_a\eta^a+f_{ab} \eta^a\eta^b + ... + f_{1...m}\eta^1...\eta^m
\end{equation}
donde puede verse que s\'{o}lo es relevante la parte antisim\'{e}trica de los
coeficientes.

Si en $f$ s\'{o}lo aparecen t\'{e}rminos con un n\'{u}mero par de generadores
diremos que $f$ es {\it bos\'{o}nico} o que tiene paridad de Grassmann igual a
cero: $\varepsilon(f)=0$. Si s\'{o}lo aparecen t\'{e}rminos con un n\'{u}mero
impar de generadores diremos que $f$ es {\it fermi\'{o}nico} o que tiene
paridad de Grassmann igual a uno: $\varepsilon(f)=1$. En el caso general, $f$
ser\'{a} una suma de un elemento par con uno impar.

Si $f_1$ y $f_2$ tienen paridad bien definida entonces:
$f_1f_2=(-1)^{\varepsilon_1\varepsilon_2}f_2f_1$

\subsubsection{Derivaci\'{o}n}

Saber derivar un elemento de $G_m$ es equivalente a saber derivar un monomio:
\begin{equation}
{\partial \over \partial \eta^a}(\eta^{a_1}...\eta^{a_i})= \delta^{a_1}_a
\eta^{a_2}...\eta^{a_i}-\delta^{a_2}_a \eta^{a_1}\eta^{a_3}...\eta^{a_i}+...+
(-1)^{i-1}\delta^{a_i}_a \eta^{a_1}...\eta^{a_{i-1}}
\end{equation}

De esta forma se define la derivada ``left'' como

\begin{equation}\label{derivleft}
\delta f=\delta \eta^a{\partial^L f\over \partial \eta^a}
\end{equation}

Se introduce adem\'{a}s la derivada ``right'' de manera que

\begin{equation}\label{derivright}
\delta f=\delta \eta^a{\partial^L f\over \partial \eta^a}={\partial^R f\over
\partial \eta^a}\delta \eta^a
\end{equation}

Ambas derivadas ser\'{a}n iguales s\'{o}lo para elementos fermi\'{o}nicos. En
general si $f$ tiene paridad bien definida:
\begin{equation}
{\partial^R f\over \partial \eta^a}=-(-1)^{\varepsilon(f)}{\partial^L f\over
\partial \eta^a}
\end{equation}

\medskip

\noindent
 \underline{Propiedades:}
\begin{description}

\item[(i)] Regla de la cadena
$${df(\eta^a(t))\over dt}={d\eta^a\over dt}{\partial f\over\partial \eta^a}$$
$${\partial f(\eta^{b'}(\eta^a))\over\partial \eta^a}={\partial \eta^{b'}
\over\partial \eta^a} {\partial f\over\partial \eta^{b'}}$$

\item[(ii)] Si $f_1$ y $f_2$ tienen paridad bien definida
$${\partial (f_1f_2)\over\partial \eta^a}={\partial f_1\over\partial
\eta^a}f_2 +(-1)^{\varepsilon(f_1)}f_1 {\partial f_2\over\partial \eta^a}$$

\item[(iii)] $${\partial \over\partial \eta^a}\left({\partial f\over\partial
\eta^b}\right)=-{\partial \over\partial \eta^b}\left({\partial f\over\partial
\eta^a}\right)$$

\end{description}

\subsubsection{Corchetes de Poisson generalizados}

A partir del estudio de una mec\'{a}nica que incluya variables fermi\'{o}nicas
y de su correspondencia con los conmutadores cu\'{a}nticos , se puede deducir
que la estructura de corchetes de Poisson resulta
\begin{equation}\label{cpg}
\{F,G\}=\left({\partial F\over\partial q^i}{\partial G\over\partial
p_i}-{\partial F\over\partial p_i}{\partial G\over\partial q^i}\right) +
(-1)^{\varepsilon_F} \left({\partial^L F\over\partial \eta^a}{\partial^L
G\over\partial {\cal P}_a}+{\partial^L F\over\partial {\cal P}_a}{\partial^L
G\over\partial \eta^a}\right)
\end{equation}

donde $(\eta^a,{\cal P}_a)$ son pares can\'{o}nicamente conjugados como puede
verificarse calculando el corchete entre ellos.

\medskip

\noindent \underline{Propiedades:}
\begin{itemize}
\item $\{F,G\}=-(-1)^{\varepsilon_F\varepsilon_G}\{G,F\}$
\item $\{F,G_1+G_2\}=\{F,G_1\}+\{F,G_2\}$
\item $\{F,G_1G_2\}=\{F,G_1\}G_2+(-1)^{\varepsilon_F\varepsilon_{G_1}}G_1
\{F,G_2\}$
\item $\varepsilon(\{F, G\})=\varepsilon_F+\varepsilon_G$
\item $\{F,G\}^{\ast}=-\{G^{\ast},F^{\ast}\}$
\item$\{\{F_1,F_2\},F_3\}+(-1)^{\varepsilon_{F_1}(\varepsilon_{F_2}+
\varepsilon_{F_3})}\{\{F_2,F_3\},F_1\}+(-1)^{\varepsilon_{F_3}
(\varepsilon_{F_1}+\varepsilon_{F_2})}\{\{F_3,F_1\},F_2\} =0. $ (Identidad de
Jacobi generalizada).
\end{itemize}

\newpage
\thispagestyle{empty}

~\newpage


\fancyhead{} \fancyhead[LE]{\bf \thepage} \fancyhead[RE]{\sl Ap\'{e}ndice B}
\fancyhead[RO]{\bf \thepage}\fancyhead[LO]{\sl Demostraci\'{o}n de $\hat\Omega^2=0$:
Modelos con un super-Hamiltoniano}

\chapter{Demostraci\'{o}n de $\hat\Omega^2=0$: Modelos con un super-Hamiltoniano}

\normalsize

~~~~ En este ap\'{e}ndice, demostraremos la nilpotencia del operador BRST para
el caso de sistemas con un v\'{\i}nculo super-Hamiltoniano cuyo potencial es
invariante de gauge (tipo 1) (la demostraci\'{o}n para el caso con potencial
definido positivo (tipo 2) no es necesaria, pues el operador BRST en ese caso
es obtenido a trav\'{e}s de una transformaci\'{o}n unitaria del primero,
preservando por lo tanto la nilpotencia).

Proponiendo los ordenamientos para las partes lineal y cuadr\'{a}tica de
$\hat\Omega=\hat{\Omega}^{cuad}+\hat{\Omega}^{lineal}$:

\begin{equation}\label{aordlin}
\hat{\Omega}^{lineal}=\sum_{i=-1}^0f^{1\over   2}\hat\Omega^{c_i} \hat{\cal
P}_{c_i}f^{-{1\over 2}},
\end{equation}

\begin{equation}\label{aordcuad}
\hat{\Omega}^{cuad}={1\over 2} \sum_{k,j=-1}^0 f^{-{1\over  2}}\hat{\cal
P}_{a_k} f \hat\Omega^{a_kb_j} \hat{\cal P}_{b_j}f^{-{1\over
2}}+\hat\eta^o\lambda,
\end{equation}

queremos demostrar que
\begin{equation}\label{nilpoa}
[\hat{{\Omega}},\hat{{\Omega}}]=[\hat\Omega^{cuad},
\hat\Omega^{cuad}]+2[\hat\Omega^{cuad},\hat\Omega^{lineal}]+
[\hat\Omega^{lineal},\hat\Omega^{lineal}]=0.
\end{equation}

El t\'{e}rmino $[\hat\Omega^{cuad},\hat\Omega^{cuad}]$ se anula trivialmente
porque ${\eta^o}^2=0$ (n\'{o}tese que $\hat\Omega$ no depende de $\hat{\cal
P}_o$). El \'{u}ltimo t\'{e}rmino es cero porque $\hat\Omega^{lineal}$ es
efectivamente nilpotente. Por lo tanto, s\'{o}lo tenemos que demostrar que el
ordenamiento propuesto satisface $[\hat\Omega^{cuad},\hat\Omega^{lineal}]=0$.

Para los desarrollos ulteriores es necesario conocer las paridades de los
objetos geom\'{e}tricos intervinientes:
\begin{eqnarray}
\varepsilon (\hat\eta^{a_k})=k , &  \varepsilon (\Pb)=j, & \varepsilon (\oc)=i
\\ \varepsilon(\oab)=k+j+1,  & \varepsilon\ds{(\frac{\pl\oc}{\pb})}=j+i+1, &
\varepsilon (\frac{\pl\ocb}{\pa})=i+j+k.
\end{eqnarray}

Luego, calculando expl\'{\i}citamente
\begin{eqnarray}\label{ac}
\lefteqn{[\hat\Omega^{cuad},\hat\Omega^{lineal}]=
\fm\ds\sum_{kj}\frac{1}{2}f^{-1}\Pa \oab\Pb\sum_i\oc\Pc\fmm+}\nonumber\\&&+
\fm\sum_{i}\oc\Pc\sum_{kj}f^{-1}\Pa f\oab\Pb\fmm=
\fm\ds[\sum_{kj}\frac{1}{2}\fmm\Pa f\oab\Pb, \sum_i\oc\Pc]\fmm=\nonumber\\
&&=\fm[\sum_{kj}\frac{1}{2}\oab\Pa\Pb-\ds\frac{i}{2}\sum_{kj}\frac{\pl(\ln
f)}{\pa}\oab\Pb-\frac{i}{2}\sum_{kj}\frac{\pl\oab}{\pa},\sum_i\oc\Pc]\fmm
\end{eqnarray}

Desarrollando cada t\'{e}rmino,
\begin{eqnarray}\label{e1}
\lefteqn{[\sum_{kj}\frac{1}{2}\fmm\oab\Pa\Pb,\sum_i\oc\Pc]=}\nonumber\\&&=
-\frac{1}{2}\oab\frac{{\pl}^2\oc}{\pab}\Pc
-\frac{i}{2}(-1)^{j(k+1)}\oab\frac{\pl\oc}{\pb}\Pc\Pa\nonumber\\&&
-\frac{i}{2}(-1)^{j+1}\oab\frac{\pl\oc}{\pa}\Pc\Pa
-\frac{i}{2}\oc\frac{\pl\oab}{\pc}\Pa\Pb
\end{eqnarray}

\begin{equation}\label{e2}
[-\frac{i}{2}\sum_{kj}\frac{\pl\oab}{\pa},\oc\Pc]=-\frac{1}{2}\frac{\pl\oab}
{\pa}\frac{\pl\oc}{\pb}\Pc-\frac{1}{2}\oc\frac{{\pl}^2\oab}{\pc\pa}\Pb
\end{equation}

\begin{eqnarray}\label{e3}
\lefteqn{[-\frac{i}{2}\sum_{kj}\frac{\pl(\ln f)}{\pa}\oab\Pb,\oc\Pc]=}
\nonumber\\&&=-\frac{1}{2}\sum_{ikj}\frac{\pl\ln
f}{\pa}\oab\frac{\pl\oc}{\pb}\Pc
-\frac{1}{2}\sum_{ikj}\oc\frac{\pl}{\pc}[\frac{\pl(\ln f)}{\pa}\oab]\Pb
\end{eqnarray}

El desarrollo del conmutador tendr\'{a} t\'{e}rminos cuadr\'{a}ticos y
lineales en los momentos $\Pa$ que deber\'{a}n anularse por separado. Los
t\'{e}rminos cuadr\'{a}ticos provienen todos de (\ref{e1}):
\[ \sum_{ijk}[-\frac{i}{2}(-1)^{j(k+1)}\oab\frac{\pl\oc}{\pb}\Pc\Pa
-\frac{i}{2}(-1)^{j+1}\oab\frac{\pl\oc}{\pa}\Pc\Pa
-\frac{i}{2}\oc\frac{\pl\oab}{\pc}\Pa\Pb], \] su anulaci\'{o}n
est\'{a}garantizada pues se corresponden exactamente a los que surgen de
calcular el corchete cl\'{a}sico $\{\Omega,\Omega\}$, que se anula por
construcci\'{o}n. N\'{o}tese que los dos primeros t\'{e}rminos son iguales; en
efecto, como $\Omega^{a_ob_o}=0$ entonces $(-1)^{jk}\oab=(-1)^{j+k+1}\oab$.
Luego, permutando los roles de $a_k$ y $b_j$ (\'{\i}ndices mudos) en el primer
t\'{e}rmino:
\begin{equation}
\sum_{ikj}-i[(-1)^{j+1}\oab\frac{\pl\oc}{\pa}+\frac{1}{2}\oa\frac{\pl\ocb}
{\pa}]\Pc\Pb=0.
\end{equation}
Es decir que debe anularse la parte sim\'{e}trica (o antisim\'{e}trica si
$i=j=0$) del corchete:
\begin{equation}\label{anucuad}
\sum_{k}[(-1)^{j+1}\oab\frac{\pl\oc}{\pa}+(-1)^{i+1}(-1)^{(j+1)(i+1)}
\oac\frac{\pl\ob}{\pa}+ \oa\frac{\pl\ocb}{\pa}]=0.
\end{equation}

Los t\'{e}rminos restantes en (\ref{e1}),(\ref{e2}) y (\ref{e3}) son todos
lineales:
\begin{eqnarray}\label{tlin}
\lefteqn{-\frac{1}{2}\sum_{ikj}[\oab\frac{{\pl}^2\oc}{\pab}+
\frac{\pl\oab}{\pa} \frac{\pl\oc}{\pb}+ \oa\frac{{\pl}^2\obc}{\pab}
+}\nonumber\\&&+ \frac{\pl\ln f} {\pa}\oab\frac{\pl\oc}{\pb}+
\oa\frac{\pl}{\pa}(\frac{\pl(\ln f)} {\pb}\obc)]\Pc
\end{eqnarray}

Debemos probar que la suma sobre los \'{\i}ndices $kj$ es cero. Con este fin,
multipliquemos la relaci\'{o}n hallada a partir de los t\'{e}rminos
cuadr\'{a}ticos, Ec. (\ref{anucuad}), por $(-1)^{j+1}$, derivando luego
respecto de $\eta^{b_j}$ y sumando sobre $j$, obtenemos:
\begin{eqnarray}\label{cualin}
\lefteqn{\sum_{kj}[\frac{\pl\oab}{\pb}\frac{\pl\oc}{\pa}
+(-1)^{(k+j+1)(j+1)}\oab\frac{{\pl}^2\oc}{\pba}
+(-1)^{(ij+1)}\frac{\pl\oac}{\pb}\frac{\pl\ob}{\pa}+}
\nonumber\\&&+(-1)^{ij+1+(k+i+1)(j+1)}\oac\frac{{\pl}^2\ob}{\pba}
+(-1)^{j+1}\frac{\pl\oa}{\pb}\frac{\pl\ocb}{\pa}\nonumber\\
&&+(-1)^{(j+1)+k(j+1)}\oa\frac{{\pl}^2\ocb}{\pba}]=0
\end{eqnarray}
Veamos que el tercer y quinto t\'{e}rmino de esta expresi\'{o}n se cancelan
mutuamente. En efecto, permutando $a_k$ con $b_j$ en el tercero:
\[(-1)^{ik+1}\frac{\pl\obc}{\pa}\frac{\pl\oa}{\pb}=
(-1)^{ik+1)(-1)^{(k+j+i}(k+j+1)} \frac{\pl\oa}{\pb}\frac{\pl\obc}{\pa}=
(-1)^{j}\frac{\pl\oa}{\pb}\frac{\pl\ocb}{\pa}\]

Permutando $a_k$ con $b_j$ en el resto de los t\'{e}rminos, obtenemos:
\begin{equation}
\sum_{kj}[\frac{\pl\oab}{\pa}\frac{\pl\oc}{\pb}+\oab\frac{{\pl}^2\oc}{\pab}
+(-1)^{i+1}\oac\frac{{\pl}^2\ob}{\pab}+\oa\frac{{\pl}^2\ocb}{\pab}]=0,
\end{equation}
y vemos que aparecen los tres primeros t\'{e}rminos lineales de (\ref{tlin}),
que resultan iguales a:
\begin{eqnarray}\label{cuat}
\lefteqn{-\sum_{kj}(-1)^{i+1}\oac\frac{{\pl}^2\ob}{\pab}=-\sum_k
(-1)^{i+1}\oac\frac{\pl}{\pa}(\sum_j\frac{\pl\ob}{\pb})}\nonumber\\&&=
\sum_{kj}(-1)^{i+1}\oac[\frac{\pl\ob}{\pa}\frac{\pl (\ln f)}{\pb}+
\frac{{\pl}^2 (\ln f)}{\pba}\ob]
\end{eqnarray}
Donde la \'{u}ltima igualdad surge de considerar que
$\sum_j\frac{\pl\ob}{\pb}=\eta^b(\xi^j_{b,j}-C^{a}_{ba})$, y que, seg\'{u}n se
demostr\'{o} en \S 4.1, $C^{a}_{ba}=f^{-1}(f\xi^j_b)_{,j}$, con lo cual
$\sum_{j}\frac{\pl\ob}{\pb}=\eta^b\xi^j_b(\ln f)_{,j}=\sum_j\ob\frac{\pl(\ln
f)}{\pb}$.

Adem\'{a}s, n\'{o}tese que debido al factor $\frac{\pl (\ln f)}{\pb}$, el
\'{u}nico valor que contribuye en la relaci\'{o}n (\ref{cuat}) es $j=-1$.
Luego, de la condici\'{o}n (\ref{anucuad}) con $j=-1$ tenemos:
\[\sum_{k}(-1)^{i+1}\oac\frac{\pl\ob}{\pa}=-\sum_k[\oab\frac{\pl\oc}{\pa} +
\oa\frac{\pl\ocb}{\pa}] \]
Finalmente
\begin{eqnarray}\label{condf}
\lefteqn{\sum_{kj}(-1)^{i+1}\oac\frac{{\pl}^2\ob}{\pab}}\nonumber\\ &&
=\sum_{kj}[(\oab\frac{\pl\oc}{\pa} + \oa\frac{\pl\ocb}{\pa})\frac{\pl (\ln
f)}{\pb}+ (-1)^{i} \oac\frac{\pl (\ln f)}{\pba}\ob].
\end{eqnarray}

N\'{o}tese que  debido al factor $\frac{\pl (\ln f)}{\pba}$, el \'{u}ltimo
t\'{e}rmino s\'{o}lo es no nulo si $j=k=-1$. Como \'{u}ltimo paso,
desarrollando y reescribiendo apropiadamente los dos \'{u}ltimos t\'{e}rminos
lineales de (\ref{tlin}): \goodbreak
\begin{eqnarray}\label{fin}
\lefteqn{\sum_{kj}[\frac{\pl(\ln
f)}{\pa}\oab\frac{\pl\oc}{\pb}+\oa\frac{\pl}{\pa}(\frac{\pl(\ln
f)}{\pb}\obc)]} \nonumber\\ &&=\sum_{kj}[\oab\frac{\pl\oc}{\pa}\frac{\pl (\ln
f)}{\pb}+\oa\frac{\pl\obc}{\pa}\frac{\pl(\ln
f)}{\pb}+(-1)^i\obc\frac{{\pl}^2(\ln f)}{\pba}\oa]
\end{eqnarray}
que se cancelan exactamente con (\ref{condf}), lo cual completa la
demostraci\'{o}n de la nilpotencia de $\hat\Omega$. \hfill $\Box$

\newpage

\thispagestyle{empty}

 ~\newpage


\renewcommand{\baselinestretch}{1}

\small
\fancyhead{} \fancyhead[LE]{\bf \thepage} \fancyhead[RE]{\sl Bibliograf\'{\i}a}
\fancyhead[RO]{\bf \thepage}

\addcontentsline{toc}{chapter}{Bibliograf\'{\i}a}

\renewcommand{\baselinestretch}{1.5}

\newpage


\renewcommand{\baselinestretch}{1}

\addcontentsline{toc}{chapter}{Agradecimientos}

\noindent {\Huge{\bf Agradecimientos}}

\thispagestyle{empty}

\bigskip
\bigskip

\normalsize

Einstein sosten\'{\i}a que la \'{u}nica sociedad de la cual se
enorgullec\'{\i}a de pertenecer era la de ``los buscadores de la verdad''; la
cual, agregaba, tiene muy pocos miembros en cualquier \'{e}poca. Sin duda
alguna, el Dr. Rafael Ferraro es un miembro destacado de ella. Esto unido a
sus extraordinarias cualidades humanas e intelectuales, hacen que me sienta
sumamente afortunado de ser su disc\'{\i}pulo y haber podido desarrollar este
trabajo de investigaci\'{o}n bajo su direcci\'{o}n.

Esta tesis se vi\'{o} favorecida por esclarecedoras conversaciones con Marc
Henneaux. Como as\'{\i} tambi\'{e}n por las estad\'{\i}as de trabajo que
realice en: Tata Institute of Fundamental Research, Bombay, India, durante
diciembre de 1997, en donde debo agradecer la hospitalidad y calidez del Prof.
P. Joshi y del Dr. Sanjay Shingan (quien adem\'{a}s me posibilit\'{o} el
acceso a bibliograf\'{\i}a de sumo inter\'{e}s); Istituto di Fisica Nucleare,
Sezione di Firenze, Italia, durante octubre de 1999, en donde pude mantener
provechosas discusiones con el Prof. Luca Lusanna, quien me brind\'{o}
tambi\'{e}n valiosa bibliograf\'{\i}a.

Este trabajo no hubiese sido posible sin el financiamiento brindado por la
Universidad de Buenos Aires y la Fundaci\'{o}n Antorchas.

No puedo dejar de mencionar al Dr. Fabian Gaioli, Lic. Marc Thibeault y el
Lic. Edgardo Garc\'{\i}a Alvarez, por las interesantes discusiones
cient\'{\i}ficas que mantuvimos y por el excelente clima de trabajo que
siempre propiciaron en el IAFE. Tampoco puedo olvidar, en este sentido al Lic.
Gerardo Milesi, y a la ``nueva generaci\'{o}n'' de doctorandos, Lic.
Gast\'{o}n Giribet, Lic. Mauricio Leston y Lic. Gabriel Catren.

A mi vida le faltar\'{\i}a color si no fuese por la compa\~{n}ia de un grupo
muy especial de amigos: Lic. Claudio Simeone, Lic. Hern\'{a}n Decicco, Lic.
Enzo Speranza, Lic. Daniel Rodriguez Sierra y particularmente Fernando Camelli
(por su presencia a pesar de la distancia).

Quiero agradecer muy especialmente al (pr\'{a}cticamente) Dr. Fabio Kalesnik,
con quien compartimos ideales, buenos y malos momentos.

A mis amigos: Leonardo Shimizukawa, Fernando Alonso, Adri\'{a}n y Carina
Otero, Santiago Beluardi, por todo lo que puede asociarse con la palabra {\it
amistad}.

A Eliana, por hacer que mis proyectos sean tambi\'{e}n los suyos, por ayudarme
a ser una mejor persona, por todo su {\it amor}.

\end{document}